\documentclass[smallextended, final]{svjour3}

\usepackage[latin1]{inputenc}
\usepackage[T1]{fontenc}
\usepackage[english]{babel}

\usepackage[usenames,dvipsnames,table]{xcolor}
\usepackage{hyperref}
\hypersetup{bookmarksnumbered=true,  
            hypertexnames=false,     
            colorlinks=true,
            citecolor=black,
            linkcolor=black,
            breaklinks,
            pdfborder=0 0 0,
            pdftitle=Bayesian multi-ojective optimization,
            pdfauthor=Paul Feliot; Julien Bect; Emmanuel Vazquez
            pdfkeywords=}
\usepackage{array, multirow, enumerate}
\usepackage{dsfont}  \let \mathbb=\mathds
\usepackage{amsmath, amssymb, amsfonts, amsmath, bm}
\usepackage{graphicx, tikz}
\usepackage{afterpage}
\graphicspath{{./Figures/}}
\usetikzlibrary{positioning,calc,patterns}
\usepackage[round]{natbib}
\usepackage[font={small},labelfont={bf}]{caption}
\usepackage[ruled,vlined,linesnumbered,english]{algorithm2e}
\DontPrintSemicolon
\renewcommand{\hat}{\widehat}

\newcommand{\X}{\mathbb{X}}
\newcommand{\Y}{\mathbb{Y}}
\newcommand{\B}{\mathbb{B}}
\newcommand{\R}{\mathbb{R}}
\renewcommand{\P}{\mathbb{P}}
\newcommand{\XX}{\mathcal{X}}
\newcommand{\YY}{\mathcal{Y}}
\newcommand{\E}{\mathbb{E}}
\newcommand{\one}{\mathds{1}}

\newcommand \PP   {\mathcal{P}}
\newcommand \PPt  {\PP^\star}       
\newcommand \tiPP {\widetilde \PP}  

\newcommand \EI     {\rho}
\newcommand \EIFeas {\EI^{\text{feas}}}
\newcommand \EIUnf  {\EI^{\text{unf}}}

\newcommand \dy {\mathrm{d}y}
\DeclareMathOperator \argmax {{\rm argmax}}
\DeclareMathOperator \argmin {{\rm argmin}}
\DeclareMathOperator \Pareto {Pareto}

\newcommand \Xv  {\underline{X\mskip -1.5mu} \mskip 1mu}

\newcommand \obj  {{\rm o}}
\newcommand \Yo   {\Y_{\obj}}
\newcommand \Bo   {\B_{\obj}}

\newcommand \cons  {{\rm c}}
\newcommand \Yc    {\Y_{\cons}}

\newcommand \yanchor {y_{\rm anchor}}

\newcommand \refHV {{\rm ref}}  
\newcommand \try   {{\rm try}}
\newcommand \upp   {{\rm upp}}

\newcommand \Bc    {\B_{\cons}}
\newcommand \BcNeg {\Bc^-}

\newcommand \xic  {\xi_{\cons}}
\newcommand \xio  {\xi_{\obj}}

\newcommand \yc   {y_{\cons}}
\newcommand \yo   {y_{\obj}}

\newcommand \dyc   {\mathrm{d}\yc}

\newcommand \yUpp {y^{\text{upp}}}
\newcommand \yLow {y^{\text{low}}}

\newcommand \lambdaObj  {\lambda_\obj}
\newcommand \lambdaCons {\lambda_\cons}

\newcommand \Hno {H_{n, \obj}}
\newcommand \Hnc {H_{n, \cons}}

\journalname{Author-generated version. %
  Published under DOI:\texttt{10.1007/s10898-016-0427-3} by Springer}

\begin{document}

\title{%
  A Bayesian approach to constrained single-
  and multi-objective optimization}

\author{%
  Paul FELIOT \and
  Julien BECT \and
  Emmanuel VAZQUEZ}

\institute{%
  Paul \textsc{Feliot} (Ph.D. student) --  \email{paul.feliot@irt-systemx.fr}
  \at Institut de Recherche Technologique SystemX, Palaiseau, France.\\
  Julien \textsc{Bect} \& Emmanuel  \textsc{Vazquez} --
  \email{firstname.lastname@centralesupelec.fr}
  \at Laboratoire des Signaux et Systèmes, Gif-sur-Yvette, France
}

\date{}

\maketitle

\begin{abstract}
  This article addresses the problem of derivative-free (single- or
  multi-objective) optimization subject to multiple inequality constraints. Both
  the objective and constraint functions are assumed to be smooth, non-linear
  and expensive to evaluate. As a consequence, the number of evaluations that
  can be used to carry out the optimization is very limited, as in complex
  industrial design optimization problems. The method we propose to overcome
  this difficulty has its roots in both the Bayesian and the multi-objective
  optimization literatures.  More specifically, an extended domination rule is
  used to handle objectives and constraints in a unified way, and a corresponding expected
  hyper-volume improvement sampling criterion is proposed. This new criterion
  is naturally adapted to the search of a feasible point when none is available,
  and reduces to existing Bayesian sampling criteria---the classical
  Expected Improvement (EI) criterion and some of its constrained/multi-objective
  extensions---as soon as at least one feasible point is available.
  The calculation and optimization of the criterion are performed using
  Sequential Monte Carlo techniques.  In particular, an algorithm similar to the
  subset simulation method, which is well known in the field of structural
  reliability, is used to estimate the
  criterion. The method, which we call BMOO (for Bayesian Multi-Objective
  Optimization), is compared to state-of-the-art algorithms for single-
  and multi-objective constrained optimization.

\end{abstract}

\section{Introduction}

This article addresses the problem of derivative-free multi-objective
optimization of real-valued functions subject to multiple inequality
constraints. The problem consists in finding an approximation of the set
\begin{equation}
\label{eq:Gamma}
  \Gamma = \{x \in \X : c(x) \leq 0 \text{ and } \nexists\, x'\in\X
  \text{\ such that\ } c(x') \leq 0 \text{ and } f(x') \prec f(x) \}  
\end{equation}
where $\X \subset \R^d$ is the search domain, $c = (c_i)_{1 \leq i \leq q}$ is a
vector of constraint functions ($c_i:\X\to\R$), $c(x) \leq 0$ means that
$c_{i}(x) \leq 0$ for all $1\leq i \leq q$, $f = (f_j)_{1 \leq j \leq p}$ is a
vector of objective functions to be minimized ($f_j:\X\to\R$), and $\prec$
denotes the Pareto domination rule \citep[see,
e.g.,][]{fonseca1998multiobjective}.  Both the objective functions~$f_j$ and the
constraint functions~$c_i$ are assumed to be continuous. The search domain~$\X$
is assumed to be compact---typically, $\X$ is a hyper-rectangle defined by
bound constraints.  Moreover, the objective and constraint functions are
regarded as black boxes and, in particular, we assume that no gradient
information is available.  Finally, the objective and the constraint functions
are assumed to be expensive to evaluate, which arises for instance when the
values $f(x)$ and~$c(x)$, for a given~$x \in \X$, correspond to the outputs of a
computationally expensive computer program. In this setting, the emphasis is on
building optimization algorithms that perform well under a very limited budget
of evaluations (e.g., a few hundred evaluations).

We adopt a Bayesian approach to this optimization problem. The
essence of Bayesian optimization is to choose a prior model for the
expensive-to-evaluate function(s) involved in the optimization
problem---usually a Gaussian process model \citep{santner2003design,
  williams2006gaussian} for tractability---and then to select the
evaluation points sequentially in order to obtain a small average error
between the approximation obtained by the optimization algorithm and the
optimal solution, under the selected prior. See, e.g.,
\cite{kushner1964new}, \cite{mockus75}, \cite{mockus78}, \cite{archetti79}
and~\cite{mockus1989bayesian} for some of the earliest references in
the field. Bayesian optimization research was first focused
on the case of single-objective bound-constrained optimization:
the Expected Improvement~(EI) criterion~\citep{mockus78,
  jones1998efficient} has emerged in this case as one of the most
popular criteria for selecting evaluation points. Later, the EI
criterion has been extended to handle constraints
\citep{schonlau1998global, sasena2002exploration, gramacy2011,
  gelbart2014bayesian, gramacy2015modeling} and to address bound-constrained
multi-objective problems \citep{emmerich2006single, jeong2006optimization,
  wagner2010expected, svenson2010multiobjective}.

The contribution of this article is
twofold. The first part of the contribution is  the proposition
of  a new sampling criterion that  handles 
multiple objectives and non-linear constraints simultaneously. This criterion
corresponds to a one-step look-ahead Bayesian strategy, using the dominated
hyper-volume as a utility function \citep[following in this
respect][]{emmerich2006single}. More specifically, the dominated hyper-volume is defined using an
extended domination rule, which handles objectives and constraints in a unified way
\citep[in the spirit of][]{fonseca1998multiobjective,
  ray2001multiobjective, oyama2007new}.
This new criterion is naturally adapted to the search of
a feasible point when none is available, and several criteria from the
literature---the EI criterion and some of its
constrained/multi-objective extensions---are recovered as
special cases when at least one feasible point is known.
The second part of the contribution lies in the numerical
methods employed to compute and optimize the sampling criterion. Indeed, this
criterion takes the form of an integral over the space of constraints and
objectives, for which no analytical expression is available in the general
case. Besides, it must be optimized at each iteration of the algorithm to
determine the next evaluation point. In order to compute the integral, we use an
algorithm similar to the subset simulation method \citep{au2001estimation,
  cerou2012sequential}, which is a well known Sequential Monte Carlo~(SMC)
technique \citep[see][and references therein] {del2006sequential, liu2008monte}
from the field of structural reliability and rare event estimation.
For the optimization of the criterion, we resort to an SMC method as well,
following earlier work by
\cite{benassi2012bayesian} for single-objective bound-constrained
problems. The resulting algorithm is called BMOO (for Bayesian
multi-objective optimization).

The structure of the article is as follows. In
Section~\ref{sec:background}, we recall the framework of Bayesian
optimization based on the expected improvement sampling criterion,
starting with the unconstrained single-objective setting.
Section~\ref{sec:method} presents our new sampling criterion for
constrained multi-objective optimization.  The calculation and the
optimization of the criterion are discussed in Section~\ref{sec:smc}.
Section~\ref{sec:exp} presents experimental results. An illustration on
a two-dimensional toy problem is proposed for visualization
purpose. Then, the performances of the method are compared to those of
reference methods on both single- and multi-objective
constrained optimization problems from the literature. Finally, future
work is discussed in Section~\ref{sec:conclusion}.

\section{Background literature}
\label{sec:background}

\subsection{Expected Improvement}
\label{sec:expected-improvement}

Consider the single-objective unconstrained optimization problem
\begin{equation*}
  x^{\star} = \argmin_{x\in\X} f(x)\,,
\end{equation*}
where $f$ is a continuous real-valued function defined over $\X\subset
\R^d$. Our objective is to find an approximation of $x^{\star}$ using a sequence
of evaluation points $X_1,\, X_2,\, \ldots \in\X$. Because the choice of a new
evaluation point $X_{n+1}$ at iteration $n$ depends on the evaluation results of
$f$ at $X_1,\,\ldots,\,X_n$, the construction of an optimization strategy $\Xv :
f \mapsto (X_1,\,X_2,\, X_3\ldots)$ is a sequential decision problem.

The Bayesian approach to this decision problem originates from the early work
of~\cite{kushner1964new} and~\cite{mockus78}. Assume that a loss
function~$\varepsilon_n(\Xv, f)$ has been chosen to measure the performance of
the strategy~$\Xv$ on~$f$ after~$n$ evaluations, for instance the classical
loss function
\begin{equation}
  \label{eq:loss-fun-EI}
  \varepsilon_n(\Xv, f) = m_n - m\,,
\end{equation}
with $m_n = f(X_1) \wedge \cdots \wedge f(X_n)$ and $m = \min_{x\in\X} f(x)$.
Then, a good strategy in the Bayesian sense is a strategy that achieves,
on average, a small value of $\varepsilon_{n}(\Xv, f)$ when $n$ increases, where
the average is taken with respect to a stochastic process model~$\xi$ (defined
on a probability space $(\Omega, \mathcal{A}, \P_0)$, with parameter in $\X$)
for the function~$f$. In other words, the Bayesian approach assumes that~$f =
\xi(\omega,\cdot)$ for some $\omega\in\Omega$. The probability distribution
of~$\xi$ represents prior knowledge about the function~$f$---before actual
evaluations are performed. The reader is referred to \cite{vazquez2014new} for
a discussion of other possible loss functions in the context of Bayesian
optimization.

Observing that the Bayes-optimal strategy for a budget of~$N$ evaluations is
intractable for $N$ greater than a few units, \cite{mockus78} proposed to use a
one-step look-ahead strategy (also known as a myopic strategy). Given~$n < N$
evaluation results, the next evaluation point~$X_{n+1}$ is chosen in order to
minimize the conditional expectation of the future loss~$\varepsilon_{n+1}(\Xv,
\xi)$ given available evaluation results:
\begin{eqnarray}
  \label{eq:oneStepLookAhead}
  X_{n+1} &=& \argmin_{x\in\X} \E_n\bigl( \varepsilon_{n+1}(\Xv, \xi)
  \mid X_{n+1}=x \bigr)\,,
\end{eqnarray}
where $\E_{n}$ stands for the conditional expectation with respect to $X_1,\,
\xi(X_1),\,\ldots,\,X_n,\,\xi(X_n)$. Most of the work produced in the field of
Bayesian optimization since then has been focusing, as the present paper will,
on one-step look-ahead (or similar) strategies\footnote{
  \citet[][Section~2.5]{mockus1989bayesian} heuristically introduces a
  modification of~\eqref{eq:oneStepLookAhead} to compensate for the fact that
  subsequent evaluation results are not taken into account in the myopic
  strategy and thus enforce a more global exploration of the search domain. In
  this work, we consider a purely myopic strategy as in
  \cite{jones1998efficient}.}; the reader is referred to~\cite{GinsbLeRiche2009}
and~\cite{benassi2013nouvel} for discussions about two-step look-ahead
strategies.

When \eqref{eq:loss-fun-EI} is used as a loss function, the right-hand side
of~\eqref{eq:oneStepLookAhead} can be rewritten as
\begin{eqnarray}
  \argmin \E_n\bigl( \varepsilon_{n+1}(\Xv, \xi) 
  \mid X_{n+1}=x \bigr)
  &=& \argmin \E_n \left( m_{n+1}       \bigm| X_{n+1}=x\right)
  \nonumber \\
  &=& \argmax  \E_n \bigl( (m_n - \xi(x))_{+} \bigr)\,,
\end{eqnarray}
with $z_+ = \max\left(z,\, 0\right)$. The function
\begin{equation}
  \label{eq:EI_mono}
  \EI_n(x) : x \mapsto \E_n \bigl( (m_n - \xi(x))_{+} \bigr)  
\end{equation}
is called the Expected Improvement~(EI) criterion
\citep{schonlau1998global, jones1998efficient}.
When $\xi$ is a Gaussian process with known mean and covariance
functions,  $\EI_n(x)$ has a closed-form expression:
\begin{equation}
  \label{eq:rho-n-rewritten}
  \EI_n(x) \;=\;
  \gamma \left( \, m_n -  \hat \xi_n(x) ,\, \sigma_n^2(x) \, \right),
\end{equation}
where
\begin{equation*}
  \label{eq:gamma-def}
  \gamma(z,s) \;=\;
  \begin{cases}
    \sqrt{s}\, \varphi \left( \frac{z}{\sqrt{s}} \right) + z\, \Phi \left(
      \frac{z}{\sqrt{s}} \right)
    & \text{if } s > 0, \\
    \max \left( z, 0 \right) & \text{if } s = 0,
  \end{cases}
\end{equation*}\\[0.5em]
with  $\Phi$ standing  for the normal cumulative distribution function,
$\varphi = \Phi^{\prime}$ for the normal probability density function,
$\hat \xi_n(x) = \E_n \left( \xi(x) \right)$ for the kriging
predictor at~$x$ (the posterior mean of~$\xi(x)$ after $n$~evaluations) and
$\sigma_n^2(x)$ for the kriging variance at~$x$ (the posterior variance
of~$\xi(x)$ after $n$~evaluations). See, e.g., the books of~\cite{stein:99},
\cite{santner2003design}, and~\cite{williams2006gaussian} for more information on Gaussian
process models and kriging (also known as Gaussian process interpolation).

Finally, observe that the one-step look-ahead
strategy~\eqref{eq:oneStepLookAhead} requires to solve an auxiliary global
optimization problem on~$\X$ for each new evaluation point to be selected. The
objective function~$\EI_n$ is rather inexpensive to evaluate when~$\xi$ is a
Gaussian process, using~\eqref{eq:rho-n-rewritten}, but it is typically severely
multi-modal. A simple method to optimize~$\EI_n$ consists in choosing a fixed
finite set of points that covers~$\X$ reasonably well and then performing a discrete
search. Recently, sequential Monte Carlo techniques \citep[see][and
references therein]{del2006sequential, liu2008monte} have been shown to be a
valuable tool for this task \citep{benassi2012bayesian}. A review of other
approaches is provided in the PhD thesis
of~\citet[][Section~4.2]{benassi2013nouvel}.

\subsection{EI-based  multi-objective optimization without constraints}
\label{sec:ei-multi}

We now turn to the case of unconstrained multi-objective optimization.  Under
this framework, we consider a set of objective functions $f_j:\X\to\R$, $j =
1,\, \ldots,\, p$, to be minimized, and the objective is to build an
approximation of the Pareto front and of the set of corresponding
solutions
\begin{equation} 
  \Gamma = \{x \in \X : \nexists\, x'\in\X \text{\ such that \ }
  f(x') \prec f(x) \}\,, 
\end{equation}
where $\prec$ stands for the Pareto domination rule defined by
\begin{equation}
  \label{eq:pareto-dom}
  y=(y_1,\,\ldots,\,y_p) \prec z=(z_1,\,\ldots,\,z_p) \Longleftrightarrow \left\{
    \begin{array}{l  l}
      \forall i \le p, &\; y_i \leq z_i\,, \\[0.6em]
      \exists j \le p, &\; y_j < z_j\,.
    \end{array}
  \right.  
\end{equation}
Given evaluation results $f(X_1) = (f_1(X_1),\,\ldots,\,f_p(X_1))$, $\ldots$,
$f(X_n) = (f_1(X_n),\,\ldots,\,f_p(X_n))$, define
\begin{equation}
  H_n = \{y \in \B; \exists i\le n,\, f(X_i) \prec y\}\,,
\end{equation}
where $\B \subset \R^p$ is a set of the form~$\B = \left\{ y \in \R^p;\; y \le
  \yUpp \right\}$ for some~$\yUpp \in \R^p$, which is introduced to
ensure that the volume of~$H_n$ is finite.  $H_n$ is the subset of $\B$ whose
points are dominated by the evaluations.

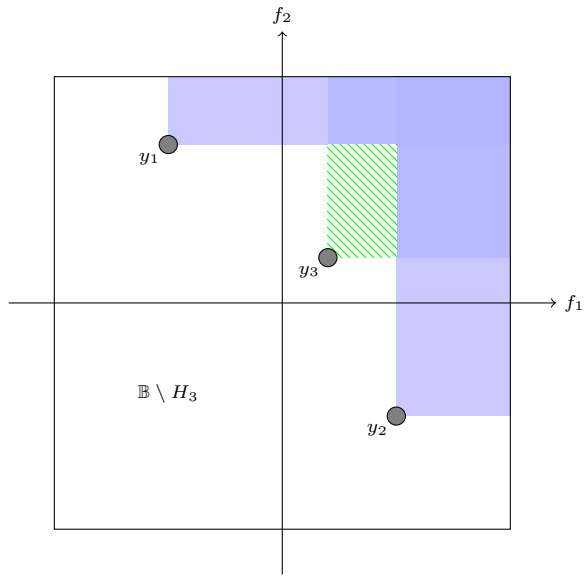
\begin{figure}[]
  \begin{center}
    \begin{tikzpicture}[scale=3]
    \fill[fill=blue!30, opacity=0.7] (-0.5,0.7) rectangle (1,1);
    \filldraw[fill=black!50] (-0.5,0.7) circle (0.04) node [below left] {\scriptsize$y_1$};

    \fill[fill=blue!30, opacity=0.7] (0.5,-0.5) rectangle (1,1);
    \filldraw[fill=black!50] (0.5,-0.5) circle (0.04) node [below left] {\scriptsize$y_2$};

    \fill[fill=blue!30, opacity=0.7] (0.2,0.2) rectangle (1,1);
    \fill[fill=white, opacity=1] (0.2,0.2) rectangle (0.5,0.7);
    \fill[pattern=north west lines, pattern color=green, opacity=1] (0.2,0.2) rectangle (0.5,0.7);
    \filldraw[fill=black!50] (0.2,0.2) circle (0.04) node [below left] {\scriptsize$y_3$};

    \draw[->] (-1.2,0)--(1.2,0) node [right] {\scriptsize$f_1$}; 
    \draw[->] (0,-1.2)--(0,1.2) node [above] {\scriptsize$f_2$};
    \draw (-1,-1) rectangle (1,1);
    \node[] at (-0.5,-0.4) {\scriptsize $\B\setminus H_3$};
\end{tikzpicture}%
  \end{center}
  \caption{Example of an improvement of the dominated region. The regions
    dominated by~$y_1$ and~$y_2$ are represented in shaded areas, with darker
    shades indicating overlapping regions. The hatched area
    corresponds to the improvement of the dominated region resulting from the
    observation of~$y_3$.}
  \label{fig:hypervolume}
\end{figure}

A natural idea, to extend the EI sampling criterion~\eqref{eq:EI_mono} to the
multi-objective case, is to use the volume of the non-dominated region as loss
function:
\begin{equation*}
  \varepsilon_n(\Xv, f) = \left| H \setminus H_n \right|\,,  
\end{equation*}
where $H = \{y \in \B; \exists x \in \X, f(x) \prec y\}$ and $\left| \,\cdot\,
\right|$ denotes the usual (Lebesgue) volume in~$\R^p$. The improvement
yielded by a new evaluation result $f(X_{n+1}) = \left(f_1(X_{n+1}), \ldots,
  f_p(X_{n+1})\right)$ is then the increase of the volume of the dominated
region (see Figure~\ref{fig:hypervolume}):
\begin{equation}
  I_n\left( X_{n+1} \right) 
  = \left| H \setminus H_n \right| - \left| H \setminus H_{n+1} \right|
  = \left| H_{n + 1} \setminus H_n \right|
  = \left| H_{n + 1} \right| - \left| H_n \right|,
\end{equation}
since $H_n \subset H_{n+1} \subset H$.
Given a vector-valued Gaussian random
process model~$\xi = \left( \xi_1, \ldots, \xi_p \right)$ of $f = \left( f_1,
  \ldots, f_p \right)$, defined on a probability space $\left( \Omega,
  \mathcal{A}, \P_0 \right)$, a multi-objective EI criterion can then be derived
as
\begin{eqnarray}
  \label{eq:EI-multi}
  \EI_{n}(x) &=& \displaystyle \E_{n}\left( I_{n}(x)\right) \nonumber \\ 
  &=& \displaystyle \mathbb{E}_{n}\left(\int_{\B \setminus H_n}
    \mathds{1}_{\xi (x)\prec y}\, \dy\right) \nonumber \\
  &=& \displaystyle \int_{\B \setminus H_n}
  \mathbb{E}_{n} \left(\mathds{1}_{\xi (x)\prec y}\right)\, \dy
  \nonumber \\
  &=& \displaystyle \int_{\B \setminus H_n} \P_n\left(\xi(x)\prec y\right)\,
  \dy\,,
  \label{eq:emmerich-crit}
\end{eqnarray}
where $\P_n$ stands for the probability $\P_0$ conditioned on $X_1,\,
\xi(X_1),\, \ldots,\, X_n,\, \xi(X_n)$.  The multi-objective sampling
criterion~\eqref{eq:emmerich-crit}, also called Expected Hyper-Volume
Improvement~(EHVI), has been proposed by Emmerich and
coworkers~\citep{emmerich2005, emmerich2006single, emmerich2008computation}.

\begin{remark}
  \label{rem:alternative-MOO-crit}
  A variety of alternative approaches have been proposed to extend the EI
  criterion to the multi-objective case, which can be roughly classified into
  aggregation-based techniques \citep{knowles2006parego,
    knowles2005multiobjective, zhang2010expensive} and domination-based
  techniques \citep{jeong2005, keane2006statistical, ponweiser2008,
    bautista2009, svenson2010multiobjective, wagner2010expected}. We consider
  these approaches are heuristic extensions of the EI criterion, in the sense
  that none of them emerges from a proper Bayesian formulation (i.e., a myopic
  strategy associated to some well-identified loss function). A detailed
  description of these approaches is out of the scope of this paper. The
  reader is referred to \cite{wagner2010expected}, \cite{couckuyt2014fast}
  and~\cite{horn2015model} for some comparisons and discussions.
  See also~\cite{picheny2014MO} and~\cite{hernandez2015predictive} for
  other approaches not directly related to the concept of expected improvement.
\end{remark}

\begin{remark}
  \label{rem:EVHIdonneEI}
  The multi-objective sampling criterion~\eqref{eq:emmerich-crit} reduces to the
  usual EI criterion~\eqref{eq:EI_mono} in the single-objective case (assuming
  that $f(X_i) \le y^{\rm upp}$ for at least one~$i \le n$).
\end{remark}

Under the assumption that the components $\xi_i$ of $\xi$ are mutually
independent\footnote{This is the most common modeling assumption in the Bayesian
  optimization literature, when several objective functions, and possibly also
  several constraint functions, have to be dealt with. See the VIPER algorithm
  of~\cite{williams2010} for an example of an algorithm based on correlated
  Gaussian processes.}, $\P_n\left(\xi(x)\prec y\right)$ can be expressed in
closed form: for all $x \in \X$ and $y \in \B \setminus H_n$,
\begin{equation}
  \label{eq:closedFormEmmerichIntegrand}
  \displaystyle \P_n\left(\xi(x) \prec y\right) = \displaystyle
  \prod_{i=1}^p
  \Phi \left( \frac{y_i- \hat\xi_{i,n}(x)}{\sigma_{i,n}(x)} \right)\,,
\end{equation}
where~$\hat\xi_{i,n}(x)$ and~$\sigma^2_{i,n}(x)$ denote respectively the kriging
predictor and the kriging variance at~$x$ for the~$i^\text{th}$ component of~$\xi$.

The integration of~\eqref{eq:closedFormEmmerichIntegrand} over~$\B \setminus
H_n$, in the expression~\eqref{eq:emmerich-crit} of the multi-objective EI
criterion, is a non-trivial problem. Several authors
\citep{emmerich2008computation, bader2011hype, hupkens2014faster, couckuyt2014fast} have proposed
decomposition methods to carry out this computation, where the integration
domain~$\B \setminus H_n$ is partitioned into hyper-rectangles, over which the
integral can be computed analytically.  The computational complexity of these
methods, however, increases exponentially with the number of objectives%
\footnote{See, e.g., \cite{beume2009}, \cite{hupkens2014faster}, 
  \cite{couckuyt2014fast} and references therein for decomposition algorithms
and complexity results.}, which
makes the approach impractical in problems with more than a few objective
functions. The method proposed in this work also encounters this type of
integration problem, but takes a different route to solve it (using SMC
techniques; see Section~\ref{sec:smc}). Our approach will make it possible to
deal with more objective functions.

\begin{remark}
  \label{rem:multiei-other}
  Exact and approximate implementations of the EHVI criterion are available,
  together with other Gaussian-process-based criteria for bound-constrained
  multi-objective optimization, in the Matlab/Octave toolbox
  STK~\citep{stktoolbox} and in the R packages GPareto~\citep{gpareto} and
  mlrMBO~\citep{horn2015model}. Note that several  approaches discussed in
  Remark~\ref{rem:alternative-MOO-crit} maintain an affordable computational
  cost when the number of objectives grows, and therefore constitute possible
  alternatives to the SMC technique proposed in this paper for many-objective
  box-constrained problems.
\end{remark}

\subsection{EI-based optimization with constraints}
\label{sec:ei-constraints}

In this section, we discuss extensions of the expected improvement criterion for
single- and multi-objective constrained optimization.

Consider first the case of problems with a single objective and several
constraints:
\begin{equation}
  \left\{
    \begin{array}{l}
      \min_{x\in\X} f(x)\,, \\[0.6em]
      c(x) \leq 0\,,
    \end{array}\right.  
\end{equation}
where $c = (c_1,\, \ldots,\, c_q)$ is a vector of continuous constraints. The
set $C = \{x\in\X;\, c(x) \leq 0\}$ is called the \emph{feasible domain}.  If it
is assumed that at least one evaluation has been made in $C$, it is natural to
define a notion of improvement with respect to the best objective value $m_n =
\min \left\{ f(x) ;\, x \in \{X_1,\, \ldots, X_n\} \cap C \right\}$:
\begin{eqnarray}
  \label{eq:improvementConstrained}
  I_n(X_{n+1}) %
  &=& m_n - m_{n+1} \nonumber\\
  &=& \mathds{1}_{c(X_{n+1}) \leq 0} \cdot \bigl( m_n -
  f(X_{n+1})\bigr)_{+} \nonumber \\
  &=& \left\{
    \begin{array}{ll}
      m_{n} - f(X_{n+1})~~ 
      & \text{if }  X_{n+1} \in C \text{ and } f(X_{n+1}) < m_n, \\[0.8em]
      0 & \text{otherwise}\,.
    \end{array}\right.  
\end{eqnarray}
In other words, a new observation makes an improvement if it is feasible and
improves upon the best past value \citep{schonlau1998global}. The corresponding
expected improvement criterion follows from taking the expectation:
\begin{equation}
  \label{eq:schonlauCrit-generalform}  
  \EI_n(x) = \displaystyle \E_{n}\left(
    \mathds{1}_{\xic(x) \leq 0} \cdot \bigl(m_n - \xio(x)\bigr)_+
  \right)\,.
\end{equation} 

If $f$ is modeled by a random process $\xio$ and $c$ is modeled by a
vector-valued random process $\xic=(\xi_{\cons, 1},\, \ldots,\,\xi_{\cons, q})$
independent of~$\xio$, then the sampling criterion~\eqref{eq:schonlauCrit-generalform}
simplifies to \citeauthor{schonlau1998global}'s criterion:
\begin{equation}
  \label{eq:schonlauCrit}  
  \EI_n(x) = %
  \P_n(\xic(x) \leq 0)\;
  \E_{n}\bigl((m_n-\xio(x))_+\bigr)\,.  
\end{equation} 
In other words, the expected improvement is equal in this case to the product of
the unconstrained expected improvement, with respect to $m_n$, with the
probability of feasibility. The sampling criterion~\eqref{eq:schonlauCrit} is
extensively discussed, and compared with other Gaussian-process-based constraint
handling methods, in the PhD thesis of~\cite{sasena2002flexibility}. More
generally, sampling criteria for constrained optimization problems have been
reviewed by~\cite{parr2012infill} and~\cite{gelbart2015phd}.

In the general case of constrained multi-objective problems, the aim is to build
an approximation of $\Gamma$ defined by~(\ref{eq:Gamma}).  If it is assumed that
an observation has been made in the feasible set $C$, a reasoning similar to
that used in the single-objective case can be made to formulate an extension of
the EI~\eqref{eq:EI-multi}:
\begin{equation}
  \label{eq:EmmercihCritConst-generalform}  
  \EI_n(x) = \E_{n}\left( 
    \left| H_{n + 1} \right| - \left| H_n \right| \right),  
\end{equation} 
where
\begin{equation}
  \label{eq:Hn-constr}
  H_n = \{y \in \B; \exists i\le n,\, X_i \in C \text{\ and\ } f(X_i) \prec y\}  
\end{equation}
is the subset of $\B$, defined as in Section~\ref{sec:ei-multi}, whose
points are dominated by feasible evaluations.  When~$\xio$ and~$\xic$ are
assumed independent, \eqref{eq:EmmercihCritConst-generalform} boils down to the
product of a modified EHVI criterion, where only feasible points are
considered\footnote{%
  Note that this modified EHVI criterion remains well defined even when $H_n =
  \emptyset$, owing to the introduction of an upper bound~$\yUpp$ in the
  definition of~$\B$. Its single-objective counterpart introduced earlier (see
  Equation~\eqref{eq:schonlauCrit-generalform}), however, was only well defined
  under the assumption that at least one feasible point is known.  Introducing
  an upper bound~$\yUpp$ is of course also possible in the single-objective
  case.%
}, %
and the probability of feasibility,
as suggested by \cite{emmerich2006single} and~\cite{shimoyama2013updating}:
\begin{equation}
  \label{eq:EmmercihCritConst}  
  \EI_n(x) = %
  \P_n \left( \xic(x) \leq 0 \right)\, %
  \int_{\B \setminus H_n} \P_n\left(\xio(x)\prec y\right)\,
  \dy.
\end{equation}

Observe that the sampling
criterion~\eqref{eq:EmmercihCritConst-generalform} is the one-step
look-ahead criterion associated to the loss function $\varepsilon_n(\Xv,
f) = - \left| H_n \right|$, where~$H_n$ is defined
by~\eqref{eq:Hn-constr}. This loss function remains constant as long as
no feasible point has been found and, therefore, is not an appropriate
measure of loss for heavily constrained problems where finding feasible
points is sometimes the main difficulty\footnote{%
  The same remark holds for the variant \citep[see,
  e.g.,][]{gelbart2014bayesian} which consists in using the probability
  of feasibility as a sampling criterion when no feasible point is
  available. This is indeed equivalent to using the loss function
  $\varepsilon_n(\Xv, f) = - \one_{\exists i \le n, X_i \in C}$ in the
  search for feasible points.%
}. From a practical point of view, not all unfeasible points should be
considered equivalent: a point that does not satisfy a constraint by a small amount has probably more
value than one that does not satisfy the constraint by a large amount, and
should therefore make the loss smaller. Section~\ref{sec:method} will
present a generalization of the expected improvement for
constrained problems, relying on a new loss function that encodes this
preference among unfeasible solutions.

\begin{remark}
  \label{rem:other-bayesian-const}
  Other Gaussian-process-based approaches that can be used to handle constraints
  include the method by~\cite{gramacy2015modeling}, based on the
  \emph{augmented Lagrangian} approach of~\cite{conn1991globally}, and
  several recent methods \citep{picheny2014, gelbart2015phd,
    hernandez2015,hernandez2015pesc} based on stepwise uncertainty reduction
  strategies \citep[see, e.g.,][for more information on this
  topic]{villemonteix2009informational, bect2012sequential, chevalier2014fast}.
\end{remark}

\begin{remark}
  \label{rem:crit-calc-const}
  The term $\E_{n}\bigl((m_n - \xio(x))_+\bigr)$ in~\eqref{eq:schonlauCrit} can
  be computed analytically as in Section~\ref{sec:expected-improvement}, and the
  computation of the integral in~\eqref{eq:EmmercihCritConst} has been discussed
  in Section~\ref{sec:ei-multi}.  If it is further assumed that the components
  of $\xic$ are Gaussian and independent, then the probability of feasibility
  can be written as
  \begin{equation}
    \label{eq:seperableProbFeas}
    \P_n(\xic(x) \leq 0) = \prod_{j=1}^q
    \Phi\left(- \frac{\hat\xi_{\cons,\,j,\,n}(x)}{\sigma_{\cons,\,j,\,n}(x)}\right)  
  \end{equation}
  where $\hat\xi_{\cons,\,j,\,n}(x)$ and $\sigma^2_{\cons,\,j,\,n}(x)$ stand
  respectively for the kriging predictor and the kriging variance of
  $\xi_{\cons,\,j}$ at~$x$.
\end{remark}

\section{An EI criterion for constrained multi-objective optimization}
\label{sec:method}

\subsection{Extended domination rule}
\label{sec:extend-domin-rule}

In a constrained multi-objective optimization setting, we propose to handle the
constraints using an extended Pareto domination rule that takes both objectives
and constraints into account, in the spirit of
\cite{fonseca1998multiobjective}, \cite{ray2001multiobjective}
and~\cite{oyama2007new}.  For ease of presentation, denote by~$\Yo = \R^p$
and~$\Yc = \R^q$ the objective and constraint spaces respectively, and let $\Y =
\Yo \times \Yc$.

We shall say that $y_1 \in \Y$ dominates $y_2 \in \Y$, which will be written as
$y_1 \lhd y_2$, if $\psi(y_1) \prec \psi(y_2)$, where~$\prec$ is the usual
Pareto domination rule recalled in Section~\ref{sec:ei-multi} and, denoting
by~$\overline\R$ the extended real line,
\begin{equation}
  \label{eq:extended-dom-rule}
  \begin{array}{r c c l}
    \psi:\; & \Yo \times \Yc & \;\rightarrow\; & \overline\R^p \times \R^q\\
    & (\yo,\,\yc) & \mapsto &
    \begin{cases}
      (\yo,\,0) & \text{if\ } \yc \leq 0,\\[0.6em]
      \bigl(+\infty,\,\max (\yc,0)\bigr) & \text{otherwise.}
    \end{cases}
  \end{array}
\end{equation}

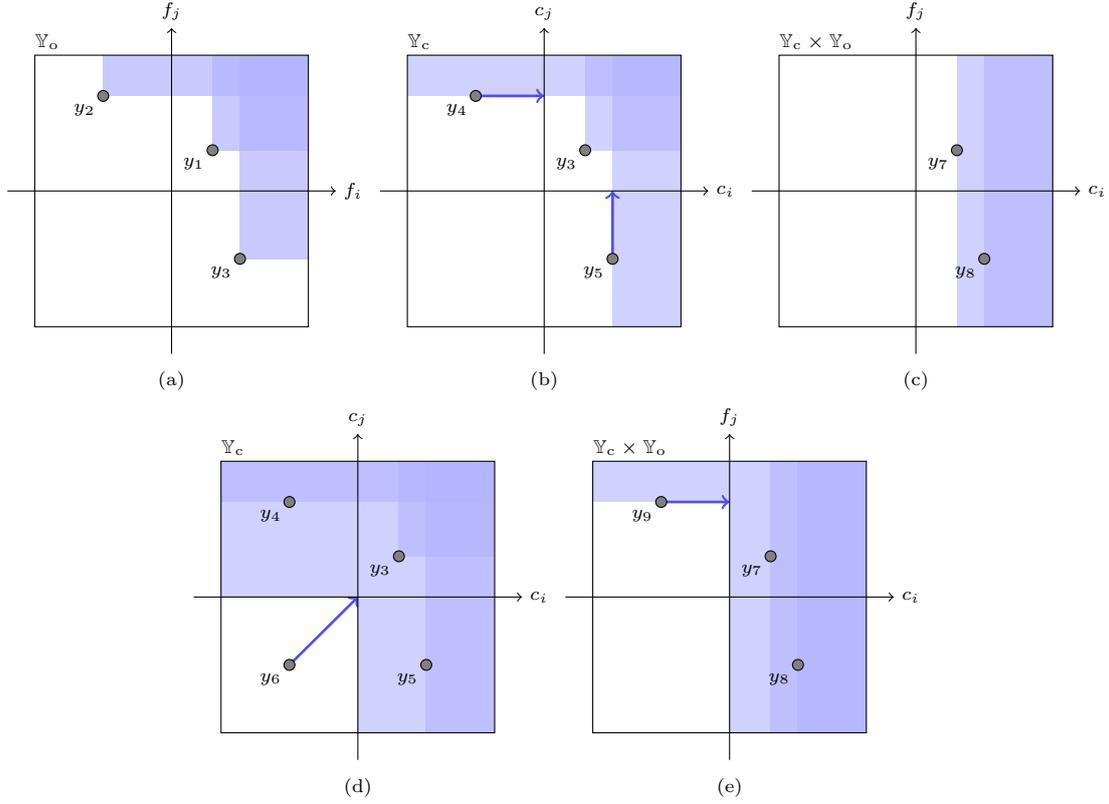
\begin{figure}
  \begin{center}
    \begin{tikzpicture}[scale=1.8]
    \fill[fill=blue!30, opacity=0.7] (0.3,0.3) rectangle (1,1);
    \filldraw[fill=black!50] (0.3,0.3) circle (0.04) node [below left] {\scriptsize$y_1$};

    \fill[fill=blue!30, opacity=0.7] (-0.5,0.7) rectangle (1,1);
    \filldraw[fill=black!50] (-0.5,0.7) circle (0.04) node [below left] {\scriptsize$y_2$};

    \fill[fill=blue!30, opacity=0.7] (0.5,-0.5) rectangle (1,1);
    \filldraw[fill=black!50] (0.5,-0.5) circle (0.04) node [below left] {\scriptsize$y_3$};

    \node[right,inner sep=0] at (-1,1.1) {\scriptsize $\Yo$};
    \draw[->] (-1.2,0)--(1.2,0) node [right] {\scriptsize$f_i$}; 
    \draw[->] (0,-1.2)--(0,1.2) node [above] {\scriptsize$f_j$};
    \draw (-1,-1) rectangle (1,1);
    \node[] at (0,-1.4) {\scriptsize {(a)}};
\end{tikzpicture}\hspace{1ex}%
\begin{tikzpicture}[scale=1.8]

  \fill[fill=blue!30, opacity=0.6] (0.3,0.3) rectangle (1,1);
  \filldraw[fill=black!50] (0.3,0.3) circle (0.04) node [below left] {\scriptsize$y_3$}; 

  \fill[fill=blue!30, opacity=0.6] (-1,0.7) rectangle (1,1);
  \draw[->,blue!70,line width=1] (-0.5,0.7)--(0,0.7);
  \filldraw[fill=black!50] (-0.5,0.7) circle (0.04) node [below left] {\scriptsize$y_4$};

  \fill[fill=blue!30, opacity=0.6] (0.5,-1) rectangle (1,1);
  \draw[->,blue!70,line width=1] (0.5,-0.5)--(0.5,0);
  \filldraw[fill=black!50] (0.5,-0.5) circle (0.04) node [below left] {\scriptsize$y_5$};

  \node[right,inner sep=0] at (-1,1.1) {\scriptsize $\Yc$};
  \draw[->] (-1.2,0)--(1.2,0) node [right] {\scriptsize$c_i$}; 
  \draw[->] (0,-1.2)--(0,1.2) node [above] {\scriptsize$c_j$};
  \draw (-1,-1) rectangle (1,1);
  \node[] at (0,-1.4) {\scriptsize {(b)}};
\end{tikzpicture}\hspace{1ex}%
\begin{tikzpicture}[scale=1.8]

  \fill[fill=blue!30, opacity=0.6] (0.3,-1) rectangle (1,1);
  \filldraw[fill=black!50] (0.3,0.3) circle (0.04) node [below left] {\scriptsize$y_7$}; 
  \fill[fill=blue!30, opacity=0.6] (0.5,-1) rectangle (1,1);
  \filldraw[fill=black!50] (0.5,-0.5) circle (0.04) node [below left] {\scriptsize$y_8$};

  \node[right,inner sep=0] at (-1,1.1) {\scriptsize $\Yc\times\Yo$};
  \draw[->] (-1.2,0)--(1.2,0) node [right] {\scriptsize$c_i$}; 
  \draw[->] (0,-1.2)--(0,1.2) node [above] {\scriptsize$f_j$};
  \draw (-1,-1) rectangle (1,1);
    \node[] at (0,-1.4) {\scriptsize {(c)}};
\end{tikzpicture}\\
\begin{tikzpicture}[scale=1.8]
  \fill[fill=blue!30, opacity=0.6] (-1,0) rectangle (0,1);
  \fill[fill=blue!30, opacity=0.6] (0,-1) rectangle (1,1);
  \draw[->,blue!70,line width=1] (-0.5,-0.5)--(0,0);
  \filldraw[fill=black!50] (-0.5,-0.5) circle (0.04) node [below left] {\scriptsize$y_6$};

  \fill[fill=blue!30, opacity=0.6] (0.3,0.3) rectangle (1,1);
  \filldraw[fill=black!50] (0.3,0.3) circle (0.04) node [below left] {\scriptsize$y_3$}; 

  \fill[fill=blue!30, opacity=0.6] (-1,0.7) rectangle (1,1);
  \filldraw[fill=black!50] (-0.5,0.7) circle (0.04) node [below left] {\scriptsize$y_4$};

  \fill[fill=blue!30, opacity=0.6] (0.5,-1) rectangle (1,1);
  \filldraw[fill=black!50] (0.5,-0.5) circle (0.04) node [below left] {\scriptsize$y_5$};

  \node[right,inner sep=0] at (-1,1.1) {\scriptsize $\Yc$};
  \draw[->] (-1.2,0)--(1.2,0) node [right] {\scriptsize$c_i$}; 
  \draw[->] (0,-1.2)--(0,1.2) node [above] {\scriptsize$c_j$};
  \draw (-1,-1) rectangle (1,1);
  \node[] at (0,-1.4) {\scriptsize {(d)}};
\end{tikzpicture}\hspace{1ex}%
\begin{tikzpicture}[scale=1.8]
  \fill[fill=blue!30, opacity=0.6] (-1,0.7) rectangle (1,1);
  \fill[fill=blue!30, opacity=0.6] (0,-1) rectangle (1,0.7);
  \draw[->,blue!70,line width=1] (-0.5,0.7)--(0,0.7);
  \filldraw[fill=black!50] (-0.5,0.7) circle (0.04) node [below left] {\scriptsize$y_9$};

  \fill[fill=blue!30, opacity=0.6] (0.3,-1) rectangle (1,1);
  \filldraw[fill=black!50] (0.3,0.3) circle (0.04) node [below left] {\scriptsize$y_7$}; 

  \fill[fill=blue!30, opacity=0.6] (0.5,-1) rectangle (1,1);
  \filldraw[fill=black!50] (0.5,-0.5) circle (0.04) node [below left] {\scriptsize$y_8$};

  \node[right,inner sep=0] at (-1,1.1) {\scriptsize $\Yc\times\Yo$};
  \draw[->] (-1.2,0)--(1.2,0) node [right] {\scriptsize$c_i$}; 
  \draw[->] (0,-1.2)--(0,1.2) node [above] {\scriptsize$f_j$};
  \draw (-1,-1) rectangle (1,1);
  \node[] at (0,-1.4) {\scriptsize {(e)}};
\end{tikzpicture}  
\end{center}
\caption{Illustration of the extended domination rule in different
  cases.  The region dominated by each point is represented by a shaded
  area. Darker regions indicate overlapping regions.  (a) Feasible
  solutions are compared with respect to their objective values using
  the usual domination rule in the objective space---see properties~(i)
  and~(ii).  (b--c) Non-feasible solutions are compared using the Pareto
  domination rule applied to the vectors of constraint violations
  according to property~(iii). Note that $y_4$ dominates points having a
  higher value of $c_j$ regardless of the corresponding value of $c_i$,
  and, likewise, $y_5$ dominates points with higher values of
  $c_i$. (d--e) Feasible solutions always dominate non-feasible
  solutions: $y_{6}$ is feasible and hence dominates $y_{3}$, $y_{4}$
  and $y_{5}$; $y_9$ is feasible and dominates both $y_7$ and
  $y_8$ as stated in (iv).}
  \label{fig:domination}
\end{figure}

The extended domination rule~\eqref{eq:extended-dom-rule} has the following
properties:
\begin{enumerate}[(i)]
\item For unconstrained problems ($q=0$, $\Yc = \emptyset$), the extended
  domination rule boils down to the Pareto domination rule on $\Y = \Yo$.
\item Feasible solutions (corresponding to $\yc \leq 0$) are compared
  using the Pareto domination rule applied in the objective space (in other
  words, using the Pareto domination rule with respect to the objective values
  $\yo$).
\item Non-feasible solutions (corresponding to $\yc \not\le 0$) are
  compared using the Pareto domination rule applied to the vector of constraint
  violations.
\item  Feasible solutions always dominate non-feasible solutions.
\end{enumerate}
These properties are illustrated on Figure~\ref{fig:domination}.

\subsection{A new EI criterion}
\label{sec:expected-improvement-mult-cons}

The extended domination rule presented above makes it possible to define a
notion of expected hyper-volume improvement as in Section~\ref{sec:ei-multi} for
the \emph{constrained} multi-objective setting. Given evaluation results
$(f(X_1), c(X_1))$, $\ldots$, $(f(X_n), c(X_n))$, define
\begin{equation*}
  H_n = \{y \in \B;\; \exists i \le n,\; (f(X_i),\, c(X_i))\, \lhd\, y\}\,
\end{equation*}
with $\B = \Bo \times \Bc$, where $\Bo \subset \Yo$ and~$\Bc \subset \Yc$ are
two bounded hyper-rectangles that are introduced to ensure, as in
Section~\ref{sec:ei-multi}, that $\left| H_n \right| < +\infty$ (see
Appendix~\ref{sec:lowerBounds}). Then, define the improvement
yielded by a new evaluation $(f(X_{n+1}),\,c(X_{n+1}))$ by
\begin{equation*}
  I_n\left( X_{n+1} \right)
  = \left| H_{n + 1} \setminus H_n \right|\,
  = \left| H_{n + 1} \right| - \left| H_n \right|\,
\end{equation*}
as in Section~\ref{sec:ei-multi}. In order to get a meaningful concept of
improvement both before and after the first feasible point has been found, we
assume without loss of generality that $\left( 0,\, \ldots,\, 0 \right) \in
\R^q$ is in the interior of~$\Bc$.

If $\left( f, c \right)$ is modeled by a vector-valued random process $\xi =
(\xio,\, \xic)$, with $\xio = \left( \xi_{\obj, 1},\,\ldots,
  \xi_{{\obj}, p} \right)$ and $\xic = \left( \xi_{\cons, 1},\,,\ldots
  \xi_{\cons, q} \right)$, then the expected improvement for the constrained
multi-objective optimization problem may be written as
\begin{equation}
  \label{eq:ei-multi-cons}
  \EI_{n}(x) = \displaystyle \E_{n}\bigl((I_{n}(x)\bigr)
  = \displaystyle \mathbb{E}_{n}\left(\int_{G_n}\mathds{1}_{\xi(x)\lhd y}\, \dy\right)
  = \displaystyle \int_{G_n}\P_n(\xi(x)\lhd y)\, \dy\,,
\end{equation}
where $G_n = \B \setminus H_n$ is the set of all non-dominated points in $\B$.

As in Section~\ref{sec:ei-multi}, under the assumption that the components of
$\xi$ are mutually independent and Gaussian, $\P_n\left(\xi(x)\lhd y\right)$ can
be expressed in closed form: for all $x \in \X$ and $y = (\yo,\, \yc)
\in G_n$,
\begin{equation}
  \P_n(\xi(x) \lhd y) =
  \left\{
    \begin{array}{l l}
      \displaystyle \left(\prod_{i=1}^p
        \Phi\biggl(\frac{y_{\obj,\,i}-\hat\xi_{\obj,\,i,\,n}(x)}{\sigma_{\obj,\,i,\,n}(x)}\biggr)\right)\left(\prod_{j=1}^q
        \Phi\biggl(-\frac{\hat\xi_{\cons,\,j,\,n}(x)}{\sigma_{\cons,\,j,\,n}(x)}\biggr)\right)
      &\quad \text{if\ } \yc \leq 0\,,\\[3.2em]
      \displaystyle \prod_{j=1}^q
      \Phi\biggl(\frac{\max(y_{\cons,\,j},\,0)-\hat\xi_{\cons,\,j,\,n}(x)}{\sigma_{\cons,\,j,\,n}(x)}\biggr)
      &\quad  \text{otherwise}\,.
    \end{array}
  \right.
\end{equation}

The EI-based constrained multi-objective optimization procedure may be written
as~\eqref{eq:oneStepLookAhead}. In practice, the computation of each new
evaluation point requires to solve two numerical problems: a) the computation of
the integral in~\eqref{eq:ei-multi-cons}; b) the optimization of $\EI_n$ in the
procedure~\eqref{eq:oneStepLookAhead}. These problems will be addressed in
Section~\ref{sec:smc}.

\begin{remark}
  When there are no constraints ($q=0$, $\Y_{\cons}=\emptyset$), the extended
  domination rule~$\lhd$ corresponds to the usual Pareto domination
  rule~$\prec$. In this case, the sampling criterion~\eqref{eq:ei-multi-cons}
  simplifies to
  \begin{equation}
    \label{eq:critSpecCase1}
    \EI_{n}(x) =   \int_{\Bo \setminus \Hno} %
    \P_n\left(\xio(x)\prec \yo\right)\, {\rm d}\yo,
  \end{equation}
  with
  \begin{equation*}
    \Hno = \{\yo\in \Bo;\; \exists i \le n,\, f(X_i) \prec \yo\}\,.  
  \end{equation*}
  Denote by~$\yLow_{\obj}$, $\yUpp_{\obj} \in \Yo$ the lower and upper corners
  of the hyper-rectangle~$\Bo$. Then, the sampling
  criterion~\eqref{eq:critSpecCase1} is equivalent to the multi-objective EI
  criterion presented in Section~\ref{sec:ei-multi} in the limit~$\yLow_{\obj}
  \to -\infty$.  If, moreover, the problem has only one objective function, then
  the criterion~\eqref{eq:ei-multi-cons} boils down to the original Expected
  Improvement criterion as soon as the best evaluation dominates~$\yUpp_{\obj}$
  (see Remark~\ref{rem:EVHIdonneEI}).
\end{remark}

\subsection{Decomposition of the expected improvement: %
  feasible and unfeasible components}
\label{sec:critDecomposition}

Assume  that there is at least one constraint ($q \ge 1$). Then, the expected
improvement~$\EI_n(x)$ can be decomposed as
\begin{equation}
  \EI_n(x) = \EIFeas_n(x) + \EIUnf_n(x),
\end{equation}
by splitting the integration domain in the right-hand side
of~\eqref{eq:ei-multi-cons} in two parts: $\EIFeas_n(x)$ corresponds to the
integral on $G_n \cap \left\{ \yc \le 0 \right\}$, while
$\EIUnf_n(x)$ corresponds to the integral on $G_n \cap \left\{
  \yc \not\le 0 \right\}$.

More explicit expressions will now be given for both terms. First,
\begin{equation}
  \begin{array}{l c l}
    \EIUnf_n(x) &=& %
    \displaystyle \int_{G_n \cap \left\{ y^{\cons} \not\le 0 \right\}}
    \P_n \left( (\xio(x),\xic(x)) \lhd (\yo,\yc) \right)\, %
    \mathrm{d}(\yo,\yc) \\
    &=& %
    \displaystyle \left| \Bo \right| \cdot %
    \int_{\Bc \setminus \Hnc}
    \P_n \left( \xic^+(x) \prec \yc^+ \right)\, %
    \one_{\yc \not\le 0}\, \dyc
  \end{array}
\end{equation}
where $\yc^+ = \max \left( \yc, 0 \right)$ and
\begin{equation*}
  \Hnc = \left\{ %
    \yc \in \Bc \mid %
    \exists i \le n,\, \xic^+(X_i) \prec \yc^+ %
  \right\}.
\end{equation*}

Let  $\BcNeg = \Bc \cap \left]-\infty,\,0\right]^q$ denote the feasible
subset of~$\Bc$. Then, assuming that~$\xic$ and~$\xio$ are independent,
\begin{equation}
  \label{eq:EIFeas}
  \begin{array}{l c l}
    \EIFeas_n(x) &=& %
    \displaystyle \int_{ G_n \cap \left\{ y^{\cons} \le 0 \right\}}
    \P_n \left( (\xio(x),\xic(x)) \lhd (\yo,\yc) \right)\, %
    \mathrm{d}(\yo,\yc) \\
    &=& %
    \displaystyle \left| \BcNeg \right| %
    \,\cdot\, \P_n(\xic(x) \leq 0) %
    \,\cdot\, \int_{\Bo \setminus \Hno} %
    \P_n\left(\xio(x)\prec \yo\right)\, {\rm d}\yo \,,
  \end{array}
\end{equation}
where
\begin{equation*}
  \Hno = \left\{ \yo \in \Bo \mid %
    \exists i \le n,\, %
    \xic(X_i) \le 0 \text{ and } \xio(X_i) \prec \yo
  \right\}\,.
\end{equation*}

\begin{remark}
  \label{rem:SpecialCaseSchonlau}
  The set~$\Bc \setminus \Hnc$ is empty as soon as a feasible point has
  been evaluated. As a consequence, the component~$\EIUnf$ of the expected
  improvement vanishes and therefore, according to~\eqref{eq:EIFeas},
  \begin{equation*}
    \EI_n(x) \propto \P_n(\xic(x) \leq 0) %
    \,\cdot\, \int_{\Bo \setminus \Hno} %
    \P_n\left(\xio(x)\prec \yo\right)\, {\rm d}\yo \,.
  \end{equation*}
  In other words, up to a multiplicative constant, the expected improvement
  is equal, in this case, to the product of the probability of feasibility
  with a modified EHVI criterion in the objective space, where only feasible
  points are used to define the dominated region. In particular, in 
  constrained single-objective problems,  the criterion of
  \citeauthor{schonlau1998global} (see
  Section~\ref{sec:ei-constraints}) is recovered as the
  limit case~$\yLow_{\obj} \to -\infty$, as soon as the best evaluation
  dominates~$\yUpp_{\obj}$.
\end{remark}

\begin{remark}
  In our numerical experiments, $\Bo$ and~$\Bc$ are defined using estimates of the range of the
  objective and constraint functions (see
  Appendix~\ref{sec:annexe:adaptBoBc}). Another natural choice for the~$\Bo$ would
  have been to use (an estimate of) the range of the objective functions
  \emph{restricted to the feasible subset $C \subset \X$} for~$\Bo$. Further
  investigation of this idea is left for future work.
\end{remark}

\section{Sequential Monte Carlo techniques to compute and optimize the
  expected improvement}
\label{sec:smc}

\subsection{Computation of the expected improvement}
\label{sec:crit-calc}

Since the dimension of $\Y$ is likely to be high in practical problems (say,
$p+q\geq 5$), the integration of $y \mapsto \P_n(\xi(x) \lhd y)$ over $G_n$
cannot be carried out using decomposition methods
\citep{emmerich2008computation, bader2011hype, hupkens2014faster} because, as
mentioned in Section~\ref{sec:ei-multi}, the computational complexity of these
methods increases exponentially with the dimension of $\Y$.

To address this difficulty, we propose to use a Monte Carlo approximation of the
integral~\eqref{eq:ei-multi-cons}:
\begin{equation}
  \label{eq:EI-smc-approx}
  \EI_n(x) \approx \frac{1}{m}\sum_{k=1}^m  \P_n(\xi(x) \lhd y_{n,k}),
\end{equation}
where $\YY_n = \left(y_{n,k} \right)_{1 \leq k \leq m}$ is a set of
\emph{particles} distributed according to the uniform density $\pi_n^\Y \propto
\mathds{1}_{G_n}$ on~$G_n$.  In principle, sampling uniformly over $G_n$ could
be achieved using an accept-reject method \citep[see, e.g.,][]{robert2013monte},
by sampling uniformly over $\B$ and discarding points in $H_n$
\citep{bader2011hype}. However, when the dimension of $\Y$ is high, $G_n$ will
probably have a small volume with respect to that of~$\B$.  Then, the acceptance
rate becomes small and the cost of generating a uniform sample on $G_n$ becomes
prohibitive. (As an example, consider an optimization problem with $q = 20$
constraints. If $\Bc=[-v/2,\, +v/2]^q$, then the volume of the feasible
region is $2^{20}\approx 10^6$ times smaller than that of $\Bc$.)

In this work, we use a variant of the technique called subset simulation
\citep{au2001estimation, cerou2012sequential} to achieve uniform sampling over
$G_n$. The subset simulation method is a well-known method in the field of
structural reliability and rare event estimation, which is used to estimate the
volume of small sets by Monte Carlo sampling.

Denote by $\Pi_0^{\Y}$ the uniform distribution over $\B$ and assume that the
probability $\Pi_0^{\Y}(G_n)$ becomes small when $n$ increases, so that sampling
$G_n$ using an accept-reject method is impractical. Observe that the sets $G_n$,
$n=1,\,2,\, \ldots$ form a nested sequence of subsets of~$\B$ (hence the name
subset simulation):
\begin{equation}
  \label{equ:decreasSubsets}
  \B \supset G_1 \supset G_2 \supset \cdots .
\end{equation}
Denote by $\Pi_n^{\Y}$ the uniform distribution on~$G_n$, which has the
probability density function~$\pi_n^\Y$ defined above.  Since the addition of a
single new evaluation, at iteration $n + 1$, is likely to yield only a small
modification of the set $G_n$, the probability
\begin{equation*}
  \Pi_{n}^{\Y}(G_{n+1}) %
  = \int_{G_{n + 1}} \pi_n^\Y(y)\, \dy
  \,=\, \frac{\Pi_0^{\Y}(G_{n+1})}{\Pi_0^{\Y}(G_n)}\,
\end{equation*}
is likely to be high. Then, supposing that a set of particles $\YY_n =
\left(y_{n,k} \right)_{1 \leq k \leq m}$ uniformly distributed on~$G_n$ is
already available, one obtains a sample $\YY_{n+1}$ uniformly distributed over
$G_{n+1}$ using the \emph{Remove-Resample-Move} procedure described in
Algorithm~\ref{alg:RemResMovUnif}.  (All the random variables generated in
Algorithm~\ref{alg:RemResMovUnif} are independent of~$\xi$ conditionally on
$X_1$, $\xi(X_1)$, \ldots, $X_{n + 1}$, $\xi(X_{n+1})$.)

\begin{algorithm}[t]
  \caption{Remove-Resample-Move procedure to construct~$\YY_n$}
  \label{alg:RemResMovUnif}
  \eIf{$n = 0$}{%
    Generate $m$ independent and uniformly distributed particles over~$G_0 =
    \B$.\;%
  }{%
    \emph{Remove: } Set $\YY_n^0 = \YY_{n-1} \cap G_n$ and $m_0 = \left| \YY_n^0 \right|$.\;
    \emph{Resample: } Set $\YY_n^1 = \YY_n^0 \cup \left\{ \tilde y_{n, 1},
      \ldots, \tilde y_{n, m - m_0} \right\}$, where $\tilde y_{n, 1}, \ldots, \tilde
    y_{n, m - m_0}$ are independent and uniformly distributed on~$\YY_n^0$.
    (Each~$\tilde y_{n,k}$ is a replicate of a particle
    from~$\YY_n^0$.) \;
    \emph{Move: } Move the particles using a Metropolis-Hastings algorithm
    \citep[see, e.g,][]{robert2013monte} which targets the uniform distribution
    over~$G_{n+1}$. The resulting set of particles is~$\YY_{n+1}$.\;
  }
\end{algorithm}

Algorithm~\ref{alg:RemResMovUnif} obviously requires that at least one particle
from~$\YY_n$, which belongs by construction to~$G_n$, also belongs to~$G_{n +
  1}$; otherwise, the set of surviving particles, referred to in the second step
of the algorithm, will be empty. More generally,
Algorithm~\ref{alg:RemResMovUnif} will typically fail to produce a good sample
from~$\Pi_{n+1}^\Y$ if the number of surviving particles is small, which happens
with high probability if $\Pi_{n}^{\Y}(G_{n+1})$ is small---indeed, the expected
number of particles of~$\YY_{n}$ in a
given\footnote{Equation~\eqref{equ:expectedNumber} does not hold exactly for $A
  = G_{n + 1}$ since, conditionally on $X_1$, $\xi(X_1)$, \ldots, $X_{n}$,
  $\xi(X_{n})$, the set $G_{n+1}$ is a \emph{random} set and is not independent
  of~$\YY_n$. Indeed, $G_{n+1}$ depends on~$\xi(X_{n+1})$ and $X_{n + 1}$ is
  chosen by minimization of the approximate expected improvement, which in turn
  is computed using~$\YY_n$.} set~$A \subset \B$ is
\begin{equation}
  \label{equ:expectedNumber}
  \E_n \Bigl( N(A;\, \YY_{n}) \Bigr) %
  = \E_n \left( \sum_{k=1}^m\mathds{1}_A(y_{n,k}) \right) %
  = m \cdot \Pi_n^{\Y}(A)\,,
\end{equation}
where $N(A;\, \YY)$ denotes the number of particles of $\YY$ in $A$. This
situation occurs, for instance, when a new evaluation point brings a large
improvement~$G_n \setminus G_{n + 1} = H_{n+1} \setminus H_n$.

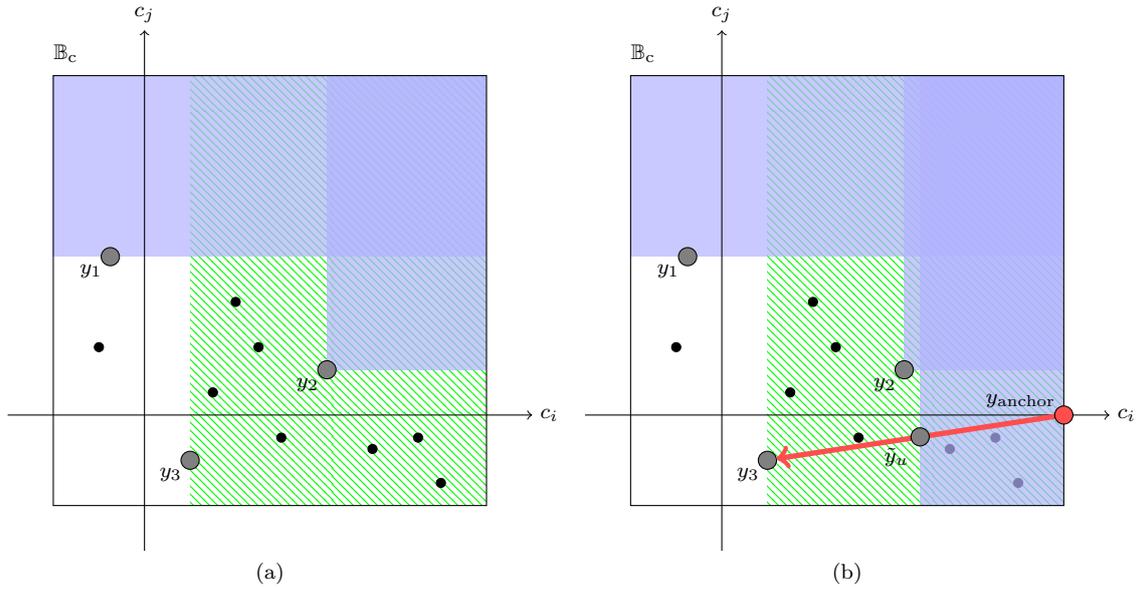
\begin{figure}
  \begin{center}
    \begin{tikzpicture}[scale=3.]

    \fill[fill=white, opacity=1] (0.2,0.2) rectangle (0.5,0.7);
    \fill[pattern=north west lines, pattern color=green, opacity=1] (0.2,-0.4) rectangle (1.5,1.5);
    \filldraw[fill=black!50] (0.2,-0.2) circle (0.04) node [below left] {\small$y_3$};

    \fill[fill=blue!30, opacity=0.7] (-0.4,0.7) rectangle (1.5,1.5);
    \filldraw[fill=black!50] (-0.15,0.7) circle (0.04) node [below left] {\small$y_1$};

    \fill[fill=blue!30, opacity=0.7] (0.8,0.2) rectangle (1.5,1.5);
    \filldraw[fill=black!50] (0.8,0.2) circle (0.04) node [below left] {\small$y_2$};
    \filldraw[fill=black] (-0.2, 0.3 ) circle (0.02);
    \filldraw[fill=black] ( 0.3, 0.1 ) circle (0.02);
    \filldraw[fill=black] ( 0.4, 0.5 ) circle (0.02);
    \filldraw[fill=black] ( 0.5, 0.3 ) circle (0.02);
    \filldraw[fill=black] ( 0.6,-0.1 ) circle (0.02);
    \filldraw[fill=black] ( 1.0,-0.15) circle (0.02);
    \filldraw[fill=black] ( 1.2,-0.1 ) circle (0.02);
    \filldraw[fill=black] ( 1.3,-0.3 ) circle (0.02);
    \node[right,inner sep=0] at (-0.4,1.6) {\small $\Bc$};
    \draw[->] (-0.6,0)--(1.7,0) node [right] {\small$c_i$}; 
    \draw[->] (0,-0.6)--(0,1.7) node [above] {\small$c_j$};
    \draw (-0.4,-0.4) rectangle (1.5,1.5);
    \node[] at (0.55,-0.7) {\small {(a)}};
\end{tikzpicture}\hspace{2ex}%
    \begin{tikzpicture}[scale=3.]

    \fill[fill=white, opacity=1] (0.2,0.2) rectangle (0.5,0.7);
    \fill[pattern=north west lines, pattern color=green, opacity=1] (0.2,-0.4) rectangle (1.5,1.5);
    \filldraw[fill=black!50] (0.2,-0.2) circle (0.04) node [below left]
    {\small$y_3$};

    \fill[fill=blue!30, opacity=0.7] (-0.4,0.7) rectangle (1.5,1.5);
    \filldraw[fill=black!50] (-0.15,0.7) circle (0.04) node [below left] {\small$y_1$};

    \fill[fill=blue!30, opacity=0.7] (0.8,0.2) rectangle (1.5,1.5);
    \filldraw[fill=black!50] (0.8,0.2) circle (0.04) node [below left] {\small$y_2$};
    \filldraw[fill=black] (-0.2, 0.3) circle (0.02);
    \filldraw[fill=black] ( 0.3, 0.1) circle (0.02);
    \filldraw[fill=black] ( 0.4, 0.5) circle (0.02);
    \filldraw[fill=black] ( 0.5,0.3)  circle (0.02);
    \filldraw[fill=black] ( 0.6,-0.1) circle (0.02);
    \filldraw[fill=black] ( 1.0,-0.15) circle (0.02);
    \filldraw[fill=black] ( 1.2,-0.1) circle (0.02);
    \filldraw[fill=black] ( 1.3,-0.3) circle (0.02);
    \coordinate (PO) at (1.5,0);
    \coordinate (P3) at ($(0.2,-0.2)+(10:0.04)$);
    \coordinate (P) at ($(PO)!0.5!(P3)$);

    \fill[fill=blue!30, opacity=0.7] ($(-0.4,-0.4)!(P)!(1.5,-0.4)$) rectangle (1.5,1.5);

    \node[right,inner sep=0] at (-0.4,1.6) {\small $\Bc$};
    \draw[->] (-0.6,0)--(1.7,0) node [right] {\small$c_i$}; 
    \draw[->] (0,-0.6)--(0,1.7) node [above] {\small$c_j$};
    \draw (-0.4,-0.4) rectangle (1.5,1.5);

    \draw[->,red!70,line width=2] (PO)--(P3);
    \filldraw[fill=red!70] (PO) circle (0.04) node [above left] {\small
    $\yanchor$};    

    \filldraw[fill=black!50] (P) circle (0.04) 
       node [below left=0.06] {\small$\tilde y_{u}$};
    \node[] at (0.55,-0.7) {\small {(b)}};
\end{tikzpicture}%
\end{center}
\caption{Illustration of the steps $10\to 12$ and $13\to15$ of
  Algorithm~\ref{alg:MultiStageUnif}. The objective is to build a
  uniform sample $\YY_{3}$ on~$G_3$ from $\YY_{2}$. The initial Pareto
  front $\PP_{0}$ is determined by evaluation results
  $y_1=(f(X_1),c(X_1))$ and $y_2=(f(X_2),c(X_2))$. $\PP_T$ corresponds
  to the Pareto front determined by $\PP_0\cup\{y_3\}$, with
  $y3=(f(X_3),c(X_3))$. At the end of steps 1--9, $y_3$ is not in $\PP$
  because the number of surviving particles in $\YY_{2}$ is too small:
  in (a), there is only one particle (black dot) in $G_{3}$ (white
  region).  Thus, intermediate subsets are needed. The main idea here is
  to build a \emph{continuous} path between $\PP_0$ and $\PPt$, which
  is illustrated in (b). Here, we pick $y^{\star} = y_{3}$ and since
  $y_3$ is not feasible, $q^{\star} < q$. Then, we set an anchor point
  $\yanchor$ on the edge of $\B$, as described in step~14, and we build
  an intermediate Pareto front $\tiPP_u$ determined by $y_1$, $y_2$
  and $\tilde y_{u}$, where $\tilde y_{u}$ lies on the segment
  ($\yanchor$--$y_3$). The intermediate Pareto front $\tiPP_u$ is
  chosen in such a way that the number of killed particles (grey dots)
  is not too large.}
\label{fig:illustr-algo-front-transition-cc}
\end{figure}

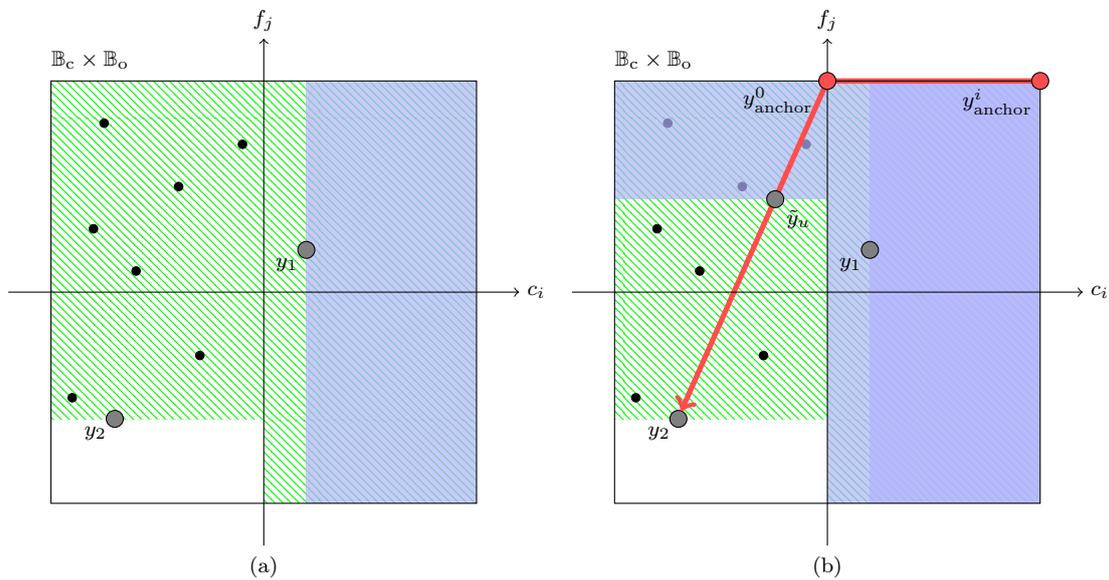
\begin{figure}
  \begin{center}
    \begin{tikzpicture}[scale=2.8]

    \fill[pattern=north west lines, pattern color=green, opacity=1] (-1,-0.6) rectangle (0,1);
    \fill[pattern=north west lines, pattern color=green, opacity=1] (0,-1) rectangle (1,1);
    \filldraw[fill=black!50] (-0.7,-0.6) circle (0.04) node [below left] {\small$y_2$};

    \fill[fill=blue!30, opacity=0.7] (0.2,-1) rectangle (1,1);
    \filldraw[fill=black!50] (0.2,0.2) circle (0.04) node [below left] {\small$y_1$};

    \filldraw[fill=black] (-0.9,-0.5) circle (0.02);
    \filldraw[fill=black] (-0.8, 0.3) circle (0.02);
    \filldraw[fill=black] (-0.75, 0.8) circle (0.02);
    \filldraw[fill=black] (-0.6, 0.1) circle (0.02);
    \filldraw[fill=black] (-0.4, 0.5) circle (0.02);
    \filldraw[fill=black] (-0.3,-0.3) circle (0.02);
    \filldraw[fill=black] (-0.1,0.7) circle (0.02);
    \node[right,inner sep=0] at (-1,1.1) {\small $\Bc\times\Bo$};
    \draw[->] (-1.2,0)--(1.2,0) node [right] {\small$c_i$}; 
    \draw[->] (0,-1.2)--(0,1.2) node [above] {\small$f_j$};
    \draw (-1,-1) rectangle (1,1);
    \node[] at (0.0,-1.3) {\small {(a)}};
\end{tikzpicture}\hspace{2ex}%
    \begin{tikzpicture}[scale=2.8]

    \fill[pattern=north west lines, pattern color=green, opacity=1] (-1,-0.6) rectangle (0,1);
    \fill[pattern=north west lines, pattern color=green, opacity=1] (0,-1) rectangle (1,1);
    \coordinate (P2) at (-0.7,-0.6);
    \filldraw[fill=black!50] (P2) circle (0.04) node [below left] {\small$y_2$};

    \coordinate (P1) at (0.2,0.2);
    \fill[fill=blue!30, opacity=0.7] (0.2,-1) rectangle (1,1);

    \filldraw[fill=black] (-0.9,-0.5) circle (0.02);
    \filldraw[fill=black] (-0.8, 0.3) circle (0.02);
    \filldraw[fill=black] (-0.75, 0.8) circle (0.02);
    \filldraw[fill=black] (-0.6, 0.1) circle (0.02);
    \filldraw[fill=black] (-0.4, 0.5) circle (0.02);
    \filldraw[fill=black] (-0.3,-0.3) circle (0.02);
    \filldraw[fill=black] (-0.1,0.7) circle (0.02);

    \coordinate (PO) at (0,1);
    \coordinate (P) at ($(PO)!0.35!(P2)$);
    \coordinate (Pi) at (1,1);

    \fill[fill=blue!30, opacity=0.7] ($(-1,-1)!(P)!(-1,1)$) rectangle (0,1);
    \fill[fill=blue!30, opacity=0.7] (0,-1) rectangle (1,1);
    \filldraw[fill=black!50] (P1) circle (0.04) node [below left] {\small$y_1$};
   
    \draw[->,red!70,line width=2] (Pi)--(PO)--($(P2)+(70:0.04)$);

    \filldraw[fill=black!50] (P) circle (0.04) 
       node [below right=0.06] {\small$\tilde y_{u}$};
    \node[right,inner sep=0] at (-1,1.1) {\small $\Bc\times\Bo$};
    \draw[->] (-1.2,0)--(1.2,0) node [right] {\small$c_i$}; 
    \draw[->] (0,-1.2)--(0,1.2) node [above] {\small$f_j$};
    \draw (-1,-1) rectangle (1,1);
    \filldraw[fill=red!70] (PO) circle (0.04) node [xshift=-0.65cm, yshift=-0.25cm] {\small$\yanchor^{0}$};
    \filldraw[fill=red!70] (Pi) circle (0.04) node [below left] {\small$\yanchor^{i}$};
    \node[] at (0.0,-1.3) {\small {(b)}};
\end{tikzpicture}%
\end{center}
\caption{Illustration of the steps $10\to 12$ and $16\to20$ of
  Algorithm~\ref{alg:MultiStageUnif}. The setting is the same as that
  described in Figure~\ref{fig:illustr-algo-front-transition-cc}, except
  that the new observation ($y_2$ in this case) is feasible. Hence, $q^{\star}=q$. As above, the main idea
  is to construct a continuous path between $\PP_0$ and $\PPt$,
  as illustrated in (b).}
\label{fig:illustr-algo-front-transition-fc}
\end{figure}

When the number of surviving particles is smaller than a prescribed fraction
$\nu$ of the population size, that is, when $N(G_{n+1};\, \YY_{n}) < m\nu$,
intermediate subsets are inserted in the decreasing
sequence~\eqref{equ:decreasSubsets} to ensure that the volume of the subsets
does not decrease too fast. The corrected version of
Algorithm~\ref{alg:RemResMovUnif} is described in
Algorithms~\ref{alg:AdaptRemResMovUnif} and~\ref{alg:MultiStageUnif}. The method
used in Algorithm~\ref{alg:MultiStageUnif} to construct the intermediate subsets 
is illustrated on Figures~\ref{fig:illustr-algo-front-transition-cc}
and~\ref{fig:illustr-algo-front-transition-fc}.

\begin{algorithm}[t]
  \caption{Modified procedure to construct~$\YY_n$}
  \label{alg:AdaptRemResMovUnif}
  \SetKwInput{KwNotation}{Notation}%
  \KwNotation{Given a set~$A$ in $\Y$, denote by $\Pareto(A)$ the set of points
    of $A$ that are not dominated by any other point of~$A$}%
  \eIf{$n = 0$}{%
    Generate $m$ independent and uniformly distributed particles over~$G_0 =
    \B$.\;%
  }{%
    Set $\PP_{n-1} = \Pareto \left( \left\{ \xi(X_1),\, \ldots, \xi(X_{n - 1}) \right\}
    \right)$.

    Set $\PP_n = \Pareto \left( \left\{ \xi(X_1),\, \ldots, \xi(X_n) \right\}
    \right) = \Pareto \left( \PP_{n-1} \cup \left\{ \xi(X_n) \right\} \right)$.

    Construct $\YY_{n}$ using the adaptive multi-level splitting procedure
    described in Algorithm~\ref{alg:MultiStageUnif}, with $\YY_{n-1}$,
    $\PP_{n-1}$, $\PP_n$ and $\B$ as inputs.\; }
\end{algorithm}

\begin{algorithm}
  \caption{Adaptive multi-level splitting in the~$\Y$-domain}
  \label{alg:MultiStageUnif}
  \newcommand\mycommfont[1]{\footnotesize{#1}}
  \SetCommentSty{mycommfont}
  \SetKwComment{Comment}{}{}
  \SetKwInput{KwNotations}{Notations}
  \KwNotations{Given a set~$A$ in $\Y$, denote by
    \begin{itemize}
    \item $\Pareto(A)$ the set of points of $A$ that are not dominated
      by any other point of~$A$,
    \item $G(A) := \B \setminus \left\{y \in \B;\, \exists y' \in A
        \text{ such that } y' \lhd y\right\}$ the region of $\B$ not
      dominated by $A$.
    \end{itemize}}
  \medskip%
  \SetKwInput{KwIn}{Inputs}
  \KwIn{$\YY_0$, $\PP_0$, $\PPt$ and $\B$ such that\;
    \begin{itemize}
    \item $\PP_0 = \Pareto \left( \PP_0 \right)$, i.e., no point
      in~$\PP_0$ is dominated by another point in~$\PP_0$, and similarly $\PPt
      = \Pareto \left( \PPt \right)$,
    \item $ G \left( \PPt \right) \subset G \left( \PP_0
      \right)$,
    \item $\YY_0 = \left( y_{0,k} \right)_{1 \le k \le m} \in \Y^m$
      is uniformly distributed on~$G \left( \PP_0 \right)$. Note that~$\YY_0$
      may contain replicated values.
    \item $\yLow_\obj, \yUpp_\obj, \yLow_\cons$ and $\yUpp_\cons$ such
      that  $\Bo = \bigl\{ y \in \Yo;\; \yLow_\obj \le y \le
      \yUpp_\obj \bigr\}$, $\Bc = \bigl\{ y \in \Yc;\; \yLow_\cons
      \le y \le \yUpp_\cons \bigr\}$, and $\B = \Bo \times \Bc$ contains $\PP_0$ and
      $\PPt$.
    \end{itemize}
  }
  \SetKwInput{KwOut}{Output}
  \KwOut{A set of particles $\YY_t = \left( y_{t,k} \right)_{1 \le k \le
      m} \in \Y^m$ uniformly distributed on~$G(\PPt)$.}
  $t \leftarrow 0$\;
  \While{$\PP_t \neq \PPt$}
  {
  	Initialize:  $\PP \leftarrow \PP_t$.\; 
  	$\PP$ is the front that we will build upon. First we try to add
        the points of $\PPt$ into $\PP$:\;
  	\For{$y \in \PPt$}
  	{
  		$\PP_{\try} \leftarrow \Pareto\left( \PP \cup \{ y \} \right)$\;
  		Compute the number~$N(G(\PP_{\try}); \YY_t)$ of
                particles of $\YY_t$ in 		$G(\PP_{\try})$\;
  		\If{$N(G(\PP_{\try}); \YY_t) \geq \nu m$}
  		{
  			$\PP \leftarrow \PP_{\try}$\;
  		}
  	}
        \Comment{At the end of this first step, either $\PP = \PPt$ or 
         $\PPt \setminus \PP $  contains points
        that cannot be added without killing a large number of
        particles, in which case we insert intermediate fronts.}

        \If{$\left( \PPt \setminus \PP \right) \neq \emptyset$}
  	{

          Randomly choose a point $y^\star = (\yo^\star, \yc^\star) \in
          \left( \PPt \setminus \PP \right)$ toward which we will try to
          augment the front $\PP$.\;

          Count the number $q^\star$ of constraints satisfied by $y^\star$.\;
          \eIf{$q^\star < q$}
  		{
                  $\yanchor \leftarrow (\yo^{\rm upp},\yc)\in\Bo\times\Bc$, where
                  $y_{\cons,j} = y_{\cons,j}^{\upp}$ if $y^\star_{\cons,j} > 0$ and
                  zero otherwise, $1 \leq j \leq q$.\; 
                  Find $\tiPP_u$ such
                  that~$N(G(\tiPP_u); \YY_t) \approx \nu m$ using a
                  dichotomy on $u\in[0,1]$, where $\tiPP_u = \Pareto(\PP \cup \{ 
                  \yanchor + u(y^{\star}-\yanchor) \})$.\;
  		}
  		{
  			$\yanchor^0 \leftarrow
                        (\yo^{\upp},0)\in\Bo\times\Bc$\;

                        $\yanchor^k \leftarrow
                        (\yo^{\upp},y^k_\cons)\in\Bo\times\Bc$, where
                        $y^k_{\cons,j} = y_{\cons,j}^{\upp}$ if $j = k$
                        and zero otherwise, for $1 \leq j \leq q$ and~$1
                        \leq k \leq q$.\;

  			\eIf{$N(G(\{\yanchor^0\}); \YY_t) \geq \nu m$}
  			{
  			Find $\tiPP_u$ such that~$N(G(\tiPP_u); \YY_t)
                        \approx \nu m$ using a dichotomy on $u\in[0,1]$,
                        where \vspace{-2ex}$$\tiPP_u = \Pareto(\PP \cup \{ \yanchor^0 +
                        u(y^{\star}-\yanchor^0) \}).$$\;\vspace{-2.5ex}
  			}
  			{
  				Find $\tiPP_u$ such  that~$N(G(\tiPP_u); \YY_t) \approx
                                \nu m$ using a dichotomy on $u\in[0,1]$, where \vspace{-2ex}
                                $$\tiPP_u = \Pareto(\PP \cup \{ \yanchor^1
                                + u(\yanchor^0-\yanchor^1) \} \cup \cdots \cup \{
                                \yanchor^q + u(\yanchor^0-\yanchor^q)
                                \}).$$\;\vspace{-2.5ex}
                        } 
               }
          $\PP \leftarrow \tiPP_u$\; 
        }
    $\PP_{t + 1} \leftarrow \PP$\;
    Generate~$\YY_{t+1} = \left( y_{t+1, k} \right)_{1 \le k \le m}$
    uniformly distributed on~$G\left( \PP_{t+1} \right)$ using the
    ``Remove-Resample-Move'' steps described in
    Algorithm~\ref{alg:RemResMovUnif}.\;%
    $t\leftarrow t+1$\; }
\end{algorithm}

\begin{remark}
  \label{rmk:critCalcSpecialCases}
  The algorithms presented in this section provide a general numerical method
  for the approximate computation of the expected improvement criterion, that
  can be used with multiple objectives, multiples constraints and possibly
  correlated Gaussian process models. When the objectives and constraints are
  independent, the decomposition introduced in
  Section~\ref{sec:critDecomposition} makes it possible to compute two integrals
  over spaces of lower dimension (over~$\Bc \setminus \Hnc$ and~$\Bo \setminus
  \Hno$, respectively) instead of one integral over~$G_n = \B \setminus H_n$. In
  fact, only one of the two integrals actually needs to be approximated
  numerically: indeed, the term~$\EIFeas$ of the decomposition can be calculated
  in closed form prior to finding feasible solutions, and the term~$\EIUnf$
  vanishes once a feasible observation has been made. We have taken advantage of
  this observation for all the numerical results presented in
  Section~\ref{sec:exp}.
\end{remark}

\subsection{Maximization of the sampling criterion}
\label{sec:crit-optim-using}

The optimization of the sampling criterion~\eqref{eq:ei-multi-cons} is a
difficult problem in itself because, even under the unconstrained
single-objective setting, the EI criterion is very often highly multi-modal. Our
proposal is to conduct a discrete search on a small set of good candidates
provided at each iteration by a sequential Monte Carlo algorithm, in the spirit
of \cite{benassi2012bayesian}, \cite{li2012bayesian}, \cite{li2012thesis}
and~\cite{benassi2013nouvel}.

The key of such an algorithm is the choice of a suitable sequence~$\left(
  \pi_n^\X \right)_{n \ge 0}$ of probability density functions on~$\X$, which
will be the targets of the SMC algorithm.  Desirable but antagonistic properties
for this sequence of densities are stability---$\pi^\X_{n+1}$ should not differ
too much from~$\pi^\X_n$---and concentration of the probability mass in regions
corresponding to high values of the expected improvement.  We propose,
following~\cite{benassi2012bayesian}, to consider the sequence defined by
\begin{equation*}\left\{
  \begin{array}{ll}
    \pi_n^\X(x)  \propto 1  &\text{if } n = 0,\\[1ex]
    \pi_n^\X(x) \propto \P_n(\xi(x) \in G_n)~~  &\text{for } n = 1, 2, \ldots    
  \end{array}\right.
\end{equation*}
In other words, we start from the uniform distribution on~$\X$ and then we use
the probability of improvement $x \mapsto \P_n(\xi(x) \in G_n)$ as an
un-normalized probability density function.

A procedure similar to that described in Algorithms~\ref{alg:RemResMovUnif} is
used to generate particles distributed from the target densities~$\pi_n^\X$. At
each step $n$ of the algorithm, our objective is to construct a set of weighted
particles
\begin{equation}
  \label{eq:particles-X}
  \XX_n = \left( x_{n,k},w_{n,k} \right)_{k=1}^{m} \in (\X \times \R)^{m}  
\end{equation}
such that the empirical distribution~$\sum_{k} w_{n,k} \delta_{x_{n,k}}$ (where
$\delta_x$ denotes the Dirac measure at $x\in\X$) is a good approximation, for
$m$ large enough, of the target distribution with density~$\pi_n^\X$. The main
difference with respect to Section~\ref{sec:crit-calc} is the introduction of
weighted particles, which makes it possible to deal with non-uniform target
distributions.  When a new sample is observed at step~$n$, the weights of the
particles are updated to fit the new density~$\pi_{n+1}^\X$:
\begin{equation}
  \label{eq:update-weights-X}
  {w}_{n+1, k}^0 \propto \frac{\pi_{n+1}^\X(x_{n, k})}{\pi_n^\X(x_{n,k})}\; {w}_{n, k}.
\end{equation}
The weighted sample $\XX_{n+1}^0 = ( x_{n,k}, w_{n+1, k}^0 )_{1 \leq k \leq m}$
is then distributed from $\pi_{n+1}^\X$. Since the densities $\pi_0,
\pi_1,\ldots$ become more and more concentrated as more information is obtained
about the functions $f$ and $c$, the regions of high values for $\pi_{n+1}^\X$
become different from the regions of high values for~$\pi_n^\X$. Consequently,
the weights of some particles degenerate to zero, indicating that those
particles are no longer good candidates for the optimization. Then, the
corresponding particles are killed, and the particles with non-degenerated
weights are replicated to keep the size of the population constant. All
particles are then moved randomly using an MCMC transition kernel targeting
$\pi_{n+1}^\X$, in order to restore some diversity. The corresponding procedure,
which is very similar to that described in Algorithm~\ref{alg:RemResMovUnif}, is
summarized in Algorithm~\ref{alg:RewRemMovePi}.

When the densities~$\pi_n^\X$ and~$\pi_{n+1}^\X$ are too far apart, it may
happen that the number of particles with non-degenerated weights is very small
and that the empirical distribution $\sum_{k} w_{n+1,k}\, \delta_{x_{n,k}}$ is not
a good approximation of~$\pi_{n+1}^{\X}$. This is similar to the problem
explained in Section~\ref{sec:crit-calc}, except that in the case of non uniform
target densities, we use the Effective Sample Size (ESS) to detect degeneracy
\citep[see, e.g., ][]{del2006sequential}, instead of simply counting the
surviving particles. When the ESS falls below a prescribed fraction of the
population size, we insert intermediate densities, in a similar way to what was
described in Section~\ref{sec:crit-calc}. The intermediate densities are of the
form $\tilde \pi_{u}(x) \propto \P_n(\xi(x) \in \tilde G_{u})$, with $G_{n+1}
\subset \tilde G_u \subset G_n$. The corresponding modification of
Algorithm~\ref{alg:RewRemMovePi} is straightforward. It is very similar to the
procedure described in Algorithms~\ref{alg:AdaptRemResMovUnif}
and~\ref{alg:MultiStageUnif} and is not repeated here for the sake of brevity.

\begin{algorithm}[t]
  \caption{Reweight-Resample-Move procedure to construct~$\XX_n$}
  \label{alg:RewRemMovePi}
  \eIf{$n = 0$}{
    Set $\XX_{0}= \left( x_{0,k}, \frac{1}{m} \right)_{1 \le k \le m}$ %
    with $x_{0,1}, \ldots, x_{0,m}$ independent and uniformly distributed
    on~$\X$. \;
  }{
    \emph{Reweight}~$\XX_{n-1}$ according to
    Equation~\eqref{eq:update-weights-X} to obtain $\XX_n^0$. \;
    \emph{Resample} with a residual resampling scheme \citep[see,
    e.g.,][]{douc2005comparison} to obtain a set of particles $\XX_n^1 = \left(
      x^1_{n,k},\frac{1}{m} \right)_{1 \le k \le m}$. \;
    \emph{Move} the particles with an MCMC transition kernel to obtain $\XX_n =
    \left( x_{n,k}, \frac{1}{m} \right)_{1 \le k \le m}$. \;
  }
\end{algorithm}

\begin{remark} 
  \label{rem:MC-approx}
  A closed form expression of the probability of improvement is available in the
  single-objective case, as soon as one feasible point has been found.  When no
  closed form expression is available, we estimate the probability of
  improvement using a Monte Carlo approximation: $1/N \sum_{k=1}^N
  \mathds{1}_{G_n}(Z_k)$, where $(Z_k)_{1 \leq k \leq N}$ is an $N$-sample of
  Gaussian vectors, distributed as~$\xi(x)$ under~$\P_n$. A rigorous
  justification for the use of such an unbiased estimator inside a
  Metropolis-Hastings transition kernel (see the \emph{Move} step of
  Algorithm~\ref{alg:RewRemMovePi}) is provided by \cite{andrieu2009pseudo}.
\end{remark}

\begin{remark}
  It sometimes happens that a new evaluation result---say, the $n$-th evaluation
  result---changes the posterior so dramatically that the ESS falls below the
  threshold~$\nu m$ (see Algorithm~\ref{alg:MultiStageUnif}) for the
  \emph{current} region~$G_{n - 1}$.  When that happens, we simply restart the
  sequential Monte Carlo procedure using a sequence of transitions from~$\PP_0 =
  \varnothing$ to the target front~$\PPt$
  (notation introduced in Algorithm~\ref{alg:MultiStageUnif}).
\end{remark}

\begin{remark}
  For the sake of clarity, the number of particles used in the SMC approximation
  has been denoted by~$m$ both in Section~\ref{sec:crit-calc} and in
  Section~\ref{sec:crit-optim-using}. Note that the two sample sizes are,
  actually, not tied to each other. We will denote them respectively by~$m_\XX$
  and~$m_\YY$ in Section~\ref{sec:exp}.
\end{remark}

\section{Experiments}
\label{sec:exp}

\subsection{Settings}
\label{sec:exp-settings}

The BMOO algorithm has been written in the Matlab/Octave programming language,
using the Small Toolbox for Kriging (STK) \citep{stktoolbox} for the
Gaussian process modeling part. All simulation results have been obtained using
Matlab~R2014b.

In all our experiments, the algorithm is initialized with a maximin Latin
hypercube design consisting of $N_{\rm init}= 3d$ evaluations. This is an
arbitrary rule of thumb.  A dedicated discussion about the size of initial
designs can be found in \cite{loeppky2009choosing}.  The objective and
constraint functions are modeled using independent Gaussian processes, with a
constant but unknown mean function, and a Mat\'ern covariance function with
regularity parameter $\nu=5/2$ \citep[these settings are described, for
instance, in][]{bect2012sequential}.  The variance parameter $\sigma^{2}$ and
the range parameters $\theta_i$, $1 \leq i \leq d$, of the covariance functions
are (re-)estimated at each iteration using a maximum a posteriori (MAP)
estimator.
Besides, we assume that the observations are slightly noisy to improve the
conditioning of the covariance matrices, as is usually done in kriging
implementations.

In Sections~\ref{sec:so_bench} and~\ref{sec:mo_bench}, the computation of the
expected improvement is carried out using the SMC method described in
Section~\ref{sec:crit-calc}. Taking advantage of
Remark~\ref{rmk:critCalcSpecialCases}, the integration is performed only on the
constraint space (prior to finding a feasible point) or the objective space
(once a feasible point is found). In the case of single-objective problems
(Section~\ref{sec:so_bench}), we perform exact calculation
using~\eqref{eq:schonlauCrit} once a feasible point has been observed.  The
parameter $\nu$ of Algorithm~\ref{alg:MultiStageUnif} is set to $0.2$ and we
take $m = m_\YY = 1000$ particles.
The bounding hyper-rectangles $\Bo$ and $\Bc$ are determined using the adaptive
procedure described in Appendix~\ref{sec:annexe:adaptBoBc} with $\lambdaObj =
\lambdaCons = 5$.

For the optimization of the sampling criterion, we use the SMC method of
Section~\ref{sec:crit-optim-using}, with~$m = m_{\XX} = 1000$ particles,
residual resampling \citep{douc2005comparison}, and an adaptive anisotropic
Gaussian random walk Metropolis-Hastings algorithm to move the particles
\citep{andrieu2008tutorial, roberts2009examples}. When the probability of
improvement cannot be written under closed-form, a Monte Carlo approximation
(see Remark~\ref{rem:MC-approx}) with $N = 100$ simulations is used.

\subsection{Illustration on a constrained multi-objective problem}

In this section, the proposed method is illustrated  on a two-dimensional two-objective toy
problem, which allows for easy visualization. The optimization problem is as
follows:
\begin{equation*}
  \begin{array}{p{2cm}l}
    \text{minimize}   \quad & f_1 \text{ and } f_2\,,\\
    \text{subject to}\quad  & c(x) \leq 0 \text{ and } x = (x_1,x_2) \in [-5,10]\times[0,15]\,,
  \end{array}
\end{equation*}
where
\begin{equation*}
  \left\{
    \begin{array}{l c l}
      f_1:(x_1,x_2)  &\mapsto& -(x_1-10)^2-(x_2-15)^2,\\[2em]
      f_2:(x_1,x_2)  &\mapsto& -(x_1+5)^2-x_2^2,\\[1em]
      c:(x_1,x_2)    &\mapsto& \displaystyle \left(x_2 - \frac{5.1}{4\pi^2}x_1^2
        + \frac{5}{\pi}x_1 - 6 \right)^ 2
      + 10\left(1 - \frac{1}{8\pi}\right)\cos (x_1) + 9.
    \end{array}
  \right.
\end{equation*}

\begin{figure}[tp]
  \centering
  \newlength{\figW} \setlength{\figW}{0.295\linewidth}
  \newlength{\vSki} \setlength{\vSki}{1mm}
  \begin{minipage}[t]{\figW} \centering
    \includegraphics[width=\figW]{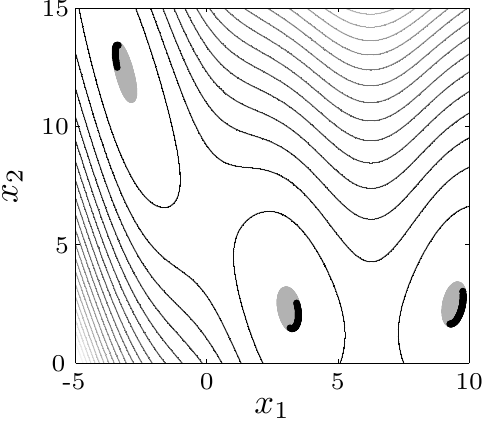}\\[\vSki]
    \hspace{3mm} (a) Constraint function
  \end{minipage}
  \hspace{1em}
  \begin{minipage}[t]{\figW} \centering
    \includegraphics[width=\figW]{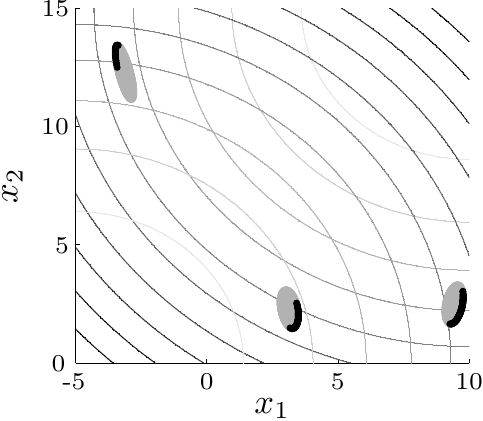}\\[\vSki]
    \hspace{3mm} (b) Objective functions
  \end{minipage}
  \hspace{1em}
  \begin{minipage}[t]{\figW} \centering
    \includegraphics[width=\figW]{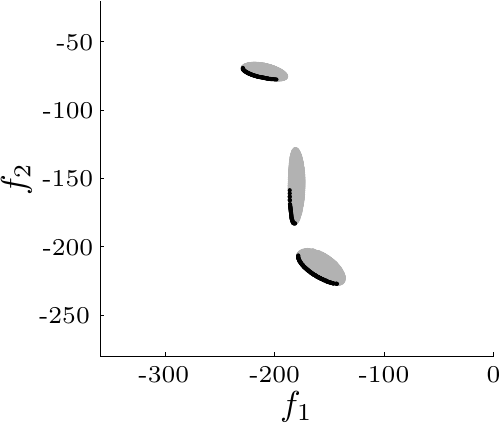}\\[\vSki]
    \hspace{5mm} (c) Objective space
  \end{minipage}
  \caption{Figure~(a) represents contour lines of the constraint function, and
    Figure~(b) corresponds to contour lines of the two  objective functions. The
    three gray areas correspond to the feasible region on Figures~(a) and~(b),
    and the image of the feasible region by the objective functions
    on Figure~(c). Thick dark curves correspond to the set of feasible and
    non-dominated solutions on Figures~(a) and~(b). On Figure~(c), thick dark curves
    correspond to the Pareto front.}
  \label{fig:branin2}
\end{figure}

The set of solutions to that problem is represented on
Figure~\ref{fig:branin2}. The feasible subset consists of three disconnected
regions of relatively small size compared to that of the search space. The
solution Pareto front consists of three corresponding disconnected fronts in the
space of objectives. (The visualization is achieved by evaluating the objectives
and constraints on a fine grid, which would not be affordable in the case of
truly expensive-to-evaluate functions.)

\begin{figure} \centering
  \setlength{\figW}{51mm} \setlength{\vSki}{1mm} \def\hh{\hspace{5mm}}
  \includegraphics[width=\figW]{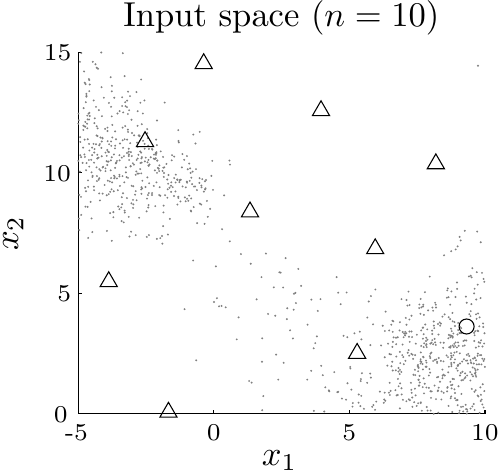} \hh
  \includegraphics[width=\figW]{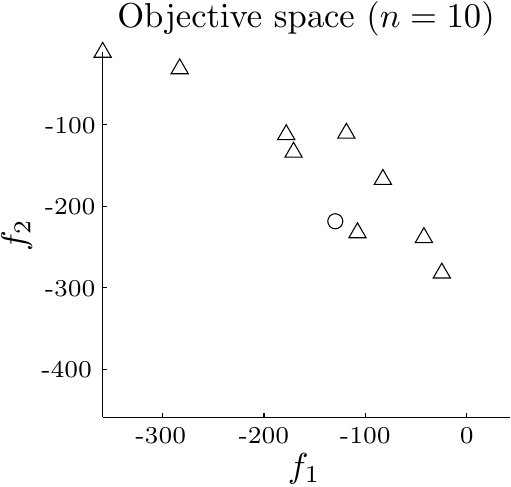}\\[\vSki]
  \includegraphics[width=\figW]{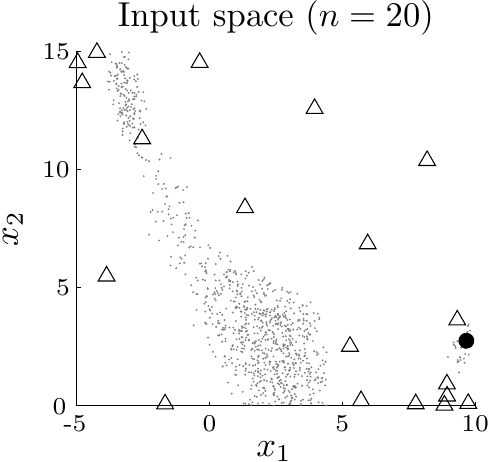} \hh
  \includegraphics[width=\figW]{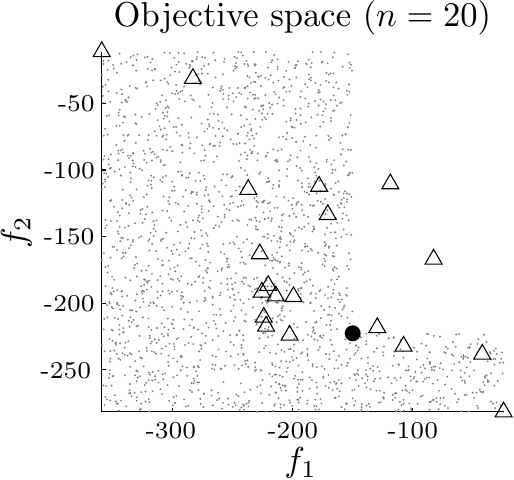}\\[\vSki]
  \includegraphics[width=\figW]{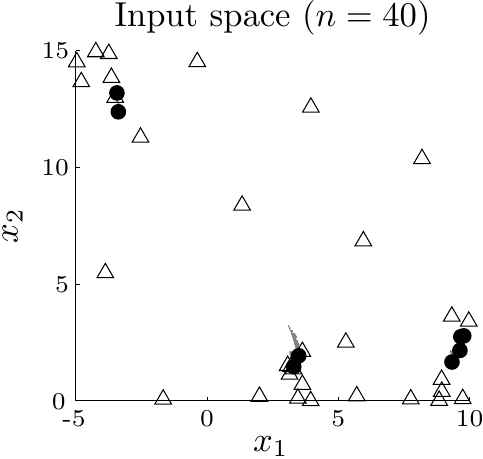} \hh
  \includegraphics[width=\figW]{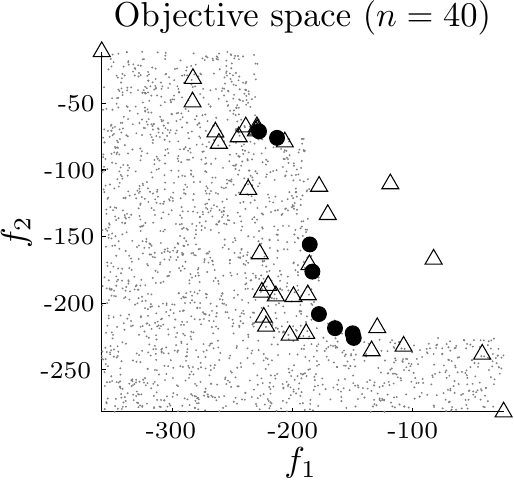}\\[\vSki]
  \includegraphics[width=\figW]{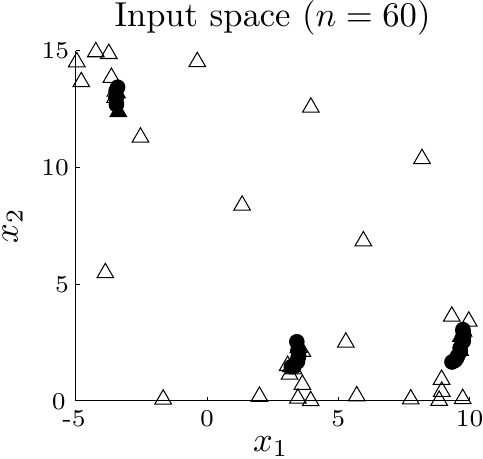} \hh
  \includegraphics[width=\figW]{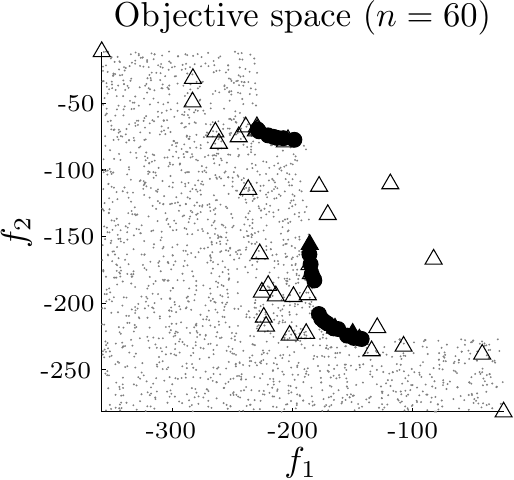}\\[\vSki]
  \caption{Convergence of the algorithm after $n = 10, 20, 40$ and $60$
    evaluations. The left column shows the input space~$\X$, whereas the
    right one shows the objective space $\Bo$. Dominated observations
    are represented by triangles (filled or empty), and non-dominated
    ones by circles (or disks). The symbols are filled for feasible
    points and empty otherwise. On the left column, the small dots
    represent the particles used for the optimization of the expected
    improvement (see Section~\ref{sec:crit-optim-using}). On the right
    column, the small dots represent the particles used for the
    computation of the expected improvement (see
    Section~\ref{sec:crit-calc}). Note in particular that they appear
    only when a feasible point has been observed: before that, the term
    $\EIFeas_n$ (see Section~\ref{sec:critDecomposition}) can be
    computed analytically.}
  \label{fig:illustration}
\end{figure}

The behavior of BMOO is presented in Figure~\ref{fig:illustration}. The algorithm 
is initialized with $5d =10$ function evaluations.  
Figure~\ref{fig:illustration} shows that the
algorithm correctly samples the three feasible regions, and achieves good
covering of the solution Pareto front after only a few iterations. Note that no
feasible solution is given at the beginning of the procedure and that the
algorithm finds one after 10 iterations.

\subsection{Mono-objective optimization benchmark}
\label{sec:so_bench}

The first benchmark that we use to assess the performance of BMOO consists of a
set of sixteen constrained single-objective test problems proposed
by~\cite{regis2014constrained}.  Table~\ref{tab:problem_so} summarizes the main
features of these problems.  The input dimension~$d$ varies from~$2$ to~$20$,
and the number~$q$ of constraints from~$1$ to~$38$.  The problems may have
linear or non-linear constraints but this information is not used by the
algorithms that we use in our comparisons (all functions are regarded as black
boxes). Column $\Gamma(\%)$ gives the ratio in percents of the volume of
feasible region~$C$ to the volume of the search space~$\X$.  This ratio has been
estimated using Monte Carlo sampling and gives an indication on the difficulty
of the problem for finding a feasible point. Note that problems g1, g3mod, g6,
g7, g10, g19 and in particular problem g18 have very small feasible regions. The
last two columns correspond respectively to the best known feasible objective
value and to target values for the optimization. The target values are the ones
used in the work of \cite{regis2014constrained}.

\begin{table}
  \centering
  \renewcommand{\arraystretch}{1.2}
  \rowcolors{1}{white}{gray!10}
  \captionsetup{justification=centering}
  \begin{tabular}{|l|c|c|c|c|c|}
    \hline
    Pbm & $d$ & $q$ & $\Gamma(\%)$ & Best & Target \\
    \hline
    g1 & 13 & 9 & $4\cdot10^{-4}$ & -15 & -14.85 \\    
    g3mod & 20 & 1 & $10^{-4}$ & -0.693 & -0.33 \\
    g5mod & 4 & 5 & $8.7\cdot10^{-2}$ & 5126.2 & 5150 \\
    g6 & 2 & 2 & $6.6\cdot10^{-3}$ & -6961.8 & -6800 \\
    g7 & 10 & 8 & $10^{-4}$ & 24.3 & 25 \\
    g8 & 2 & 2 & 0.86 & -0.0958 & -0.09 \\
    g9 & 7 & 4 & 0.52 & 680.6 & 1000 \\
    g10 & 8 & 6 & $7\cdot10^{-4}$ & 7049.4 & 8000 \\
    g13mod & 5 & 3 & 4.5 & 0.0035 & 0.005 \\
    g16 & 5 & 38 & $1.3\cdot10^{-2}$ & -1.916 & -1.8 \\
    g18 & 9 & 13 & $2\cdot10^{-10}$ & -0.866 & -0.8 \\
    g19 & 15 & 5 & $3.4\cdot10^{-3}$ & 32.66 & 40 \\
    g24 & 2 & 2 & 44.3 & -5.5080 & -5 \\
    SR7 & 7 & 11 & $9.3\cdot10^{-2}$ & 2994.4 & 2995 \\
    PVD4 & 4 & 3 & $5.6\cdot10^{-1}$ & 5804.3 & 6000 \\
    WB4 & 4 & 6 & $5.6\cdot10^{-2}$ & 2.3813 & 2.5 \\
    \hline
  \end{tabular}
  \caption{Main features of the mono-objective problems of our first
    benchmark.}
  \label{tab:problem_so}
\end{table}

BMOO is compared to two classes of algorithms. The first class consists of four
local optimization algorithms: the COBYLA algorithm of \cite{powell1994direct},
using the implementation proposed by \cite{johnson2012nlopt}, and three
algorithms implemented in the Matlab function
\texttt{fmincon}\footnote{Optimization toolbox v7.1, MATLAB R2014b}, namely, an
interior-point algorithm, an active-set algorithm and an SQP algorithm. Local
optimization methods are known to perform well on a limited budget provided that
good starting points are chosen. We think that they are relevant competitors in
our context. The second class of algorithms are those proposed by
\cite{regis2014constrained}, which are state-of-the-art---to the best of our
knowledge---algorithms for constrained optimization under a limited budget of
evaluations.

Each algorithm is run 30 times on each problem of the benchmark. Note that we
use a random starting point uniformly distributed inside the search domain for
local search algorithms, and a random initial design for BMOO, as described in
Section~\ref{sec:exp-settings}. For the local search algorithms the maximum
number of evaluations is set to two hundred times the dimension $d$ of the
problem. Concerning the algorithms proposed by~\cite{regis2014constrained}, we
simply reproduce the results presented by the author; the
reader is referred to the original article for more details about the settings.  Results
are presented in Tables~\ref{tab:res1} and~\ref{tab:res2}.  In both tables, a
solution is considered as feasible when there is no constraint violation larger
than~$10^{-5}$.

In Table~\ref{tab:res1}, we measure the performance for finding feasible
solutions.  For local algorithms and Regis' algorithms, only the results of the
best scoring algorithm are reported in the table. Full results are
presented in
Appendix~\ref{sec:annexe:mono-bench-results}.  For local algorithms, the first
column indicates the best scoring algorithm: Cob for the COBYLA algorithm, IP
for the interior-point algorithm, AS for the active-set algorithm and SQP for
the SQP algorithm. Similarly, for the algorithms proposed by
\cite{regis2014constrained}, the first column indicates the best scoring
algorithm: CG for COBRA-Global, CL for COBRA-Local and Ext for
Extended-ConstrLMSRBF. The second column gives the number of successful runs---a
run being successful when at least one feasible solution has been found. The
third column gives the number of function evaluations that were required to find
the first feasible point, averaged over the successful runs. The corresponding
standard deviation is given in parentheses.

Table~\ref{tab:res2} presents convergence results. Again, for local algorithms
and for those proposed by~\cite{regis2014constrained}, the first column
indicates the best scoring algorithm. The next columns give successively the
number of successful runs (a run being considered successful when a feasible
solution with objective value below the target value of
Table~\ref{tab:problem_so} has been found), the average number---over successful
runs---of evaluations that were required to reach the target value, and the
corresponding standard deviation (in parentheses). The reader is referred to
Appendix~\ref{sec:annexe:mono-bench-results} for the full results.

\begin{table}
  \centering
  \renewcommand{\arraystretch}{1.2}
  \rowcolors{1}{white}{gray!10}
  \begin{tabular}{|l|c|c|c|c|c|c|c|c|} 
    \hline 
    Pbm & \multicolumn{3}{c|}{Local (best among 4)} &
    \multicolumn{3}{c|}{Regis (best among 3)} & \multicolumn{2}{c|}{BMOO}\\ 
    \hline 
    g1 & IP & 30 & 128.4 (27.8) & CG & 30 & $\bm{15.0}$ (0) & 30 & 44.2 (1.9)\\ 
    g3mod & IP & 30 & 342.3 (66.3) & Ext & 30 & $\bm{31.2}$ (0.3) & 30 & 63.1 (0.6)\\ 
    g5mod & AS & 30 & 35.0 (5.5) & CL & 30 & $\bm{6.4}$ (0.1) & 30 & 13.0 (1.2)\\ 
    g6 & AS & 30 & 29.7 (5.0) & CL & 30 & $\bm{10.9}$ (0.3) & 30 & $\bm{9.7}$ (0.7)\\ 
    g7 & SQP & 30 & 107.6 (9.3) & CG & 30 & 47.5 (4.6) & 30 & $\bm{38.8}$ (3.3)\\ 
    g8 & IP & 30 & 12.1 (7.7) & CL & 30 & $\bm{6.5}$ (0.2) & 30 & $\bm{7.0}$ (0.2)\\ 
    g9 & IP & 30 & 170.9 (42.9) & CG & 30 & $\bm{21.5}$ (1.9) & 30 & $\bm{21.8}$ (5.1)\\ 
    g10 & SQP & 25 & 144.6 (132.3) & CG & 30 & $\bm{22.8}$ (1.5) & 30 & 71.5 (28.1)\\ 
    g13mod & IP & 30 & 21.4 (17.1) & Ext & 30 & $\bm{8.6}$ (0.7) & 30 & 10.5 (5.6)\\ 
    g16 & Cob & 27 & 31.5 (20.4) & Ext & 30 & $\bm{19.6}$ (1.8) & 30 & $\bm{21.7}$ (7.3)\\ 
    g18 & SQP & 30 & $\bm{101.9}$ (19.8) & CL & 30 & $\bm{108.6}$ (6.5) & 0 & - (-)\\ 
    g19 & SQP & 30 & $\bm{19.7}$ (6.1) & CL & 30 & $\bm{16.5}$ (0.5) & 30 & 46.4 (3.0)\\ 
    g24 & IP & 30 & 4.0 (3.5) & CG & 30 & $\bm{1.3}$ (0.1) & 30 & 2.6 (1.6)\\ 
    SR7 & SQP & 30 & 27.1 (3.6) & CG & 30 & $\bm{9.5}$ (0.1) & 30 & 22.0 (0.2)\\ 
    WB4 & SQP & 30 & 76.6 (21.9) & CL & 30 & 37.4 (5.9) & 30 & $\bm{19.1}$ (5.8)\\ 
    PVD4 & SQP & 26 & $\bm{7.6}$ (4.8) & CG & 30 & $\bm{7.9}$ (0.4) & 30 & 15.7 (5.7)\\ 
    \hline 
  \end{tabular} 
  \caption{Number of evaluations to find a first feasible point.
    In bold, the good results in terms of average number of evaluations.
    We consider the results to be good if more than 20 runs where
    successful and the average number of evaluations is at 
    most 20\% above the best result. 
    See Tables~\ref{tab:local_feasible_full} and~\ref{tab:regis_feasible_full}
    in Appendix~\ref{sec:annexe:mono-bench-results} for more detailed results.
    Dash symbols are used when a value cannot be calculated.}
  \label{tab:res1}
  \vspace{\floatsep}

  \rowcolors{1}{white}{gray!10}
  \begin{tabular}{|l|c|c|c|c|c|c|c|c|} 
    \hline 
    Pbm & \multicolumn{3}{c|}{Local (best among 4)} &
    \multicolumn{3}{c|}{Regis (best among 3)} & \multicolumn{2}{c|}{BMOO}\\ 
    \hline 
    g1 & IP & 20 & 349.7 (57.0) & CG & 30 & 125.2 (15.3) & 30 & $\bm{57.7}$ (2.6)\\ 
    g3mod & IP & 30 & 356.9 (65.1) & Ext & 30 & $\bm{141.7}$ (8.6) & 0 & - (-)\\ 
    g5mod & AS & 30 & 35.8 (4.3) & CL & 30 & $\bm{12.9}$ (0.5) & 30 & 16.3 (0.6)\\ 
    g6 & AS & 30 & 29.7 (5.0) & CL & 30 & 53.6 (14.0) & 30 & $\bm{13.3}$ (0.8)\\ 
    g7 & SQP & 30 & 107.6 (9.3) & CG & 30 & 99.8 (5.7) & 30 & $\bm{55.8}$ (2.8)\\ 
    g8 & IP & 18 & 59.3 (87.0) & CL & 30 & $\bm{30.3}$ (2.8) & 30 & $\bm{26.3}$ (10.4)\\ 
    g9 & IP & 30 & 179.3 (42.0) & CG & 30 & 176.4 (26.3) & 30 & $\bm{61.6}$ (14.3)\\ 
    g10 & SQP & 18 & 658.3 (316.7) & CG & 29 & $\bm{193.7}$ (-) & 0 & - (-)\\ 
    g13mod & IP & 25 & $\bm{122.5}$ (70.3) & Ext & 30 & $\bm{146.4}$ (29.2) & 30 & 180.3 (84.6)\\ 
    g16 & Cob & 27 & 60.0 (65.2) & Ext & 30 & 38.4 (3.6) & 30 & $\bm{30.3}$ (12.3)\\ 
    g18 & SQP & 21 & $\bm{97.5}$ (23.8) & CL & 24 & 195.9 (-) & 0 & - (-)\\ 
    g19 & SQP & 30 & $\bm{61.3}$ (12.4) & CL & 30 & 698.5 (75.3) & 30 & 133.3 (6.2)\\ 
    g24 & IP & 16 & 10.4 (5.3) & CG & 30 & $\bm{9.0}$ (0) & 30 & $\bm{9.9}$ (1.0)\\ 
    SR7 & SQP & 30 & $\bm{27.1}$ (3.6) & CG & 30 & $\bm{33.5}$ (1.6) & 30 & $\bm{29.3}$ (0.7)\\ 
    WB4 & SQP & 30 & 78.3 (18.0) & CL & 30 & 164.6 (12.2) & 30 & $\bm{44.5}$ (13.3)\\ 
    PVD4 & SQP & 23 & $\bm{54.7}$ (27.5) & CG & 30 & 155.4 (38.2) & 2 & 151.0 (21.2)\\ 
    \hline 
  \end{tabular} 
  \caption{Number of evaluations to reach specified target. See Table~\ref{tab:res1} for conventions. See Tables~\ref{tab:local_target_full} and~\ref{tab:regis_target_full}
    in Appendix~\ref{sec:annexe:mono-bench-results} for more detailed results.}
  \label{tab:res2}
\end{table} 

BMOO achieves very good results on most test problems. It very often comes close
to the best algorithm in each of the two classes of competitors, and sometimes
significantly outperforms both of them---see, in particular, the results for g1,
g6, g7, g9, g16 and WB4 in Table~\ref{tab:res2}. However, BMOO stalls on test
problems g3mod, g10, g18 and PVD4. We were able to identify the causes of theses
problems and to propose remedies, which are presented in the following
paragraphs. It can also be observed that BMOO is sometimes slower than the best
algorithm of \cite{regis2014constrained} to find a first feasible point. In
almost all cases (except for g10, g18 and PVD4, which are discussed separately
below), this is easily explained by the size of the initial design which is
$N_{\rm init} = 3d$ in our experiments (see
Section~\ref{sec:exp-settings}). Further work on this issue is required to make
it possible to start BMOO with a much smaller set of evaluations.

Regarding g3mod, g10 and PVD4, the difficulty lies in the presence of functions,
among the objective or the constraints, which are not adequately modeled using a
Gaussian process with a stationary covariance function. However, as we can see in Table~\ref{tab:modified-problems-results}, the
performances of BMOO are greatly improved in all three
cases if a transformation of the form $f
\rightarrow f^\lambda$  (for $\lambda>0$) is applied to the functions that cause the
problem (see Appendix~\ref{sec:annexe:modified-problems} for more details).
Thus, we think that the theoretical foundations of BMOO are not being questioned
by these tests problems, but further work is needed on the Gaussian process
models for a proper treatment of these cases.  In light of the results of our
experiments, one promising direction would be to consider models of the form
$\xi^\lambda$, where $\xi$ is a Gaussian process and $\lambda$ is a parameter to
be estimated from the evaluation results \citep[see, e.g.,][]{boxcox1964,
  snelson2004warped}.

Regarding the g18 test problem, the difficulty stems from our choice of a
sampling density derived from the probability of improvement for optimizing the
expected improvement. When the number of constraints is high ($q = 13$ for the
g18 test problem) and no feasible point has yet been found, the expected number
of particles in the feasible region~$C$ is typically very small with this
choice of density.  Consequently, there is a high probability that none of the
particles produced by the SMC algorithm are good candidates for the optimization
of the expected improvement.  To illustrate this phenomenon, consider the
following idealized setting. Suppose that $q = d$, $\X = \left[ -1/2, 1/2
\right]^q$ and $c_j:x = (x_1,\ldots,x_q) \mapsto |x_j| - \frac{\varepsilon}{2}$,
$j = 1, \ldots, q$, for some~$\varepsilon \in \left(0; 1\right]$. Thus, the
feasible domain is $C = \left[ -\varepsilon/2, \varepsilon/2 \right]^{q}$ and
the volume of the subset of~$\X$ where exactly~$k$ constraints are satisfied is
\begin{equation*}
  V_k \approx \left( \begin{smallmatrix} q\\k \end{smallmatrix} \right)\,
  \varepsilon^k\, \left( 1 - \varepsilon \right)^{q-k}.
\end{equation*}
Assume moreover that the
Gaussian process models are almost perfect, i.e.,
\begin{equation}
  \P_n \left( \xi_{\cons, j}(x) \leq 0 \right) \approx
  \begin{cases}
    1, & \text{if } c_j(x) \le 0,\\ 
    0, & \text{otherwise,}
  \end{cases}
\end{equation}
for $j = 1, \ldots, q$. Further assume $n = 1$ with $X_1 = \left( \frac{1}{2},
  \ldots, \frac{1}{2} \right)$ and observe that $\xi(X_1)$ is dominated
by~$\xi(x)$ for any~$x \in \X$ (except at the corners) so that the probability
of improvement~$\P_n \left( \xi(x) \in G_1 \right)$ is close to one everywhere
on~$\X$. As a consequence, the sampling density~$\pi_1^\X$ that we use to
optimize the expected improvement is (approximately) uniform on~$\X$ and the
expected number of particles satisfying exactly $k$ constraints is~$m\,V_k$.  In
particular, if~$q$ is large, the particles thus tend to concentrate in regions
where~$k \approx q \varepsilon$, and the expected number~$m\, V_q$ of particles
in~$C$ is small. To compensate for the decrease of~$V_k$, when $k$ is close
to~$q$, we suggest using the following modified sampling density:
\begin{equation*}
  \pi_n^\X \propto \E_n \left( K(x)!\, \one_{\xi(x) \in G_n} \right),
\end{equation*}
where $K(x)$ is the number of constraints satisfied by~$\xi$
at~$x$. Table~\ref{tab:g18-mpi} shows the promising results obtained with this
modified density on~g18. Further investigations on this particular issue are
left for future work.

\begin{table}
  \centering
  \renewcommand{\arraystretch}{1.2}
  \rowcolors{1}{white}{gray!10}
  \begin{tabular}{|l|c|c|c|c|} 
    \hline 
    Pbm & \multicolumn{2}{c|}{Feasible} & \multicolumn{2}{c|}{Target}\\ 
    \hline 
    modified-g3mod & 30 & 63.3 (0.8) & 30 & 151.8 (12.2)\\ 
    modified-g10 & 30 & 48.4 (8.0) & 30 & 63.1 (10.4)\\
    modified-PVD4 & 30 & 12.9 (1.6) & 30 & 32.9 (13.2)\\ 
    \hline 
  \end{tabular} 
  \caption{Number of evaluations to find a first feasible point and to
    reach the target on transformed versions of the
    g3mod, g10 and PDV4 problems, using the BMOO algorithm.} 
  \label{tab:modified-problems-results} 
  \vspace{\floatsep}

  \rowcolors{1}{white}{gray!10}
  \begin{tabular}{|l|c|c|c|c|} 
    \hline 
    Pbm & \multicolumn{2}{c|}{Feasible} & \multicolumn{2}{c|}{Target}\\ 
    \hline 
    g18 & 30 & 75.5 (11.5) & 30 & 83.6 (9.1)\\ 
    \hline 
  \end{tabular} 
  \caption{\label{tab:g18-mpi} Number of evaluations to find a first
    feasible point and to reach the target using a modified probability
    density function for the criterion optimization.} 
\end{table}

\subsection{Multi-objective optimization benchmark}
\label{sec:mo_bench}

\begin{figure}
  \centering \setlength{\vSki}{3mm}
  \newcommand{\mysubfig}[3][1mm]{\begin{minipage}[b]{7cm} \centering
      \includegraphics[width=55mm]{#2}\\[#1] \hspace{3mm} #3 \end{minipage}}
  \mysubfig{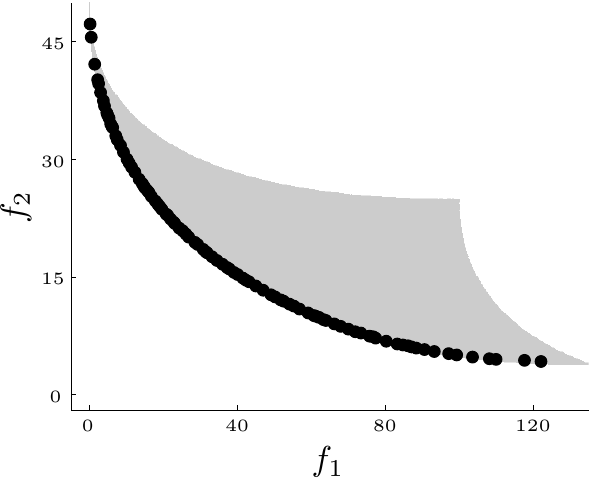}{(a) BNH}
  \mysubfig{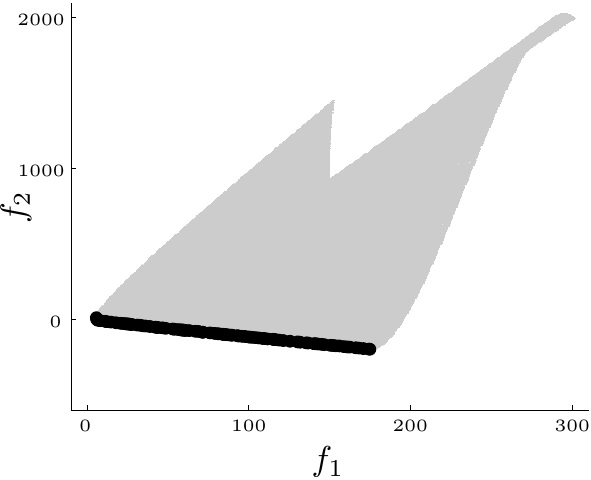}{(b) SRN}\\[\vSki]
  \mysubfig{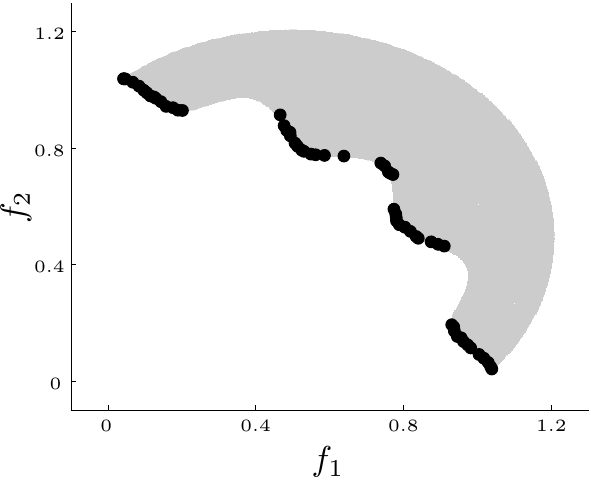}{(c) TNK}
  \mysubfig{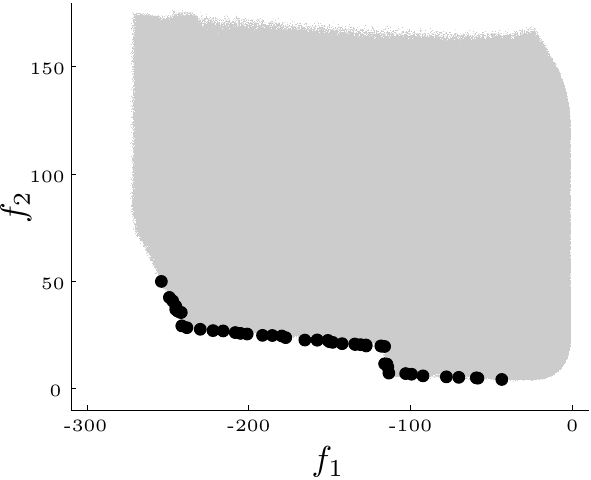}{(d) OSY}\\[\vSki]
  \mysubfig{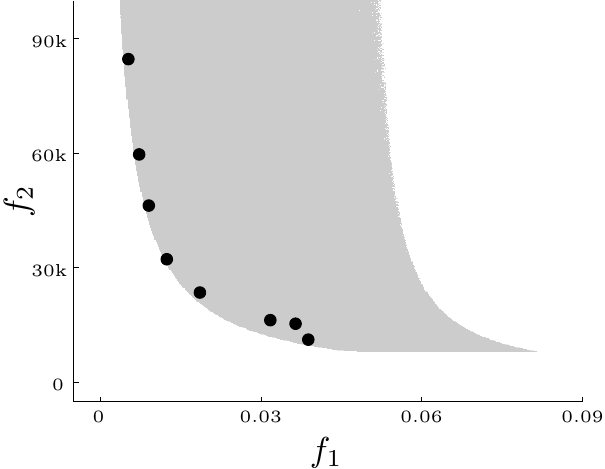}{(f) TwoBarTruss}
  \mysubfig{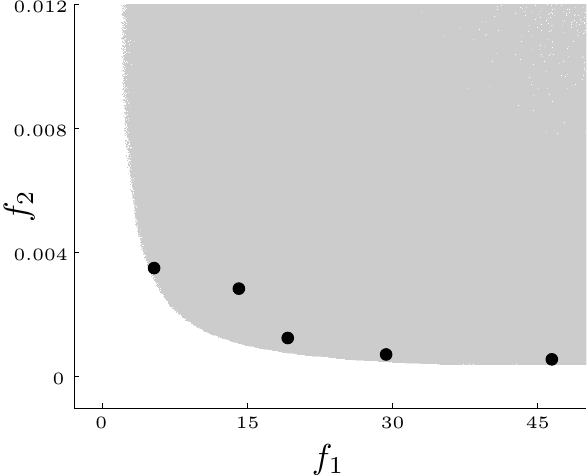}{(g) WeldedBeam}\\[\vSki]
  \mysubfig{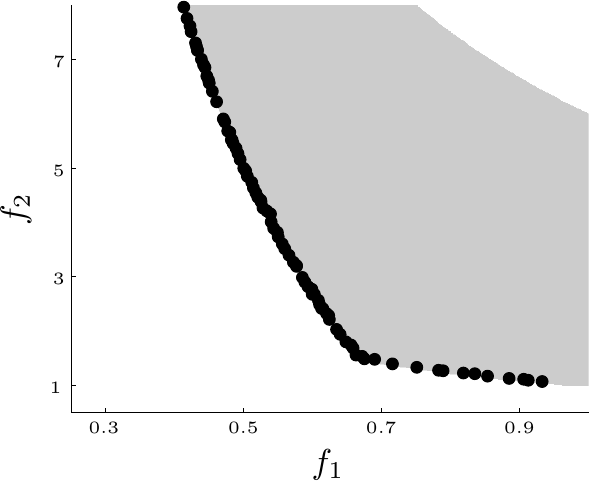}{(h) CONSTR}
  \mysubfig[4mm]{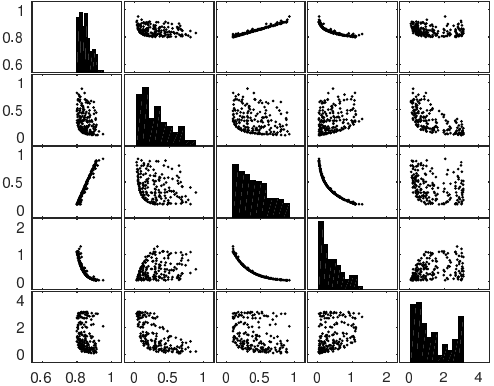}{(i) WATER}\\[\vSki]
  \caption{Result of one run of the BMOO algorithm on the problems of
    Table~\ref{tab:problem_mo}, with $n=100$ evaluations on the bi-objective
    problems and $n=200$ evaluations on the WATER problem. Black dots represent
    non-dominated solutions.  For bi-objective problems, the set of feasible
    objective values is shown in gray.  On the subfigure corresponding to the
    WeldedBeam problem, a zoom has been made to improve visualization.}
  \label{fig:pareto}
\end{figure}

The second benchmark consists of a set of eight constrained multi-objective test
problems from \cite{chafekar2003constrained} and \cite{deb2002fast}. The main
features of these problems are given in Table~\ref{tab:problem_mo}. The input
dimension~$d$ varies from~two to~six, and the number~$q$ of constraints from~one
to~seven. All problems have two objective functions, except the WATER test
problem, which has five. As in Table~\ref{tab:problem_so}, column $\Gamma(\%)$
gives an estimate of the ratio (in percents) of the volume of the feasible
region to that of the search space.  Column $V$ gives the volume of the region
dominated by the Pareto front\footnote{This volume has been obtained using
  massive runs of the \texttt{gamultiobj} algorithm of Matlab. It might be slightly under-estimated.}, measured using a reference point
$\yo^\refHV$, whose coordinates are specified in the last column.
As an illustration, the result of one run of BMOO is shown on
Figure~\ref{fig:pareto}, for each test problem.

\begin{table}
  \centering
  \renewcommand{\arraystretch}{1.2}
  \rowcolors{1}{white}{gray!10}
  \captionsetup{justification=centering}
  \begin{tabular}{|l|c|c|c|c|c|c|}
    \hline
    Pbm & $d$ & $q$ & $p$ & $\Gamma(\%)$ & $V$ & $\yo^\refHV$ \\
    \hline
    BNH & 2 & 2 & 2 & 93,6 & 5249 & [140; 50] \\
    SRN & 2 & 2 & 2 & 16,1 & 31820 & [200; 50] \\
    TNK & 2 & 2 & 2 & 5,1 & 0,6466 & [1,2; 1,2] \\
    OSY & 6 & 6 & 2 & 3,2 & 16169 & [0; 80] \\
    TwoBarTruss & 3 & 1 & 2 & 86,3 & 4495 & [0,06; $10^{5}$] \\
    WeldedBeam & 4 & 4 & 2 & 45,5 & 0,4228 & [50; 0,01] \\
    CONSTR & 2 & 2 & 2 & 52,5 & 3,8152 & [1; 9] \\
    WATER & 3 & 7 & 5 & 92 & 0,5138 & [1; 1; 1; 1,6; 3,2] \\
    \hline
  \end{tabular}
  \caption{Main features of the multi-objective problems in our benchmark.}
  \label{tab:problem_mo}
\end{table}

To the best of our knowledge, published state-of-the-art methods to solve multi-objective
optimization problems in the presence of non-linear constraints are based on
genetic or evolutionary approaches. The most popular algorithms are probably
NSGA2 \citep{deb2002fast} and SPEA2 \citep{zitzler2001spea2}. Such algorithms,
however, are not primarily designed to work on a limited budget of function
evaluations. Some methods that combine genetic/evolutionary approaches and
surrogate modeling techniques have been proposed in the literature~\citep[see,
e.g.,][and references therein]{emmerich2006single, jin2011surrogate}, but a
quantitative comparison with these methods would necessitate to develop 
proper implementations, which is out of the scope of this
paper.  In this section, we shall limit ourselves to emphasizing advantages and
limitations of the proposed approach.  Since the ability of the BMOO algorithm
to find feasible solutions has already been demonstrated in
Section~\ref{sec:so_bench}, we will focus here on the other contributions of the
paper: the SMC methods for the computation and optimization of the expected
improvement sampling criterion.

First, we demonstrate the effectiveness of the proposed SMC algorithm for
optimizing expected improvement based criteria. We compare our SMC approach (see
Section~\ref{sec:crit-optim-using}) with the approach used by
\cite{couckuyt2014fast}, that we shall call MCSQP (for Monte-Carlo
Sequential Quadratic Programming). This approach consists in
selecting the best point out of a population of candidates uniformly
distributed on the search space $\X$, and then running an SQP algorithm starting
from this point. In our experiments, the number of candidates is chosen equal to
the population size $m_\XX = 1000$ of the SMC method.

Table~\ref{tab:mo-smc-mcsqp-ehvi} presents experimental results obtained with
the extended EHVI criterion proposed in Section~\ref{sec:method} as a sampling
criterion.  As a preliminary remark, observe that the finest target precision is
systematically reached by our SMC method in all but three test cases (OSY,
TwoBarTruss and WeldedBeam). The OSY case will be discussed
below. On the
TwoBarTruss and WeldedBeam problems, we found out that the poor performances
are due to
Gaussian process modeling issues, similar to those encountered earlier on the
g3mod, g10 and PVD4 test problems (see Section~\ref{sec:so_bench}). The results
on these problems are thus left out of the analyses in the following, but will
motivate future work on the models, as concluded in
Section~\ref{sec:so_bench}. Regarding the optimization of the criteria, the
results show that our SMC approach compares very favorably with MCSQP. More
specifically, we note a drop of performance of the MCSQP method compared with
the SMC approach as we try to converge more finely toward the Pareto front (see,
in particular, column ``Target $99\%$'' of Table~\ref{tab:mo-smc-mcsqp-ehvi},
but this is also visible in the other columns as well for most of the test
cases). Because of its sequential nature, the SMC approach is able to track much
more efficiently the concentration of the sampling criterion in the
search domain, and thus makes it
possible to reach higher accuracy.

Tables~\ref{tab:mo-smc-mcsqp-emmi} and~\ref{tab:mo-smc-mcsqp-wcpi} provide
additional results obtained when performing the same study with respectively the
EMMI and WCPI criteria\footnote{An implementation of the EMMI criterion is
  available in the STK. An implementation of the WCPI sampling crtiterion for
  bi-objective problems is distributed alongside with
  \citeauthor{forrester2008book}'s book \citep{forrester2008book}.}
\citep[see][respectively]{svenson2010multiobjective,
  keane2006statistical}. These criteria are not primarily designed to address
constrained problems, but they can easily be extended to handle constraints by
calculating them using only feasible values of the objectives, and then
multiplying them by the probability of satisfying the constraints (as explained
in Section~\ref{sec:ei-constraints}). When no feasible solution is
available at the start of the optimization procedure, we use the probability of feasibility as a sampling criterion, as
advised by~\cite{gelbart2014bayesian}. The conclusions drawn from
Table~\ref{tab:mo-smc-mcsqp-ehvi} for the extended EHVI criterion carry through
to the results presented in Tables~\ref{tab:mo-smc-mcsqp-emmi}--\ref{tab:mo-smc-mcsqp-wcpi}.
It shows that the SMC algorithm proposed in Section~\ref{sec:crit-optim-using}
can be viewed as a contribution of independent interest for optimizing improvement-based sampling criteria.

\begin{table}
  \centering
  \renewcommand{\arraystretch}{1.4}
  \begin{tabular}{|l|l|c|c|c|c|c|c|}
    \hline
    Problem & optimizer & \multicolumn{2}{c|}{Target $90\%$}
    & \multicolumn{2}{c|}{Target $95\%$} & \multicolumn{2}{c|}{Target $99\%$} \\
    \hline
    \multirow{2}{*}{BNH}
    & SMC & 30 & 8.5 (0.6) & 30 & 12.7 (0.7) & 30 & 34.6 (1.3) \\
    & MCSQP & 30 & 8.4 (0.6) & 30 & 12.8 (0.7) & 30 & 38.9 (2.2) \\
    \hline
    \multirow{2}{*}{SRN}
    & SMC & 30 & 16.7 (0.9) & 30 & 22.4 (1.0) & 30 & 52.6 (4.1) \\
    & MCSQP & 30 & 20.5 (2.4) & 30 & 35.6 (5.9) & 0 & $> 250$ (-) \\
    \hline
    \multirow{2}{*}{TNK}
    & SMC & 30 & 35.5 (2.6) & 30 & 44.1 (2.5) & 30 & 71.1 (4.0) \\
    & MCSQP & 30 & 43.5 (4.6) & 30 & 71.6 (11.3) & 0 & $> 250$ (-) \\
    \hline
    \multirow{2}{*}{OSY}
    & SMC & 30 & 29.0 (1.7) & 30 & 38.2 (3.4) & 13 & 119.8 (53.0) \\
    & MCSQP & 0 & $> 250$ (-) & 0 & $> 250$ (-) & 0 & $> 250$ (-) \\
    \hline
    \multirow{2}{*}{TwoBarTruss}
    & SMC & 22 & 90.9 (62.0) & 1 & 234 (-) & 0 & $> 250$ (-) \\
    & MCSQP & 26 & 88.7 (68.4) & 2 & 162.0 (29.7) & 0 & $> 250$ (-) \\
    \hline 
    \multirow{2}{*}{WeldedBeam}
    & SMC & 28 & 146.5 (41.1) & 2 & 212 (33.9) & 0 & $> 250$ (-) \\
    & MCSQP & 26 & 171.3 (46.9) & 1 & 229.0 (-) & 0 & $> 250$ (-) \\
    \hline
    \multirow{2}{*}{CONSTR}
    & SMC & 30 & 12.4 (1.0) & 30 & 19.2 (1.4) & 30 & 83.5 (5.9) \\
    & MCSQP & 30 & 13.8 (1.4) & 30 & 26.3 (3.3) & 0 & $> 250$ (-) \\
    \hline
    \multirow{2}{*}{WATER}
    & SMC & 30 & 48.3 (3.6) & 30 & 80.7 (5.6) & 30 & 139.1 (8.0) \\
    & MCSQP & 30 & 53.5 (4.8) & 30 & 88.7 (7.5) & 30 & 164.3 (9.6) \\
    \hline
  \end{tabular}		
  \caption{Results achieved when using either SMC or MCSQP for the optimization
    of the extended EHVI, on the problems of Table~\ref{tab:problem_mo}. We
    measure the number of function evaluations for dominating successively
    $90\%$, $95\%$ and $99\%$ of the volume~$V$. For each target, the first
    column contains the number of successful runs over 30 runs. The second
    column contains the number of function evaluations, averaged over the
    successful runs, with the corresponding standard deviation (in
    parentheses). Dash symbols are used when a value cannot be calculated.}
  \label{tab:mo-smc-mcsqp-ehvi}
\end{table}
 
\begin{table}
  \centering
  \renewcommand{\arraystretch}{1.4}
  \begin{tabular}{|l|l|c|c|c|c|c|c|}
    \hline
    Problem & optimizer & \multicolumn{2}{c|}{Target $90\%$}
    & \multicolumn{2}{c|}{Target $95\%$} & \multicolumn{2}{c|}{Target $99\%$} \\
    \hline
    \multirow{2}{*}{BNH} 
    & SMC & 30 & 9.8 (1.1) & 30 & 15.9 (1.5) & 30 & 41.2 (2.8) \\
    & MCSQP & 30 & 9.5 (0.7) & 30 & 15.4 (1.4) & 30 & 42.6 (2.4) \\
    \hline
    \multirow{2}{*}{SRN}
    & SMC & 30 & 15.5 (1.2) & 30 & 21.0 (1.4) & 30 & 48.3 (2.8) \\
    & MCSQP & 30 & 18.6 (1.8) & 30 & 29.1 (2.7) & 30 & 90.9 (9.0) \\
    \hline
    \multirow{2}{*}{TNK}
    & SMC & 30 & 47.7 (3.5) & 30 & 61.8 (4.4) & 30 & 100.2 (5.4) \\
    & MCSQP & 30 & 60.6 (8.2) & 30 & 94.3 (13.2) & 5 & 224.2 (15.0) \\
    \hline
    \multirow{2}{*}{OSY} 
    & SMC & 30 & 32.3 (2.9) & 30 & 41.9 (3.9) & 25 & 73.6 (20.8) \\
    & MCSQP & 0 & $> 250$ (-) & 0 & $> 250$ (-) & 0 & $> 250$ (-) \\
    \hline
    \multirow{2}{*}{TwoBarTruss} 
    & SMC & 28 & 116.5 (48.5) & 3 & 199.0 (24.1) & 0 & $> 250$ (-) \\
    & MCSQP & 26 & 130.9 (63.9) & 1 & 174.0 (-) & 0 & $> 250$ (-) \\
    \hline 
    \multirow{2}{*}{WeldedBeam} 
    & SMC & 16 & 156.6 (50.5) & 4 & 177.0 (40.5) & 0 & $> 250$ (-) \\
    & MCSQP & 9 & 161.9 (60.1) & 3 & 156.0 (35.8) & 0 & $> 250$ (-) \\
    \hline
    \multirow{2}{*}{CONSTR} 
    & SMC & 30 & 22.1 (2.5) & 30 & 33.8 (3.0) & 30 & 100.9 (8.6) \\
    & MCSQP & 30 & 18.4 (2.1) & 30 & 30.9 (3.1) & 30 & 154.8 (9.0) \\
    \hline
    \multirow{2}{*}{WATER}
    & SMC & 30 & 60.4 (6.5) & 30 & 93.4 (8.8) & 30 & 153.9 (9.0) \\
    & MCSQP & 30 & 68.2 (8.1) & 30 & 103.9 (11.3) & 30 & 172.7 (13.7) \\
    \hline
  \end{tabular}		
  \caption{Results achieved when using the EMMI 
    criterion. See Table~\ref{tab:mo-smc-mcsqp-ehvi} for more information.}
  \label{tab:mo-smc-mcsqp-emmi}

  \bigskip
  
  \begin{tabular}{|l|l|c|c|c|c|c|c|}
    \hline
    Problem & optimizer & \multicolumn{2}{c|}{Target $90\%$}
    & \multicolumn{2}{c|}{Target $95\%$} & \multicolumn{2}{c|}{Target $99\%$} \\
    \hline
    \multirow{2}{*}{BNH} 
    & SMC & 30 & 20.9 (8.9) & 30 & 43.4 (7.6) & 30 & 132.4 (15.4) \\
    & MCSQP & 30 & 18.7 (8.2) & 30 & 49.0 (14.2) & 30 & 176.1 (29.1) \\
    \hline
    \multirow{2}{*}{SRN} 
    & SMC & 30 & 39.1 (6.0) & 30 & 57.53 (7.5) & 30 & 154.9 (12.8) \\
    & MCSQP & 20 & 154.5 (62.1) & 1 & 248.0 (-) & 0 & $> 250$ (-) \\
    \hline
    \multirow{2}{*}{TNK}
    & SMC & 30 & 53.3 (6.8) & 30 & 68.3 (6.9) & 30 & 120.8 (13.7) \\
    & MCSQP & 0 & $> 250$ (-) & 0 & $> 250$ (-) & 0 & $> 250$ (-) \\
    \hline
    \multirow{2}{*}{OSY} 
    & SMC & 30 & 39.7 (5.7) & 29 & 61.5 (22.0) & 14 & 123.0 (41.9) \\
    & MCSQP & 0 & $> 250$ (-) & 0 & $> 250$ (-) & 0 & $> 250$ (-) \\
    \hline
    \multirow{2}{*}{TwoBarTruss}
    & SMC & 29 & 70.1 (40.3) & 8 & 180.4 (40.0) & 0 & $> 250$ (-) \\
    & MCSQP & 29 & 69.6 (47.3) & 11 & 185.2 (53.0) & 0 & $> 250$ (-) \\
    \hline 
    \multirow{2}{*}{WeldedBeam}
    & SMC & 0 & $> 250$ (-) & 0 & $> 250$ (-) & 0 & $> 250$ (-) \\
    & MCSQP & 0 & $> 250$ (-) & 0 & $> 250$ (-) & 0 & $> 250$ (-) \\
    \hline
    \multirow{2}{*}{CONSTR}
    & SMC & 30 & 40.0 (5.6) & 30 & 60.4 (7.8) & 30 & 212.1 (15.6) \\
    & MCSQP & 30 & 42.2 (16.0) & 26 & 150.7 (42.8) & 0 & $> 250$ (-) \\
    \hline
  \end{tabular}		
  \caption{Results achieved when using the WCPI 
    criterion. See Table~\ref{tab:mo-smc-mcsqp-ehvi} for more information.}
  \label{tab:mo-smc-mcsqp-wcpi}
\end{table}

Next we study the influence on the convergence of the algorithm of the number
$m=m_\YY $ of particles used in Algorithm~\ref{alg:RemResMovUnif} for
approximating the expected improvement value. In Table~\ref{tab:mo-ehvi-approx}
we compare the number of evaluations required to dominate successively $90\%$,
$95\%$ and $99\%$ of the volume $V$ of Table~\ref{tab:problem_mo} when using
different numbers of particles.
As expected, the overall performances of the algorithm deteriorate when
the number~$m_\YY$ of particles used to approximate the expected improvement
decreases. However, the algorithm maintains satisfactory convergence properties
even with a small number of particles.  For reference, we have also included
results obtained by choosing the evaluation point randomly in the set of
candidate points.  Notice that these results are always much worse than those
 obtained using the sampling criterion with~$m_\YY = 200$. This shows
that not all candidate points are equally good, and thus confirms that the sampling
criterion, even with a rather small value of~$m_\YY$, is effectively discriminating between good and bad candidate points.

\begin{table}
  \centering
  \renewcommand{\arraystretch}{1.4}
  \begin{tabular}{|l|l|c|c|c|c|c|c|}
    \hline
    Problem & EHVI & \multicolumn{2}{c|}{Target $90\%$}
    & \multicolumn{2}{c|}{Target $95\%$} & \multicolumn{2}{c|}{Target $99\%$} \\
    \hline
    \multirow{4}{*}{BNH}
    & SMC ($m_\YY = 5000$) & 30 & 8.3 (0.7) & 30 & 12.5 (0.5) & 30 & 32.8 (1.0) \\
    & SMC ($m_\YY = 1000$) & 30 & 8.5 (0.6) & 30 & 12.7 (0.7) & 30 & 34.6 (1.3) \\
    & SMC ($m_\YY = 200$) & 30 & 8.8 (0.6) & 30 & 13.1 (0.7) & 30 & 39.2 (2.0) \\
    & random				 & 30 & 12.8 (2.7) & 30 & 29.6 (6.0) & 30 & 106.8 (13.2) \\
    \hline
    \multirow{4}{*}{SRN}
    & SMC ($m_\YY = 5000$) & 30 & 16.3 (1.0) & 30 & 21.6 (1.1) & 30 & 47.3 (2.1) \\
    & SMC ($m_\YY = 1000$) & 30 & 16.7 (0.9) & 30 & 22.4 (1.0) & 30 & 52.6 (4.1) \\
    & SMC ($m_\YY = 200$) & 30 & 16.6 (1.3) & 30 & 23.0 (1.9) & 30 & 60.9 (6.9) \\
    & random				 & 30 & 30.6 (5.2) & 30 & 51.1 (8.2) & 30 & 146.2 (13.2) \\
    \hline
    \multirow{4}{*}{TNK}
    & SMC ($m_\YY = 5000$) & 30 & 36.2 (4.4) & 30 & 43.4 (3.6) & 30 & 65.1 (3.1) \\
    & SMC ($m_\YY = 1000$) & 30 & 35.5 (2.6) & 30 & 44.1 (2.5) & 30 & 71.1 (4.0)\\
    & SMC ($m_\YY = 200$) & 30 & 37.7 (4.1) & 30 & 48.4 (5.0) & 30 & 87.3 (5.9) \\
    & random				 & 30 & 64.0 (10.3) & 30 & 94.2 (12.4) & 29 & 193.3 (27.4)\\
    \hline
    \multirow{4}{*}{OSY}
    & SMC ($m_\YY = 5000$) & 30 & 28.6 (2.0) & 30 & 36.0 (2.8) & 22 & 82.5 (33.5) \\
    & SMC ($m_\YY = 1000$) & 30 & 29.0 (1.7) & 30 & 38.2 (3.4) & 13 & 119.8 (53.0) \\
    & SMC ($m_\YY = 200$) & 30 & 32.4 (3.1) & 29 & 49 (16.0) & 5 & 164.8 (54.6) \\
    & random				 & 30 & 140.2 (21.0) & 25 & 203.4 (21.4) & 0 & $> 250$ (-)\\
    \hline
    \multirow{4}{*}{CONSTR}
    & SMC ($m_\YY = 5000$) & 30 & 12.2 (0.7) & 30 & 18.0 (1.0) & 30 & 68.8 (4.7) \\
    & SMC ($m_\YY = 1000$) & 30 & 12.4 (1.0) & 30 & 19.2 (1.4) & 30 & 83.5 (5.9) \\
    & SMC ($m_\YY = 200$) & 30 & 12.9 (1.2) & 30 & 21.0 (1.6) & 30 & 109.2 (10.7) \\
    & random				 & 30 & 31.1 (6.6) & 30 & 58.1 (8.5) & 18 & 235.1 (11.0) \\
    \hline
    \multirow{4}{*}{WATER}
    & SMC ($m_\YY = 5000$) & 30 & 45.8 (4.0) & 30 & 75.3 (6.2) & 30 & 127 (8.2) \\
    & SMC ($m_\YY = 1000$) & 30 & 48.3 (3.6) & 30 & 80.7 (5.6) & 30 & 139.1 (8.0) \\
    & SMC ($m_\YY = 200$)  & 30 & 52.5 (4.5) & 30 & 88.6 (6.0) & 30 & 154.8 (8.8) \\
    & random 				  & 14 & 223.2 (15.4) & 0 & $> 250$ (-) & 0 & $> 250$ (-) \\
    \hline
  \end{tabular}		
  \caption{Results achieved on the problems of Table~\ref{tab:problem_mo} when
    using successively $m_\YY = 200, 1000 \text{\ and\ } 5000$ particles for the
    approximate computation of the extended EHVI criterion. For reference,
    results obtained by selecting the evaluation point randomly in the pool
    of candidates points are provided (``random'' rows).
    See Table~\ref{tab:mo-smc-mcsqp-ehvi} for more information.}
  \label{tab:mo-ehvi-approx}
\end{table}

\begin{figure}
  \centering
  \includegraphics{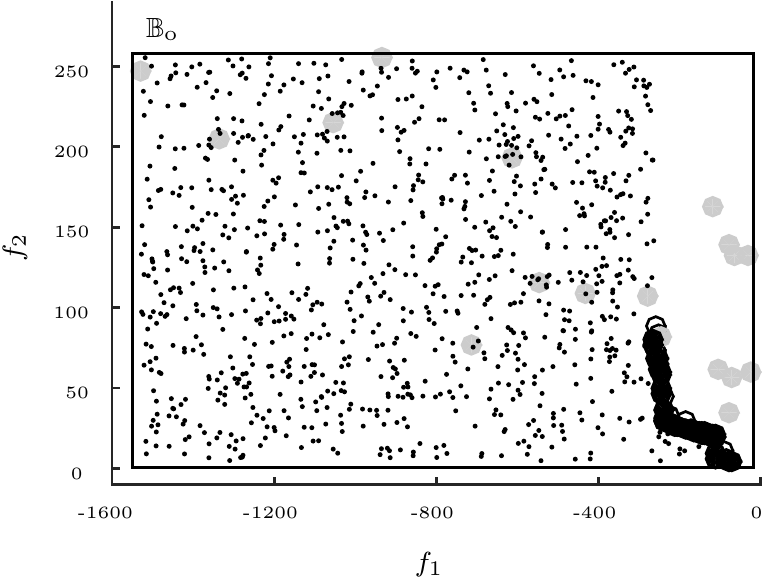}
  \caption{An illustration, in the objective domain~$\Yo$, of BMOO running on
    the OSY test problem. The small dots are the particles used for the
    computation of the expected improvement. They are uniformly distributed on
    the non-dominated subset of~$\Bo$. Dark disks indicate the non-dominated
    solutions found so far, light gray disks indicate the dominated ones.}
  \label{fig:OSY}
\end{figure}

We observe poor performances of the BMOO algorithm on the OSY test problem,
regardless of the number of particles that are used to estimate the expected
improvement. Figure~\ref{fig:OSY} reveals that this is due to the choice of a
uniform sampling density on~$\Bo \setminus H_n$ as the target density of the SMC
algorithm used for the approximate computation of the criterion. Indeed, most of
the particles do not effectively participate to the approximation of the
integral, since they lie outside the set of feasible objective values (see Figure~\ref{fig:pareto}(d)). Further
work is required on this topic to propose a better sampling density, that would
concentrate on objective values that are likely to be feasible (instead of
covering uniformly the entire non-dominated region $\Bo \setminus H_n$).

In practice, for problems with a small number of objectives, and especially for
bi-objective problems, we do not recommend the use of our SMC algorithm for the
(approximate) computation of the EHVI criterion since exact and efficient
domain-decomposition-based algorithms are available \citep[see][and references
therein]{hupkens2014faster,couckuyt2014fast}. An in-depth study of the quality
of the approximation provided by our SMC method, and a comparison with exact
methods, is therefore needed before more precise recommandations can be made.

\section{Conclusions and future work}
\label{sec:conclusion}

In this article, a new Bayesian optimization approach is proposed to solve
multi-objective optimization problems with non-linear constraints. The
constraints are handled using an extended domination rule and a new expected
improvement formulation is proposed. In particular, the new formulation makes it
possible to deal with problems where no feasible solution is available from the
start. Several criteria from the literature are recovered as special cases.

The computation and optimization of the new expected improvement criterion are
carried out using sequential Monte Carlo sampling techniques. Indeed, the
criterion takes the form of an integral over the space of objectives and
constraints, for which no closed-form expression is known. Besides, the sampling
criterion may be highly multi-modal, as is well known in the special case of
unconstrained single-objective optimization, which makes it difficult to
optimize. The proposed sampling techniques borrow ideas from the literature of
structural reliability for estimating the probability of rare events, and can be
viewed as a contribution in itself.

We show that the resulting algorithm, which we call BMOO, achieves good results
on a set of single-objective constrained test problems, with respect to
state-of-the-art algorithms. In particular, BMOO is able to effectively find
feasible solutions, even when the feasible region is very small compared to the
size of the search space and when the number of constraints is high. In the case
of multi-objective optimization with non-linear constraints, we show that BMOO
is able to yield good approximations of the Pareto front on small budgets of
evaluations.

Several questions are left open for future work.
First, our numerical studies reveal that the choice of sampling densities in the
input domain (as demonstrated by unsatisfactory results on the g18 test problem)
and in the output domain (as shown on the OSY case) could be
improved. Suggestions for improvement are proposed in the article and will be
the object of future investigations.
Second, an in-depth study of the quality of the approximation provided by our
SMC method, and a comparison with exact methods, is needed before
recommandations can be made on when to switch between exact and approximate
calculation of the expected improvement, and how to select the sample
size---possibly in an adaptive manner---used for the SMC approximation.
Last, the choice of the random processes used for modeling objective and
constraint functions deserves more attention. Stationary Gaussian process models
have been found to lack flexibility on some single- and multi-objective cases
(g3mod, g10, PVD4, TwoBarTruss and WeldedBeam).  Several types of models
proposed in the literature---warped Gaussian processes
\citep{snelson2004warped}, non-stationary Gaussian processes \citep[see][and
references therein]{toal2012}, deep Gaussian processes \citep{damianou2013deep},
etc.---provide interesting directions regarding this issue.

\begin{acknowledgements}
  This research work has been carried out within the Technological Research
  Institute SystemX, using public funds from the French Programme
  \emph{Investissements d'Avenir}.
\end{acknowledgements}

\bibliographystyle{plainnat}
\bibliography{biblio}

\begin{thebibliography}{79}
\providecommand{\natexlab}[1]{#1}
\providecommand{\url}[1]{\texttt{#1}}
\expandafter\ifx\csname urlstyle\endcsname\relax
  \providecommand{\doi}[1]{doi: #1}\else
  \providecommand{\doi}{doi: \begingroup \urlstyle{rm}\Url}\fi

\bibitem[Andrieu and Roberts(2009)]{andrieu2009pseudo}
C.~Andrieu and G.~O. Roberts.
\newblock The pseudo-marginal approach for efficient monte carlo computations.
\newblock \emph{The Annals of Statistics}, 37\penalty0 (2):\penalty0 697--725,
  2009.

\bibitem[Andrieu and Thoms(2008)]{andrieu2008tutorial}
Christophe Andrieu and Johannes Thoms.
\newblock A tutorial on adaptive mcmc.
\newblock \emph{Statistics and Computing}, 18\penalty0 (4):\penalty0 343--373,
  2008.

\bibitem[Archetti and Betr{\`o}(1979)]{archetti79}
F.~Archetti and B.~Betr{\`o}.
\newblock A probabilistic algorithm for global optimization.
\newblock \emph{CALCOLO}, 16\penalty0 (3):\penalty0 335--343, 1979.

\bibitem[Au and Beck(2001)]{au2001estimation}
S.-K. Au and J.~L Beck.
\newblock Estimation of small failure probabilities in high dimensions by
  subset simulation.
\newblock \emph{Probabilistic Engineering Mechanics}, 16\penalty0 (4):\penalty0
  263--277, 2001.

\bibitem[Bader and Zitzler(2011)]{bader2011hype}
J.~Bader and E.~Zitzler.
\newblock Hype: An algorithm for fast hypervolume-based many-objective
  optimization.
\newblock \emph{Evolutionary Computation}, 19\penalty0 (1):\penalty0 45--76,
  2011.

\bibitem[Bautista(2009)]{bautista2009}
D.~C. Bautista.
\newblock \emph{A sequential design for approximating the pareto front using
  the expected pareto improvement function}.
\newblock PhD thesis, The Ohio State University, 2009.

\bibitem[Bect et~al.(2012)Bect, Ginsbourger, Li, Picheny, and
  Vazquez]{bect2012sequential}
J.~Bect, D.~Ginsbourger, L.~Li, V.~Picheny, and E.~Vazquez.
\newblock Sequential design of computer experiments for the estimation of a
  probability of failure.
\newblock \emph{Statistics and Computing}, 22\penalty0 (3):\penalty0 773--793,
  2012.

\bibitem[Bect et~al.(2016)Bect, Vazquez, et~al.]{stktoolbox}
J.~Bect, E.~Vazquez, et~al.
\newblock {STK}: a {S}mall ({M}atlab/{O}ctave) {T}oolbox for {K}riging.
  {R}elease 2.4 (to appear), 2016.
\newblock URL \url{http://kriging.sourceforge.net}.

\bibitem[Benassi(2013)]{benassi2013nouvel}
R.~Benassi.
\newblock \emph{Nouvel algorithme d'optimisation bay{\'e}sien utilisant une
  approche Monte-Carlo s{\'e}quentielle.}
\newblock PhD thesis, Sup{\'e}lec, 2013.

\bibitem[Benassi et~al.(2012)Benassi, Bect, and Vazquez]{benassi2012bayesian}
R.~Benassi, J.~Bect, and E.~Vazquez.
\newblock {B}ayesian optimization using sequential {M}onte {C}arlo.
\newblock In \emph{Learning and Intelligent Optimization. 6th International
  Conference, LION 6, Paris, France, January 16-20, 2012, Revised Selected
  Papers}, volume 7219 of \emph{Lecture Notes in Computer Science}, pages
  339--342. Springer, 2012.

\bibitem[Beume(2009)]{beume2009}
N.~Beume.
\newblock S-metric calculation by considering dominated hypervolume as klee's
  measure problem.
\newblock \emph{Evolutionary Computation}, 17\penalty0 (4):\penalty0 477--492,
  2009.

\bibitem[Binois and Picheny(2015)]{gpareto}
M.~Binois and V.~Picheny.
\newblock \emph{GPareto: Gaussian Processes for Pareto Front Estimation and
  Optimization}, 2015.
\newblock URL \url{http://CRAN.R-project.org/package=GPareto}.
\newblock R package version 1.0.1.

\bibitem[Box and Cox(1964)]{boxcox1964}
G.~E.~P. Box and D.~R. Cox.
\newblock An analysis of transformations.
\newblock \emph{Journal of the Royal Statistical Society. Series B
  (Methodological)}, 26\penalty0 (2):\penalty0 211--252, 1964.

\bibitem[C{\'e}rou et~al.(2012)C{\'e}rou, Del~Moral, Furon, and
  Guyader]{cerou2012sequential}
F.~C{\'e}rou, P.~Del~Moral, T.~Furon, and A.~Guyader.
\newblock Sequential {M}onte {C}arlo for rare event estimation.
\newblock \emph{Statistics and Computing}, 22\penalty0 (3):\penalty0 795--808,
  2012.

\bibitem[Chafekar et~al.(2003)Chafekar, Xuan, and
  Rasheed]{chafekar2003constrained}
D.~Chafekar, J.~Xuan, and K.~Rasheed.
\newblock Constrained multi-objective optimization using steady state genetic
  algorithms.
\newblock In \emph{Genetic and Evolutionary Computation-GECCO 2003}, pages
  813--824. Springer, 2003.

\bibitem[Chevalier et~al.(2014)Chevalier, Bect, Ginsbourger, Vazquez, Picheny,
  and Richet]{chevalier2014fast}
C.~Chevalier, J.~Bect, D.~Ginsbourger, E.~Vazquez, V.~Picheny, and Y.~Richet.
\newblock Fast parallel kriging-based stepwise uncertainty reduction with
  application to the identification of an excursion set.
\newblock \emph{Technometrics}, 56\penalty0 (4):\penalty0 455--465, 2014.

\bibitem[Conn et~al.(1991)Conn, Gould, and Toint]{conn1991globally}
A.~R. Conn, N.~I.~M. Gould, and P.~Toint.
\newblock A globally convergent augmented lagrangian algorithm for optimization
  with general constraints and simple bounds.
\newblock \emph{SIAM Journal on Numerical Analysis}, 28\penalty0 (2):\penalty0
  545--572, 1991.

\bibitem[Couckuyt et~al.(2014)Couckuyt, Deschrijver, and
  Dhaene]{couckuyt2014fast}
I.~Couckuyt, D.~Deschrijver, and T.~Dhaene.
\newblock Fast calculation of multiobjective probability of improvement and
  expected improvement criteria for pareto optimization.
\newblock \emph{Journal of Global Optimization}, 60\penalty0 (3):\penalty0
  575--594, 2014.

\bibitem[Damianou and Lawrence(2013)]{damianou2013deep}
A.~Damianou and N.~Lawrence.
\newblock Deep gaussian processes.
\newblock In \emph{Proceedings of the Sixteenth International Conference on
  Artificial Intelligence and Statistics}, pages 207--215, 2013.

\bibitem[Deb et~al.(2002)Deb, Pratap, Agarwal, and Meyarivan]{deb2002fast}
K.~Deb, A.~Pratap, S.~Agarwal, and T.~Meyarivan.
\newblock A fast and elitist multiobjective genetic algorithm: {NSGA-II}.
\newblock \emph{Evolutionary Computation, IEEE Transactions on}, 6\penalty0
  (2):\penalty0 182--197, 2002.

\bibitem[Del~Moral et~al.(2006)Del~Moral, Doucet, and Jasra]{del2006sequential}
P.~Del~Moral, A.~Doucet, and A.~Jasra.
\newblock Sequential monte carlo samplers.
\newblock \emph{Journal of the Royal Statistical Society: Series B (Statistical
  Methodology)}, 68\penalty0 (3):\penalty0 411--436, 2006.

\bibitem[Douc and Capp{\'e}(2005)]{douc2005comparison}
R.~Douc and O.~Capp{\'e}.
\newblock Comparison of resampling schemes for particle filtering.
\newblock In \emph{Image and Signal Processing and Analysis, 2005. ISPA 2005.
  Proceedings of the 4th International Symposium on}, pages 64--69. IEEE, 2005.

\bibitem[Emmerich(2005)]{emmerich2005}
M.~Emmerich.
\newblock \emph{Single- and multiobjective evolutionary design optimization
  assisted by {G}aussian random field metamodels}.
\newblock PhD thesis, Technical University Dortmund, 2005.

\bibitem[Emmerich and Klinkenberg(2008)]{emmerich2008computation}
M.~Emmerich and J.~W. Klinkenberg.
\newblock The computation of the expected improvement in dominated hypervolume
  of {P}areto front approximations.
\newblock \emph{Technical report, Leiden University}, 2008.

\bibitem[Emmerich et~al.(2006)Emmerich, Giannakoglou, and
  Naujoks]{emmerich2006single}
M.~Emmerich, K.~C. Giannakoglou, and B.~Naujoks.
\newblock Single- and multi-objective evolutionary optimization assisted by
  {Ga}ussian random field metamodels.
\newblock \emph{IEEE Transactions on Evolutionary Computation}, 10\penalty0
  (4):\penalty0 421--439, 2006.

\bibitem[Fonseca and Fleming(1998)]{fonseca1998multiobjective}
C.~M. Fonseca and P.~J. Fleming.
\newblock Multiobjective optimization and multiple constraint handling with
  evolutionary algorithms. {I}. {A} unified formulation.
\newblock \emph{IEEE Transactions on Systems, Man and Cybernetics. Part A:
  Systems and Humans}, 28\penalty0 (1):\penalty0 26--37, 1998.

\bibitem[Forrester et~al.(2008)Forrester, Sobester, and
  Keane]{forrester2008book}
A.~I.~J. Forrester, A.~Sobester, and A.~J. Keane.
\newblock \emph{Engineering design via surrogate modelling: a practical guide}.
\newblock John Wiley \& Sons, 2008.

\bibitem[Gelbart(2015)]{gelbart2015phd}
M.~A. Gelbart.
\newblock \emph{Constrained Bayesian Optimization and Applications}.
\newblock PhD thesis, Harvard University, Graduate School of Arts and Sciences,
  2015.

\bibitem[Gelbart et~al.(2014)Gelbart, Snoek, and Adams]{gelbart2014bayesian}
M.~A. Gelbart, J.~Snoek, and R.~P. Adams.
\newblock Bayesian optimization with unknown constraints.
\newblock \emph{arXiv preprint arXiv:1403.5607}, 2014.

\bibitem[Ginsbouger and Le~Riche(2009)]{GinsbLeRiche2009}
D.~Ginsbouger and R.~Le~Riche.
\newblock Towards {G}aussian process-based optimization with finite time
  horizon.
\newblock In \emph{Invited talk at the 6th Autumn Symposium of the "Statistical
  Modelling" Research Training Group}, November 21st 2009.

\bibitem[Gramacy and Lee(2011)]{gramacy2011}
R.~B. Gramacy and H.~Lee.
\newblock Optimization under unknown constraints.
\newblock In \emph{Bayesian Statistics 9. Proceedings of the Ninth Valencia
  International Meeting}, pages 229--256. Oxford University Press, 2011.

\bibitem[Gramacy et~al.(just accepted)Gramacy, Gray, Le~Digabel, Lee, Ranjan,
  Wells, and Wild]{gramacy2015modeling}
R.~B. Gramacy, G.~A. Gray, S.~Le~Digabel, H.~K.~H. Lee, P.~Ranjan, G.~Wells,
  and S.~M. Wild.
\newblock Modeling an augmented lagrangian for blackbox constrained
  optimization.
\newblock \emph{Technometrics}, just accepted.

\bibitem[Hern{\'a}ndez-Lobato et~al.(2015{\natexlab{a}})Hern{\'a}ndez-Lobato,
  Hern{\'a}ndez-Lobato, Shah, and Adams]{hernandez2015predictive}
D.~Hern{\'a}ndez-Lobato, J.~M. Hern{\'a}ndez-Lobato, A.~Shah, and R.~P. Adams.
\newblock Predictive entropy search for multi-objective bayesian optimization.
\newblock arXiv preprint arXiv:1511.05467, 2015{\natexlab{a}}.

\bibitem[Hern{\'a}ndez-Lobato et~al.(2015{\natexlab{b}})Hern{\'a}ndez-Lobato,
  Gelbart, Hoffman, Adams, and Ghahramani]{hernandez2015}
J.~M. Hern{\'a}ndez-Lobato, M.~A. Gelbart, M.~W. Hoffman, R.~P. Adams, and
  Z.~Ghahramani.
\newblock Predictive entropy search for bayesian optimization with unknown
  constraints.
\newblock In \emph{Proceedings of the 32nd International Conference on Machine
  Learning, Lille, France, 2015. JMLR: W\&CP volume 37}, 2015{\natexlab{b}}.

\bibitem[Hern{\'a}ndez-Lobato et~al.(to appear)Hern{\'a}ndez-Lobato, Gelbart,
  Adams, Hoffman, and Ghahramani]{hernandez2015pesc}
J.~M. Hern{\'a}ndez-Lobato, M.~A. Gelbart, R.~P. Adams, M.~W. Hoffman, and
  Z.~Ghahramani.
\newblock A general framework for constrained bayesian optimization using
  information-based search.
\newblock \emph{TBD}, to appear.

\bibitem[Horn et~al.(2015)Horn, Wagner, Biermann, Weihs, and
  Bischl]{horn2015model}
D.~Horn, T.~Wagner, D.~Biermann, C.~Weihs, and B.~Bischl.
\newblock Model-based multi-objective optimization: Taxonomy, multi-point
  proposal, toolbox and benchmark.
\newblock In \emph{Evolutionary Multi-Criterion Optimization}, pages 64--78.
  Springer, 2015.

\bibitem[Hupkens et~al.(2014)Hupkens, Emmerich, and Deutz]{hupkens2014faster}
I.~Hupkens, M.~Emmerich, and A.~Deutz.
\newblock Faster computation of expected hypervolume improvement.
\newblock \emph{arXiv preprint arXiv:1408.7114}, 2014.

\bibitem[Jeong and Obayashi(2005)]{jeong2005}
S.~Jeong and S.~Obayashi.
\newblock Efficient global optimization (ego) for multi-objective problem and
  data mining.
\newblock In \emph{Evolutionary Computation, 2005. The 2005 IEEE Congress on},
  volume~3, pages 2138--2145, 2005.

\bibitem[Jeong et~al.(2006)Jeong, Minemura, and
  Obayashi]{jeong2006optimization}
S.~Jeong, Y.~Minemura, and S.~Obayashi.
\newblock Optimization of combustion chamber for diesel engine using kriging
  model.
\newblock \emph{Journal of Fluid Science and Technology}, 1\penalty0
  (2):\penalty0 138--146, 2006.

\bibitem[Jin(2011)]{jin2011surrogate}
Y.~Jin.
\newblock Surrogate-assisted evolutionary computation: Recent advances and
  future challenges.
\newblock \emph{Swarm and Evolutionary Computation}, 1\penalty0 (2):\penalty0
  61--70, 2011.

\bibitem[Johnson(2012)]{johnson2012nlopt}
S.~G. Johnson.
\newblock The nlopt nonlinear-optimization package (version 2.3).
\newblock \emph{URL http://ab-initio. mit. edu/nlopt}, 2012.

\bibitem[Jones et~al.(1998)Jones, Schonlau, and Welch]{jones1998efficient}
D.~R. Jones, M.~Schonlau, and W.~J. Welch.
\newblock Efficient global optimization of expensive black-box functions.
\newblock \emph{Journal of Global Optimization}, 13\penalty0 (4):\penalty0
  455--492, 1998.

\bibitem[Keane(2006)]{keane2006statistical}
A.~J. Keane.
\newblock Statistical improvement criteria for use in multiobjective design
  optimization.
\newblock \emph{AIAA journal}, 44\penalty0 (4):\penalty0 879--891, 2006.

\bibitem[Knowles(2006)]{knowles2006parego}
J.~Knowles.
\newblock Parego: a hybrid algorithm with on-line landscape approximation for
  expensive multiobjective optimization problems.
\newblock \emph{Evolutionary Computation, IEEE Transactions on}, 10\penalty0
  (1):\penalty0 50--66, 2006.

\bibitem[Knowles and Hughes(2005)]{knowles2005multiobjective}
J.~Knowles and E.~J. Hughes.
\newblock Multiobjective optimization on a budget of 250 evaluations.
\newblock In \emph{Evolutionary Multi-Criterion Optimization}, pages 176--190.
  Springer, 2005.

\bibitem[Kushner(1964)]{kushner1964new}
H.~J Kushner.
\newblock A new method of locating the maximum point of an arbitrary multipeak
  curve in the presence of noise.
\newblock \emph{Journal of Fluids Engineering}, 86\penalty0 (1):\penalty0
  97--106, 1964.

\bibitem[Li(2012)]{li2012thesis}
L.~Li.
\newblock \emph{Sequential Design of Experiments to Estimate a Probability of
  Failure.}
\newblock PhD thesis, Sup{\'e}lec, 2012.

\bibitem[Li et~al.(2012)Li, Bect, and Vazquez]{li2012bayesian}
L.~Li, J.~Bect, and E.~Vazquez.
\newblock {B}ayesian {S}ubset {S}imulation: a kriging-based subset simulation
  algorithm for the estimation of small probabilities of failure.
\newblock In \emph{Proceedings of PSAM 11 \& ESREL 2012, 25-29 June 2012,
  Helsinki, Finland}. IAPSAM, 2012.

\bibitem[Liu(2001)]{liu2008monte}
J.~S. Liu.
\newblock \emph{{M}onte {C}arlo strategies in scientific computing}.
\newblock Springer, 2001.

\bibitem[Loeppky et~al.(2009)Loeppky, Sacks, and Welch]{loeppky2009choosing}
J.~L. Loeppky, J.~Sacks, and W.~J. Welch.
\newblock Choosing the sample size of a computer experiment: A practical guide.
\newblock \emph{Technometrics}, 51\penalty0 (4), 2009.

\bibitem[Mockus(1975)]{mockus75}
J.~Mockus.
\newblock On {B}ayesian methods of optimization.
\newblock In \emph{Towards Global Optimization}, pages 166--181. North-Holland,
  1975.

\bibitem[Mockus(1989)]{mockus1989bayesian}
J.~Mockus.
\newblock \emph{{B}ayesian approach to global optimization: theory and
  applications}, volume~37.
\newblock Kluwer Academic Publishers, 1989.

\bibitem[Mockus et~al.(1978)Mockus, Tiesis, and \v{Z}ilinskas]{mockus78}
J.~Mockus, V.~Tiesis, and A.~\v{Z}ilinskas.
\newblock The application of {B}ayesian methods for seeking the extremum.
\newblock In L.~C.~W. Dixon and G{\'{a}}bor.~P. Szeg{\"{o}}, editors,
  \emph{Towards Global Optimization}, volume~2, pages 117--129, North Holland,
  New York, 1978.

\bibitem[Oyama et~al.(2007)Oyama, Shimoyama, and Fujii]{oyama2007new}
A.~Oyama, K.~Shimoyama, and K.~Fujii.
\newblock New constraint-handling method for multi-objective and
  multi-constraint evolutionary optimization.
\newblock \emph{Transactions of the Japan Society for Aeronautical and Space
  Sciences}, 50\penalty0 (167):\penalty0 56--62, 2007.

\bibitem[Parr et~al.(2012)Parr, Keane, Forrester, and Holden]{parr2012infill}
J.~M. Parr, A.~J. Keane, A.~I.~J. Forrester, and C.~M.~E. Holden.
\newblock Infill sampling criteria for surrogate-based optimization with
  constraint handling.
\newblock \emph{Engineering Optimization}, 44\penalty0 (10):\penalty0
  1147--1166, 2012.

\bibitem[Picheny(2014{\natexlab{a}})]{picheny2014}
V.~Picheny.
\newblock A stepwise uncertainty reduction approach to constrained global
  optimization.
\newblock In \emph{Proceedings of the 17th International Conference on
  Artificial Intelligence and Statistics (AISTATS), 2014, Reykjavik, Iceland.},
  volume~33, pages 787--795. JMLR: W\&CP, 2014{\natexlab{a}}.

\bibitem[Picheny(2014{\natexlab{b}})]{picheny2014MO}
V.~Picheny.
\newblock Multiobjective optimization using {G}aussian process emulators via
  stepwise uncertainty reduction.
\newblock \emph{Statistics and Computing},
  DOI:10.1007/s11222-014-9477-x:\penalty0 1--16, 2014{\natexlab{b}}.

\bibitem[Ponweiser et~al.(2008)Ponweiser, Wagner, Biermann, and
  Vincze]{ponweiser2008}
W.~Ponweiser, T.~Wagner, D.~Biermann, and M.~Vincze.
\newblock Multiobjective optimization on a limited budget of evaluations using
  model-assisted $\mathcal{S}$-metric selection.
\newblock In \emph{Parallel Problem Solving from Nature (PPSN X)}, volume 5199
  of \emph{Lecture Notes in Computer Science}, pages 784--794. Springer, 2008.

\bibitem[Powell(1994)]{powell1994direct}
M.~J.~D. Powell.
\newblock A direct search optimization method that models the objective and
  constraint functions by linear interpolation.
\newblock In \emph{Advances in optimization and numerical analysis}, pages
  51--67. Springer, 1994.

\bibitem[Ray et~al.(2001)Ray, Tai, and Seow]{ray2001multiobjective}
T.~Ray, K.~Tai, and K.~C. Seow.
\newblock Multiobjective design optimization by an evolutionary algorithm.
\newblock \emph{Engineering Optimization}, 33\penalty0 (4):\penalty0 399--424,
  2001.

\bibitem[Regis(2014)]{regis2014constrained}
R.~G. Regis.
\newblock Constrained optimization by radial basis function interpolation for
  high-dimensional expensive black-box problems with infeasible initial points.
\newblock \emph{Engineering Optimization}, 46\penalty0 (2):\penalty0 218--243,
  2014.

\bibitem[Robert and Casella(2004)]{robert2013monte}
C.~Robert and G.~Casella.
\newblock \emph{{M}onte {C}arlo statistical methods. Second Edition}.
\newblock Springer, 2004.

\bibitem[Roberts and Rosenthal(2009)]{roberts2009examples}
Gareth~O Roberts and Jeffrey~S Rosenthal.
\newblock Examples of adaptive mcmc.
\newblock \emph{Journal of Computational and Graphical Statistics}, 18\penalty0
  (2):\penalty0 349--367, 2009.

\bibitem[Santner et~al.(2003)Santner, Williams, and Notz]{santner2003design}
T.~J. Santner, B.~J. Williams, and W.~Notz.
\newblock \emph{The design and analysis of computer experiments}.
\newblock Springer, 2003.

\bibitem[Sasena(2002)]{sasena2002flexibility}
M.~J. Sasena.
\newblock \emph{Flexibility and efficiency enhancements for constrained global
  design optimization with kriging approximations}.
\newblock PhD thesis, University of Michigan, 2002.

\bibitem[Sasena et~al.(2002)Sasena, Papalambros, and
  Goovaerts]{sasena2002exploration}
M.~J. Sasena, P.~Papalambros, and P.~Goovaerts.
\newblock Exploration of metamodeling sampling criteria for constrained global
  optimization.
\newblock \emph{Engineering Optimization}, 34\penalty0 (3):\penalty0 263--278,
  2002.

\bibitem[Schonlau et~al.(1998)Schonlau, Welch, and Jones]{schonlau1998global}
M.~Schonlau, W.~J. Welch, and D.~R. Jones.
\newblock Global versus local search in constrained optimization of computer
  models.
\newblock In \emph{New Developments and Applications in Experimental Design:
  Selected Proceedings of a 1997 Joint AMS-IMS-SIAM Summer Conference},
  volume~34 of \emph{IMS Lecture Notes-Monographs Series}, pages 11--25.
  Institute of Mathematical Statistics, 1998.

\bibitem[Shimoyama et~al.(2013)Shimoyama, Sato, Jeong, and
  Obayashi]{shimoyama2013updating}
K.~Shimoyama, K.~Sato, S.~Jeong, and S.~Obayashi.
\newblock Updating kriging surrogate models based on the hypervolume indicator
  in multi-objective optimization.
\newblock \emph{Journal of Mechanical Design}, 135\penalty0 (9):\penalty0
  094503, 2013.

\bibitem[Snelson et~al.(2004)Snelson, Rasmussen, and
  Ghahramani]{snelson2004warped}
E.~Snelson, C.~E. Rasmussen, and Z.~Ghahramani.
\newblock Warped gaussian processes.
\newblock \emph{Advances in neural information processing systems},
  16:\penalty0 337--344, 2004.

\bibitem[Stein(1999)]{stein:99}
M.~L. Stein.
\newblock \emph{Interpolation of Spatial Data: Some Theory for {K}riging}.
\newblock Springer, 1999.

\bibitem[Svenson and Santner(2010)]{svenson2010multiobjective}
J.~D. Svenson and T.~J. Santner.
\newblock Multiobjective optimization of expensive black-box functions via
  expected maximin improvement.
\newblock Technical report, Tech. rep., 43210, Ohio University, Columbus, Ohio,
  2010.

\bibitem[Toal and Keane(2012)]{toal2012}
D.~J.~J. Toal and A.~J. Keane.
\newblock Non-stationary kriging for design optimization.
\newblock \emph{Engineering Optimization}, 44\penalty0 (6):\penalty0 741--765,
  2012.

\bibitem[Vazquez and Bect(2014)]{vazquez2014new}
E.~Vazquez and J.~Bect.
\newblock A new integral loss function for {B}ayesian optimization.
\newblock \emph{arXiv preprint arXiv:1408.4622}, 2014.

\bibitem[Villemonteix et~al.(2009)Villemonteix, Vazquez, and
  Walter]{villemonteix2009informational}
J.~Villemonteix, E.~Vazquez, and E.~Walter.
\newblock An informational approach to the global optimization of
  expensive-to-evaluate functions.
\newblock \emph{Journal of Global Optimization}, 44\penalty0 (4):\penalty0
  509--534, 2009.

\bibitem[Wagner et~al.(2010)Wagner, Emmerich, Deutz, and
  Ponweiser]{wagner2010expected}
T.~Wagner, M.~Emmerich, A.~Deutz, and W.~Ponweiser.
\newblock On expected-improvement criteria for model-based multi-objective
  optimization.
\newblock In \emph{Parallel Problem Solving from Nature, PPSN XI. 11th
  International Conference, Krakov, Poland, September 11-15, 2010, Proceedings,
  Part I}, volume 6238 of \emph{Lecture Notes in Computer Science}, pages
  718--727. Springer, 2010.

\bibitem[Williams et~al.(2010)Williams, Santner, Notz, and
  Lehman]{williams2010}
B.~J. Williams, T.~J. Santner, W.~I. Notz, and J.~S. Lehman.
\newblock Sequential design of computer experiments for constrained
  optimization.
\newblock In T.~Kneib and G.~Tutz, editors, \emph{Statistical Modelling and
  Regression Structures}, pages 449--472. Physica-Verlag HD, 2010.

\bibitem[Williams and Rasmussen(2006)]{williams2006gaussian}
C.~K.~I. Williams and C.~Rasmussen.
\newblock {G}aussian processes for machine learning.
\newblock \emph{the MIT Press}, 2\penalty0 (3):\penalty0 4, 2006.

\bibitem[Zhang et~al.(2010)Zhang, Liu, Tsang, and Virginas]{zhang2010expensive}
Q.~Zhang, W.~Liu, E.~Tsang, and B.~Virginas.
\newblock Expensive multiobjective optimization by {MOEA/D} with gaussian
  process model.
\newblock \emph{Evolutionary Computation, IEEE Transactions on}, 14\penalty0
  (3):\penalty0 456--474, 2010.

\bibitem[Zitzler et~al.(2002)Zitzler, Laumanns, and Thiele]{zitzler2001spea2}
E.~Zitzler, M.~Laumanns, and L.~Thiele.
\newblock {SPEA2}: Improving {S}trength {P}areto {E}volutionary {A}lgorithm.
\newblock In \emph{Evolutionary Methods for Design, Optimisation and Control
  with Application to Industrial Problems (EUROGEN 2001)}, pages 95--100.
  International Center for Numerical Methods in Engineering (CIMNE), 2002.

\end{thebibliography}

\afterpage{\clearpage}

\appendix
\section{On the bounded hyper-rectangles~$\Bo$ and~$\Bc$}
\label{sec:lowerBounds}

We have assumed in Section~\ref{sec:method} that~$\Bo$ and~$\Bc$ are \emph{bounded}
hyper-rectangles; that is,  sets of the form
\begin{align*}
  \Bo & = \bigl\{ y \in \Yo;\; \yLow_\obj \le y \le \yUpp_\obj \bigr\},\\
  \Bc & = \bigl\{ y \in \Yc;\; \yLow_\cons \le y \le \yUpp_\cons \bigr\},
\end{align*}
for some~$\yLow_\obj, \yUpp_\obj \in \Yo$ and $\yLow_\cons,\, \yUpp_\cons \in
\Yc$, with the additional assumption that $\yLow_{\cons, j} < 0 < \yUpp_{\cons,
  j}$ for all~$j \le q$. Remember that upper bounds only where required in the 
unconstrained case discussed in Section~\ref{sec:ei-multi}. To
shed some light on the role of these lower and upper bounds, let us now compute the
improvement~$I_1(X_1) = \left| H_1 \right|$ brought by a single evaluation. 

If $X_1$ is not feasible, then
\begin{equation}
  \label{eq:H1notFeas}
  \left| H_1 \right| = \left| \Bo \right| \;\cdot\; %
  \prod_{j = 1}^q
  \left( \yUpp_{\cons,j} - \yLow_{\cons,j} \right)^{\gamma_j}
  \left( \yUpp_{\cons,j} - \xi_{\cons,j}(X_1) \right)^{1 - \gamma_j}
\end{equation}
where $\gamma_j = \one_{\xi_{\cons,j}(X_1) \le 0}$. It is clear from the
right-hand side of~\eqref{eq:H1notFeas} that both~$\Bo$ and~$\Bc$ have to be
bounded if we want $\left| H_1 \right| < +\infty$ for any~$\gamma = \left(
  \gamma_1,, \ldots,\, \gamma_q \right) \in \{ 0, 1 \}^q$. Note, however, that
only the volume of~$\Bo$ actually matters in this expression, not the actual
values of~$\yLow_\obj$ and~$\yUpp_\obj$. Equation~\eqref{eq:H1notFeas} also reveals
that the improvement is a discontinuous function of the observations: indeed,
the $j^{\text{th}}$ term in the product jumps from~$\yUpp_{\cons,j}$
to~$\yUpp_{\cons,j} - \yLow_{\cons,j} > \yUpp_{\cons,j}$ when~$\xi_{\cons,j}(X_1)$
goes from~$0^+$ to~$0$. The increment $- \yLow_{\cons,j}$ can be thought of as a
reward  associated to finding a point which is feasible with respect
to the~$j^{\text{th}}$ constraint.

The value of~$\left| H_1 \right|$ when~$X_1$ is feasible is
\begin{equation}
  \label{eq:H1feas}
  \left| H_1 \right| =%
  \left| \Bo \right| \,\cdot\, \left(
    \left| \Bc \right| -  \left| \BcNeg \right|
  \right) \;+\;
  \prod_{j \le p} \left( %
    \min\left( \xi_{\obj,j}(X_1), \yUpp_{\obj,j} \right) %
    - \max\left( \xi_{\obj,j}(X_1), \yLow_{\obj,j} \right) %
  \right) \,\cdot\, \left| \BcNeg \right|,
\end{equation}
where $ \left| \BcNeg \right| = \prod_{j=1}^q \left| \yLow_{\cons,j}
\right|$ is the volume of the feasible subset of~$\Bc$, $\BcNeg = \Bc
\cap \left] -\infty; 0 \right]^q$. The first term in the right-hand side
of~\eqref{eq:H1feas} is the improvement associated to the domination of
the entire unfeasible subset of~$\B = \Bo \times \Bc$; the second term
measures the improvement in the space of objective values.

\section{An adaptive procedure to set $\Bo$ and $\Bc$}
\label{sec:annexe:adaptBoBc}

This section describes the adaptive numerical procedure that is used, in our
numerical experiments, to define the hyper-rectangles~$\Bo$
and~$\Bc$. As said in Section~\ref{sec:critDecomposition}, these
hyper-rectangles are defined using estimates of the range of the objective
and constraint functions, respectively. To this end, we will use the available
evaluations results, together with posterior quantiles provided by our Gaussian
process models on the set of candidate points~$\mathcal{X}_n$ (defined in
Section~\ref{sec:crit-optim-using}).

More precisely, assume that $n$ evaluation results~$\xi(X_i)$, $1 \le i \le n$,
are available. Then, we define the corners of~$\Bo$ by
\begin{equation}
  \label{eq:bounds}
  \left\{
    \begin{array}{l c l}
      \yLow_{\obj,i,n} &=& \min \left(
        \min_{i \le n} \xi_{\obj,i}(X_i),\;
        \min_{x \in \mathcal{X}_n} \hat\xi_{\obj,\,i,\,n}(x)-\lambdaObj \sigma_{\obj,\,i,\,n}(x)
      \right), \\
      \yUpp_{\obj,i,n} &=& \max \left(
        \max_{i \le n} \xi_{\obj,i}(X_i),\;
        \max_{x \in \mathcal{X}_n} \hat\xi_{\obj,\,i,\,n}(x)+\lambdaObj \sigma_{\obj,\,i,\,n}(x)
      \right),
    \end{array}
  \right.
\end{equation}
for $1 \leq i \leq p$, and the corners of~$\Bc$ by
\begin{equation}
  \label{eq:bounds}
  \left\{
    \begin{array}{l c l}
      \yLow_{\cons,j,n} &=& \min \left( 0,\; 
        \min_{i \le n} \xi_{\cons,j}(X_i),\;
        \min_{x \in \mathcal{X}_n} \hat\xi_{\cons,\,j,\,n}(x) - \lambdaCons \sigma_{\cons,\,j,\,n}(x)
      \right), \\
      \yUpp_{\cons,j,n} &=& \max \left( 0,\; 
        \max_{i \le n} \xi_{\cons,j}(X_i),\;
        \max_{x \in \mathcal{X}_n} \hat\xi_{\cons,\,j,\,n}(x) + \lambdaCons \sigma_{\cons,\,j,\,n}(x)
      \right),
    \end{array}	
  \right.
\end{equation}
for $1 \leq j \leq q$, where $\lambdaObj$ and $\lambdaCons$ are positive numbers.

\section{Mono-objective benchmark result tables}
\label{sec:annexe:mono-bench-results}

In Section~\ref{sec:so_bench}, only the best results for both the
``Local'' and the ``Regis'' groups of algorithms were shown. In this
Appendix, we present the full results.
Tables~\ref{tab:local_feasible_full} and \ref{tab:local_target_full},
and Tables \ref{tab:regis_feasible_full} and \ref{tab:regis_target_full}
present respectively the results obtained with the local optimization
algorithms and the results obtained by \cite{regis2014constrained} on
the single-objective benchmark test problems (see
Table~\ref{tab:problem_so}). Table~\ref{tab:local_feasible_full} and
Table~\ref{tab:local_target_full} show the performances for finding
feasible solutions and for reaching the targets specified in
Table~\ref{tab:problem_so} for the COBYLA, Active-Set, Interior-Point
and SQP algorithms. Similarly, Table~\ref{tab:regis_feasible_full} and
Table~\ref{tab:regis_target_full} show the performances for finding
feasible solutions and for reaching the targets for the COBRA-Local,
COBRA-Global and Extended-ConstrLMSRBF algorithms of
\cite{regis2014constrained}.

\begin{table}
	\rowcolors{1}{white}{gray!10}
	\centering
  \renewcommand{\arraystretch}{1.2}
\begin{tabular}{|l|c|c|c|c|c|c|c|c|} 
\hline 
Pbm & \multicolumn{2}{c|}{COBYLA} & \multicolumn{2}{c|}{active-set} & \multicolumn{2}{c|}{interior-point} & \multicolumn{2}{c|}{SQP}\\ 
\hline 
g1 & 30 & 52.3 (102.3) & 30 & 15.0 (0.0) & 30 & 128.4 (27.8) & 30 & 15.0 (0.0)\\ 
g3mod & 28 & 386.1 (645.8) & 30 & 643.2 (248.9) & 30 & 342.3 (66.3) & 30 & 794.3 (53.7)\\ 
g5mod & 22 & 30.7 (23.0) & 30 & 35.0 (5.5) & 30 & 41.3 (16.9) & 30 & 38.5 (10.5)\\ 
g6 & 26 & 39.7 (12.7) & 30 & 29.7 (5.0) & 30 & 99.7 (14.3) & 30 & 32.6 (5.4)\\ 
g7 & 28 & 162.4 (175.7) & 30 & 109.4 (11.2) & 30 & 146.0 (18.1) & 30 & 107.6 (9.3)\\ 
g8 & 28 & 53.3 (77.1) & 28 & 17.6 (5.0) & 30 & 12.1 (7.7) & 30 & 19.6 (8.5)\\ 
g9 & 25 & 95.2 (104.7) & 30 & 313.7 (84.4) & 30 & 170.9 (42.9) & 30 & 194.5 (60.2)\\ 
g10 & 2 & 14.5 (3.5) & 9 & 53.6 (41.9) & 12 & 469.8 (393.8) & 25 & 144.6 (132.3)\\ 
g13mod & 30 & 53.9 (68.8) & 30 & 74.0 (59.5) & 30 & 21.4 (17.1) & 30 & 69.4 (62.4)\\ 
g16 & 27 & 31.5 (20.4) & 30 & 38.0 (15.0) & 22 & 100.9 (160.3) & 30 & 40.7 (17.1)\\ 
g18 & 26 & 345.0 (275.7) & 30 & 114.5 (41.5) & 30 & 70.3 (22.2) & 30 & 101.9 (19.8)\\ 
g19 & 19 & 31.4 (19.5) & 30 & 21.8 (7.5) & 30 & 291.3 (57.9) & 30 & 19.7 (6.1)\\ 
g24 & 30 & 7.7 (10.2) & 30 & 5.2 (5.3) & 30 & 4.0 (3.5) & 30 & 5.1 (5.2)\\ 
SR7 & 29 & 30.0 (50.1) & 30 & 27.5 (3.9) & 30 & 78.6 (23.1) & 30 & 27.1 (3.6)\\ 
WB4 & 27 & 71.8 (82.5) & 30 & 125.7 (71.0) & 30 & 93.5 (48.9) & 30 & 76.6 (21.9)\\ 
PVD4 & 12 & 50.8 (70.2) & 3 & 51.3 (27.7) & 30 & 59.1 (43.5) & 26 & 7.6 (4.8)\\ 
\hline 
\end{tabular} 
\caption{Number of evaluations to find a first feasible point for the
  COBYLA, Active-Set, Interior-Point and SQP local optimization
  algorithms. See Table~\ref{tab:res1} for conventions.} 
\label{tab:local_feasible_full}
\end{table} 

\begin{table}[tp]
	\rowcolors{1}{white}{gray!10}
	\centering
  \renewcommand{\arraystretch}{1.2}
\begin{tabular}{|l|c|c|c|c|c|c|c|c|} 
\hline 
Pbm & \multicolumn{2}{c|}{COBYLA} & \multicolumn{2}{c|}{active-set} & \multicolumn{2}{c|}{interior-point} & \multicolumn{2}{c|}{SQP}\\ 
\hline 
g1 & 7 & 212.9 (225.8) & 6 & 22.0 (7.7) & 20 & 349.7 (57.0) & 6 & 22.0 (7.7)\\ 
g3mod & 16 & 1312.3 (1123.6) & 24 & 760.5 (79.8) & 30 & 356.9 (65.1) & 30 & 794.3 (53.7)\\ 
g5mod & 22 & 53.4 (20.3) & 30 & 35.8 (4.3) & 30 & 54.8 (11.7) & 30 & 41.8 (7.5)\\ 
g6 & 26 & 41.0 (11.1) & 30 & 29.7 (5.0) & 30 & 99.7 (14.3) & 30 & 32.6 (5.4)\\ 
g7 & 20 & 495.5 (461.3) & 30 & 109.4 (11.2) & 30 & 147.2 (18.2) & 30 & 107.6 (9.3)\\ 
g8 & 4 & 79.5 (84.6) & 2 & 30.5 (2.1) & 18 & 59.3 (87.0) & 4 & 55.8 (27.0)\\ 
g9 & 22 & 144.9 (143.7) & 30 & 334.5 (84.0) & 30 & 179.3 (42.0) & 30 & 194.5 (60.2)\\ 
g10 & 0 & - (-) & 0 & - (-) & 0 & - (-) & 18 & 658.3 (316.7)\\ 
g13mod & 23 & 191.9 (209.7) & 24 & 153.9 (46.6) & 25 & 122.5 (70.3) & 22 & 147.6 (75.1)\\ 
g16 & 27 & 60.0 (65.2) & 14 & 85.1 (41.1) & 13 & 400.0 (242.1) & 30 & 152.2 (53.2)\\ 
g18 & 14 & 383.0 (389.3) & 21 & 101.0 (30.2) & 21 & 149.1 (39.4) & 21 & 97.5 (23.8)\\ 
g19 & 16 & 912.1 (685.8) & 30 & 61.3 (12.4) & 30 & 335.5 (65.4) & 30 & 61.3 (12.4)\\ 
g24 & 18 & 17.5 (8.9) & 17 & 14.7 (3.9) & 16 & 10.4 (5.3) & 17 & 16.4 (5.3)\\
SR7 & 28 & 62.5 (52.1) & 30 & 27.5 (3.9) & 30 & 80.2 (22.1) & 30 & 27.1 (3.6)\\
WB4 & 24 & 247.1 (176.2) & 29 & 162.0 (73.1) & 30 & 168.2 (94.4) & 30 & 78.3 (18.0)\\ 
PVD4 & 2 & 58.0 (35.4) & 3 & 54.0 (25.1) & 26 & 146.7 (115.2) & 23 & 54.7 (27.5)\\ 
\hline 
\end{tabular} 
\caption{Number of evaluations to reach the target for the COBYLA,
  Active-Set, Interior-Point and SQP local optimization algorithms. See Table~\ref{tab:res1} for conventions.} 
\label{tab:local_target_full}
\end{table} 

\begin{table} 
	\rowcolors{1}{white}{gray!10}
	\centering
  \renewcommand{\arraystretch}{1.2}
\begin{tabular}{|l|c|c|c|c|c|c|} 
\hline 
Pbm & \multicolumn{2}{c|}{COBRA-Local} & \multicolumn{2}{c|}{COBRA-Global} & \multicolumn{2}{c|}{Extended-ConstrLMSRBF}\\ 
\hline 
g1 & 30 & 15.0 (0.0) & 30 & 15.0 (0.0) & 30 & 19.1 (0.4)\\ 
g3mod & 30 & 23.5 (0.2) & 30 & 23.5 (0.2) & 30 & 31.2 (0.3)\\ 
g5mod & 30 & 6.4 (0.1) & 30 & 6.4 (0.1) & 30 & 9.6 (0.3)\\ 
g6 & 30 & 10.9 (0.3) & 30 & 10.9 (0.3) & 30 & 11.9 (0.2)\\ 
g7 & 30 & 47.5 (4.6) & 30 & 47.5 (4.7) & 30 & 39.8 (2.9)\\ 
g8 & 30 & 6.5 (0.2) & 30 & 6.5 (0.2) & 30 & 5.2 (0.2)\\ 
g9 & 30 & 21.5 (1.9) & 30 & 21.5 (1.9) & 30 & 23.1 (2.3)\\ 
g10 & 30 & 22.8 (1.5) & 30 & 22.8 (1.5) & 30 & 51.1 (6.5)\\ 
g13mod & 30 & 9.4 (0.8) & 30 & 9.4 (0.8) & 30 & 8.6 (0.7)\\ 
g16 & 30 & 14.7 (2.4) & 30 & 14.7 (2.4) & 30 & 19.6 (1.8)\\ 
g18 & 30 & 108.6 (6.5) & 30 & 108.6 (6.5) & 30 & 122.0 (5.6)\\ 
g19 & 30 & 16.5 (0.5) & 30 & 16.5 (0.5) & 30 & 20.8 (0.8)\\ 
g24 & 30 & 1.3 (0.1) & 30 & 1.3 (0.1) & 30 & 1.3 (0.1)\\ 
SR7 & 30 & 9.5 (0.1) & 30 & 9.5 (0.1) & 30 & 12.4 (0.4)\\ 
WB4 & 30 & 37.4 (5.9) & 30 & 37.4 (5.9) & 30 & 25.0 (4.1)\\ 
PVD4 & 30 & 7.9 (0.4) & 30 & 7.9 (0.4) & 30 & 10.4 (0.7)\\ 
\hline 
\end{tabular} 
\caption{Number of evaluations to find a first feasible point for the
  COBRA-Local, COBRA-Global and Extended-ConstrLMSRBF optimization
  algorithms. These results are taken from
  \citep{regis2014constrained}.See Table~\ref{tab:res1} for conventions.} 
\label{tab:regis_feasible_full}
\end{table} 

\begin{table}[tp] 
	\rowcolors{1}{white}{gray!10}
	\centering
  \renewcommand{\arraystretch}{1.2}
\begin{tabular}{|l|c|c|c|c|c|c|} 
\hline 
Pbm & \multicolumn{2}{c|}{COBRA-Local} & \multicolumn{2}{c|}{COBRA-Global} & \multicolumn{2}{c|}{Extended-ConstrLMSRBF}\\ 
\hline 
g1 & 7 & 387.8 (-) & 30 & 125.2 (15.3) & 0 & > 500 (-)\\ 
g3mod & 6 & 451.1 (-) & 6 & 440.0 (-) & 30 & 141.7 (8.6)\\ 
g5mod & 30 & 12.9 (0.5) & 30 & 16.6 (1.8) & 30 & 40.3 (1.4)\\ 
g6 & 30 & 53.6 (14.0) & 30 & 62.5 (10.5) & 26 & 101.2 (-)\\ 
g7 & 30 & 199.5 (20.7) & 30 & 99.8 (5.7) & 30 & 264.5 (34.2)\\ 
g8 & 30 & 30.3 (2.8) & 30 & 31.2 (2.5) & 30 & 46.2 (6.2)\\ 
g9 & 28 & 275.5 (-) & 30 & 176.4 (26.3) & 29 & 294.0 (-)\\ 
g10 & 30 & 276.4 (43.6) & 29 & 193.7 (-) & 24 & 394.3 (-)\\ 
g13mod & 30 & 221.7 (35.6) & 30 & 169.0 (19.1) & 30 & 146.4 (29.2)\\ 
g16 & 30 & 38.8 (9.3) & 30 & 46.3 (13.5) & 30 & 38.4 (3.6)\\ 
g18 & 24 & 195.9 (-) & 23 & 212.8 (-) & 21 & 276.0 (-)\\ 
g19 & 30 & 698.5 (75.3) & 30 & 850.9 (70.6) & 0 & > 1000 (-)\\ 
g24 & 30 & 9.0 (0.0) & 30 & 9.0 (0.0) & 30 & 91.9 (6.0)\\ 
SR7 & 30 & 35.0 (2.7) & 30 & 33.5 (1.6) & 0 & > 500 (-)\\ 
WB4 & 30 & 164.6 (12.2) & 30 & 202.0 (13.0) & 30 & 238.6 (20.0)\\ 
PVD4 & 28 & 212.2 (-) & 30 & 155.4 (38.2) & 29 & 263.5 (-)\\ 
\hline 
\end{tabular} 
\caption{Number of evaluations to reach the target for the COBRA-Local,
  COBRA-Global and Extended-ConstrLMSRBF optimization algorithms. These
  results are taken from \citep{regis2014constrained}. See Table~\ref{tab:res1} for conventions.} 
\label{tab:regis_target_full}
\end{table} 

\section{Modified g3mod, g10 and PVD4 test problems}
\label{sec:annexe:modified-problems}

We detail here the modified formulations of the g3mod, g10 and PVD4
problems that were used in Section~\ref{sec:so_bench} to overcome the
modeling problems with BMOO. Our modifications are shown in
boldface. The rationale of the modifications is to smooth local jumps.

\begin{itemize}
\item modified-g3mod problem
\begin{equation*}
\left\{
\begin{array}{lcl}
	f(x) &=& -\text{plog}((\sqrt{d})^d\prod_{i=1}^d x_i)^{\bm{0.1}}\\
	c(x) &=& (\sum_{i=1}^d x_i^2) - 1
\end{array}
\right.
\end{equation*}

\item modified-g10 problem

\begin{equation*}
\left\{
\begin{array}{lcl}
	f(x) &=& x_1 + x_2 + x_3\\
	c_1(x) &=& 0.0025(x_4+x_6) - 1\\
	c_2(x) &=& 0.0025(x_5+x_7-x_4) - 1\\
	c_3(x) &=& 0.01(x_8-x_5) - 1\\
	c_4(x) &=& \text{plog}(100x_1 - x_1x_6 + 833.33252x_4 - 83333.333)^{\bm{7}}\\
	c_5(x) &=& \text{plog}(x_2x_4 - x_2x_7 -1250x_4 + 1250x_5)^{\bm{7}}\\
	c_6(x) &=& \text{plog}(x_3x_5 - x_3x_8 -2500x_5 + 1250000)^{\bm{7}}
\end{array}
\right.
\end{equation*}

\item modified-PVD4 problem

\begin{equation*}
\left\{
\begin{array}{lcl}
	f(x) &=& 0.6224x_1x_3x_4 + 1.7781x_2x_3^2 + 3.1661x_1^2x_4 + 19.84x_1^2x_3\\
	c_1(x) &=& -x_1 + 0.0193x_3\\
	c_2(x) &=& -x_2 + 0.00954x_3\\
	c_3(x) &=& \text{plog}(-\pi x_3^2x_4 - 4/3\pi x_3^3 + 1296000)^{\bm{7}}
\end{array}
\right.
\end{equation*}
\end{itemize}

Note that the above defined problems make use of the plog function defined below (see \cite{regis2014constrained}).

\begin{equation*}
\text{plog}(x) = 
\left\{
\begin{array}{ll}
	\log(1+x) & \text{if\ } x \geq 0\\
	-\log(1-x) & \text{otherwise}
\end{array}
\right.
\end{equation*}

\end{document}